\documentclass[a4paper,11pt]{article}
\pdfoutput=1 
\usepackage{jheppub} 
\usepackage[T1]{fontenc} 

\usepackage[config=altsf]{subfig}
\usepackage{xcolor}
\usepackage{makecell}
\usepackage{multirow}
\usepackage{gensymb}
\usepackage{upgreek}
\usepackage{comment}
\usepackage{lineno}

\graphicspath{{fig/}}
\newcommand{\tabincell}[2]{\begin{tabular}{@{}#1@{}}#2\end{tabular}}  

\newcommand{\Bo}{$B^{0}$}
\newcommand{\Bs}{$B^{0}_{s}$}
\newcommand{\Bos}{$B^{0}_{(s)}$}

\newcommand{\Bopio}{$B^{0} \to \pi^{0}\pi^{0}$}
\newcommand{\Bspio}{$B^{0}_{s} \to \pi^{0}\pi^{0}$}
\newcommand{\Bospio}{$B^{0}_{(s)} \to \pi^{0}\pi^{0}$}

\newcommand{\Boeta}{$B^{0} \to \eta\eta$}
\newcommand{\Bseta}{$B^{0}_{s} \to \eta\eta$}
\newcommand{\Boseta}{$B^{0}_{(s)} \to \eta\eta$}

\newcommand{\pio}{$\pi^{0}$}
\newcommand{\piogamma}{$\pi^{0}\to \gamma\gamma$}
\newcommand{\etagamma}{$\eta \to \gamma\gamma$}
\newcommand{\SigmaB}{$\sigma_{m_{B}}$}
\newcommand{\MPioPio}{$m_{\pi^{0}\pi^{0}}$}
\newcommand{\MEtaEta}{$m_{\eta\eta}$}
\newcommand{\qq}{$q\bar{q}$}
\newcommand{\bb}{$b\bar{b}$}
\newcommand{\Zqq}{$Z\to q\bar{q}$}
\newcommand{\Zbb}{$Z\to b\bar{b}$}

\title{\boldmath Prospects for $B^0_{(s)}\to\pi^0\pi^0$ and $B^0_{(s)}\to\eta\eta$ modes and corresponding $CP$ asymmetries at Tera-$Z$}
\author[a,b]{Yuexin Wang,}
\author[c]{Sébastien Descotes-Genon,}
\author[d]{Olivier Deschamps,}
\author[e]{Lingfeng Li,}
\author[a,b]{Shanzhen Chen,}
\author[a,b]{Yongfeng Zhu}
\author[a,b,1]{and Manqi Ruan\note{Corresponding author.}}

\affiliation[a]{Institute of High Energy Physics, Chinese Academy of Sciences,\\Beijing 100049, China}
\affiliation[b]{University of Chinese Academy of Sciences (UCAS),\\Beijing 100049, China}
\affiliation[c]{Universit\'e Paris-Saclay, CNRS/IN2P3, IJCLab,\\ 91405 Orsay, France}
\affiliation[d]{Universit\'e Clermont Auvergne, CNRS/IN2P3, LPC,\\ Clermont-Ferrand, France}
\affiliation[e]{Physics Department, Brown University,\\Providence, RI 02912, USA}

\emailAdd{wangyuexin@ihep.ac.cn}
\emailAdd{sebastien.descotes-genon@ijclab.in2p3.fr}
\emailAdd{olivier.deschamps@clermont.in2p3.fr}
\emailAdd{lingfeng\_li@brown.edu}
\emailAdd{shanzhen.chen@ihep.ac.cn}
\emailAdd{zhuyf@ihep.ac.cn}
\emailAdd{manqi.ruan@ihep.ac.cn}
\abstract{The physics potential of measuring $B^0_{(s)}\to\pi^0\pi^0$ and $B^0_{(s)}\to\eta\eta$ decays via four-photon final states at Tera-$Z$ phase of CEPC or FCC-ee is investigated in this paper. 
We propose an electromagnetic calorimeter (ECAL) with both high energy resolution and excellent separation power to efficiently reconstruct $\pi^0$ and $\eta$ from hadronic final states with high photon multiplicity. The resulting $B$-meson mass resolution is approximately 30\,MeV, allowing 3\,$\sigma$ separation between $B^0$ and $B_s^0$. 
With the assistance of the $b$-jet tagging, the relative sensitivities to $B^0\to\pi^0\pi^0$, $B^0_s\to\pi^0\pi^0$, $B^0\to\eta\eta$, and $B^0_s\to\eta\eta$ signal strengths at Tera-$Z$ are projected as 0.45\%, 4.5\%, 18\%, and 0.95\%, respectively. Their dependence on various detector performances is also discussed. In addition, $B^0\to\pi^0\pi^0$ and its two isospin-related modes are paid special attention due to their roles in the determination of the CKM angle $\alpha$ ($\phi_2$). The anticipated precisions of their branching-ratio and $CP$-asymmetry measurements at Tera-$Z$ are evaluated. We show that the measurement of the time-integrated $B^0\to\pi^0\pi^0$ $CP$ asymmetry at Tera-$Z$ is complementary to $B$-factory ones. The precision on $\alpha$ combining $Z$- and $B$-factory results reaches $0.4^\circ$, lower than the systematic uncertainties attached to isospin breaking.}
\arxivnumber{2208.08327}
\makeatletter
\gdef\@fpheader{}
\makeatother

\begin{document}
\maketitle
\flushbottom
\section{Introduction}
\label{sec:intro}

It has been long acknowledged that the Tera-$Z$ phase of future circular $e^+e^-$ colliders will hold a unique position for electroweak precision physics~\cite{CEPC_CDR_Phy,FCC:2018byv}. With a vast amount ($\gtrsim 10^{12}$) of on-shell $Z$ produced, the Tera-$Z$ will reach unprecedented sensitivities on $Z$ properties and corresponding electroweak precision observables. The information gathered will be crucial to test the consistency of the Standard Model (SM) and potentially to provide evidence for new physics. More recently, the outstanding flavor physics potential at Tera-$Z$ has also been investigated~\cite{Zheng:2020ult,Li:2022tov,Mingrui,Aleksan:2021gii,Aleksan:2021fbx,Amhis:2021cfy,Kwok_FCCC,Kamenik:2017ghi,Li:2020bvr,Monteil:2021ith,Chrzaszcz:2021nuk,Dam:2018rfz,Qin:2017aju,Li:2018cod,Calibbi:2021pyh}. The rich phenomenology of flavored final states offers a complementary way to understand the SM better and scrutinize the physics beyond the Standard Model (BSM) other than through electroweak precision measurements. Thanks to the large $\sigma (e^+e^-\to Z \to b\bar{b},~c\bar{c},~\tau^+\tau^-)$ and a high integrated luminosity, Tera-$Z$ will produce large statistics of flavored hadrons and $\tau$ leptons comparable to other proposed flavor physics experiments. Table~\ref{tab:BYield} summarizes the expected numbers of $b$-hadrons produced at Belle~II~\cite{BelleII2019}, LHCb Upgrade II~\cite{LHCb:2018roe}, and Tera-$Z$. In addition, $m_Z$ is significantly larger than $m_{b,c,\tau}$, creating a surplus of energy and generating various hadronic final states. Even soft decay products of flavored particles will be boosted to higher energies and larger displacements, enhancing the measurement precision. Moreover, the cleanness of a lepton collider helps to investigate rare decay modes that contain neutral/invisible particles. Therefore, understanding the flavor physics at Tera-$Z$ not only complements the studies at Belle II and LHCb, but also strengthens the physics case for future circular $e^+e^-$ colliders.  

\begin{table}[thp]
	\centering
	\begin{tabular}[t]{|c|c|c|l|}
		\hline
		$b$-hadrons
		& Belle II  & LHCb (300 fb$^{-1}$)  & Tera-$Z$ \\
		\hline
		$B^{0}$, $\bar{B}^{0}$
		& $5.4 \times 10^{10}$ (50 ab$^{-1}$ on $\Upsilon(4S)$)     & $3 \times 10^{13}$    & $1.2 \times 10^{11}$ \\
		$B^{\pm}$
		& $5.7 \times 10^{10}$ (50 ab$^{-1}$ on $\Upsilon(4S)$)     & $3 \times 10^{13}$    & $1.2 \times 10^{11}$ \\
		$B^{0}_{s}$, $\bar{B}^{0}_{s}$
		& $6.0 \times 10^{8}$ (5 ab$^{-1}$ on $\Upsilon(5S)$)       & $1 \times 10^{13}$    & $3.1 \times 10^{10}$ \\
		$B^{\pm}_{c}$
		& -                                                         & $1 \times 10^{11}$    & $1.8 \times 10^{8}$ \\
		$\Lambda_{b}^{0}$, $\bar{\Lambda}_{b}^{0}$
		& -                                                         & $2 \times 10^{13}$    & $2.5 \times 10^{10}$ \\
		\hline
	\end{tabular}
	\caption{Expected yields of $b$-hadrons at Belle II, LHCb Upgrade II, and Tera-$Z$. The cross sections for \bb\ productions at $E_{\rm cm}(\Upsilon(4S))$ and $E_{\rm cm}(\Upsilon(5S))$ are taken from~\cite{BelleII2019}. The $b$-quark production cross section in the acceptance of LHCb is taken from~\cite{LHCb_CX}. We use the production fractions of $B^0_s$ and $\Lambda^0_b$ in~\cite{LHCb_bfraction} and assume $f_u + f_d + f_s + f_{\rm baryon} = 1$, $f_u = f_d$, and $f_{\Lambda^0_b} = f_{\rm baryon}$ to estimate the production fractions of $B^0$ and $B^{\pm}$ at LHCb. The production fractions of $B^0$, $B^{\pm}$, $B^0_s$, and $\Lambda^0_b$ in $Z$ decays are taken from~\cite{HFLAV2021}. As for $B_c$ meson, its production fraction at the $Z$-pole (including the contribution from $B_c^*$ decays) is taken from~\cite{Zpole_BcFraction}, while its production fraction at LHCb is taken from~\cite{LHCb_BcFraction}. }
	\label{tab:BYield}
\end{table}

There are currently two candidates for Tera-$Z$, the Circular Electron Positron Collider (CEPC)~\cite{CEPC_CDR_Acc, CEPC_CDR_Phy} and the Future Circular Collider (FCC) with its electron-positron mode (FCC-ee)~\cite{FCC:2018byv,FCC:2018evy,Bernardi:2022hny}. 
Both of them are primarily proposed as future Higgs factories, but they also provide unique opportunities for flavor physics as $Z$-factories beyond the standard Tera-$Z$.
CEPC is initially designed to operate at the $Z$-pole with high luminosity to collect $0.7 \times 10^{12}$ $Z$ bosons~\cite{CEPC_CDR_Acc, CEPC_CDR_Phy}. With the recent updates on the design~\cite{Gao:2022lew}, CEPC is expected to deliver $\gtrsim 3 \times 10^{12}$ $Z$ bosons in two years~\cite{CEPCPhysicsStudyGroup:2022uwl}. FCC-ee is proposed to run at $\sqrt{s}$ ranging from 91.2 to 365\,GeV. During a four-year $Z$-pole run, about $5\times 10^{12}$ $Z$ bosons will be produced in total at FCC-ee. 
Detector designs and technologies for both colliders are similar, allowing us to uniformly evaluate their physics potential.

\begin{figure}[tbp]
	\centering
	\includegraphics[width=.7\textwidth]{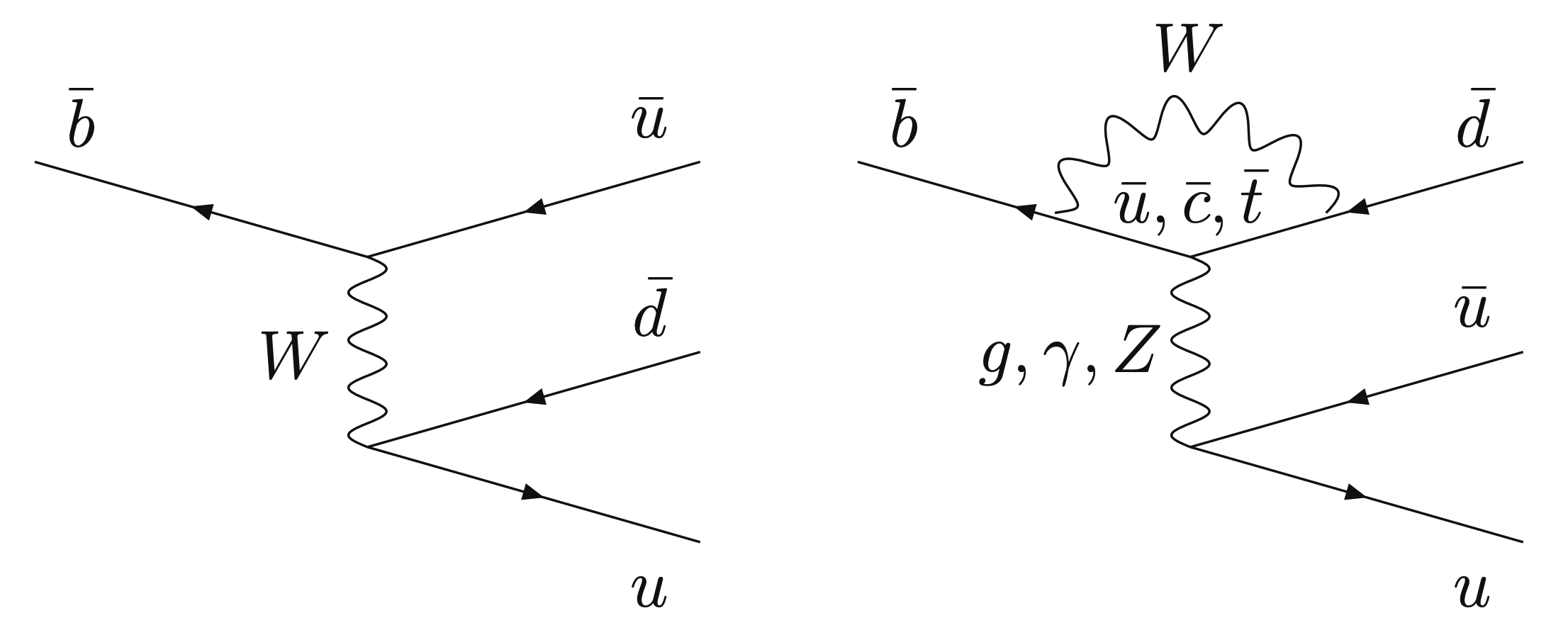}
	\caption{Tree (left) and QCD/electroweak penguin (right) diagrams for the weak transition $\bar{b} \to u\bar{u}\bar{d}$. Both plots are adapted from~\cite{IsospinAna}.}
	\label{fig:FD}
\end{figure}

In this paper, we focus on four neutral charmless $B$ decay channels, \Bospio\ and \Boseta\footnote{Charge-conjugate states $\bar{B}^{0}$ and $\bar{B}^{0}_{s}$ are implied throughout the paper.}, to exploit the physics potential of Tera-$Z$ and to get hints on the requirement for the detector performance. Since almost 98.8\% neutral pions decay into two photons~\cite{PDG2020}, and $\mathcal{B}$(\etagamma) is nearly 40\%~\cite{PDG2020}, we only focus on the di-photon decay of \pio\ and $\eta$ in this work. The precise measurements of these four channels are highly challenging due to the low branching ratios ($\lesssim \mathcal{O}(10^{-5})$) as well as the difficulty of reconstructing the fully neutral final state. The latest experimental result of $\mathcal{B}$(\Bopio) was given by Belle, with a signal significance of 6.4\,$\sigma$~\cite{Belle:2017lyb}. In contrast, \Bspio\ has not been observed in any experiment to date. An upper limit of $\mathcal{B}$(\Bspio) was given by the L3 experiment at LEP, which is $2.1 \times 10^{-4}$~\cite{L3}. Similarly, the other two modes, \Boeta\ and \Bseta, have not been observed experimentally. The SM predicted branching ratios of \Boeta\ and \Bseta\ are $\mathcal{O}(10^{-7})$~\cite{B0To2Eta_QCDF_2009, B0To2Eta_QCDF_2003, B0To2Eta_pQCD} and $\mathcal{O}(10^{-5})$~\cite{BsTo2Eta_QCDF, BsTo2Eta_pQCD}, respectively. It is hence necessary to update our knowledge on these charmless two-body decays. 
Besides, the $CP$ asymmetry of \Bopio\ mode is essential for the determination of the CKM angle $\alpha$ ($\phi_2$), which plays a fundamental role in the study of the SM flavor sector~\cite{Gronau:1990ka,IsospinAna,Descotes-Genon:2017thz}. Indeed, $\alpha$ can be determined by an isospin analysis of $B\to\pi\pi$ decay modes, which can be described by the interference between two topologies of $\bar{b} \to u\bar{u}\bar{d}$ shown in figure \ref{fig:FD}. Currently, this isospin analysis involves only a subset of the observables that could be exploited, since the mixing-induced $CP$ asymmetry of \Bopio\ ($S_{CP}^{00}$) has not been measured yet. Moreover, as can be seen from table~13 of ref.~\cite{IsospinAna}, the determination of $\alpha$ is significantly limited by our current knowledge of the direct $CP$ asymmetry of \Bopio\ ($C_{CP}^{00}$). The measurement of \Bopio\ at Tera-$Z$ will actually be the first phenomenological attempt to determine $\alpha$ at the $Z$-pole. Let us add that this mode involves electroweak loop (penguin) contributions which can be affected by new physics in $b\to dq\bar{q}$ transitions. 
In addition to the determination of $\alpha$, such two-body charmless $B$ decays also offer a chance to test and improve our approaches of hadron physics, exploiting the scale separation between $\Lambda_{\rm QCD}$ and $m_b$ with QCD factorization~\cite{Chiang:2006ih,B0To2Eta_QCDF_2009,Wang:2013fya} or Soft-Collinear Effective Theory (SCET)~\cite{SCET}, perturbative QCD~\cite{Xiao:2014uza,BsTo2Eta_pQCD,Chai:2022kmq}, as well as flavor symmetry through flavor diagram~\cite{Cheng:2014rfa} methods.
Furthermore, being rare charmless two-body decays, the companion \Bspio\ and \Boseta\ channels provide useful information on lesser known aspects of nonleptonic decays, namely the size of weak annihilation on one side and the $\eta(')$ dynamics on the other side. Their SM predictions including $CP$ asymmetries are available in~\cite{B0To2Eta_QCDF_2009, B0To2Eta_QCDF_2003, B0To2Eta_pQCD,Xiao:2014uza,BsTo2Eta_QCDF, BsTo2Eta_pQCD}. 

In this paper, we use fast simulation to investigate the Tera-$Z$ potential of measuring \Bospio\ and \Boseta\ decays. The electromagnetic calorimeter (ECAL) is the key sub-detector to reconstruct the neutral final state \piogamma\ and \etagamma. In addition, for these four channels with $B$ mesons in the initial state, the $b$-jet tagging ($b$-tagging for short hereafter) with high efficiency and purity performance is essential to suppress the background from non-\bb\ events and to increase the measurement precision of these four channels. Using the detector performance modeled from the full simulation results of the CEPC baseline detector~\cite{CEPC_CDR_Phy}, and considering the SM background corresponding to $1\times10^{12}$ $Z$ bosons, we derive the relative precisions of the signal strength measurements of these four channels with different detector parameters.

The remainder of this paper is organized as follows. Section~\ref{sec:method} introduces the simulation samples and the detector modeling. The analysis flow and measurement precision of \Bospio\ and \Boseta\ with the reference detector setup are presented in section~\ref{sec:CutChain}. In section~\ref{sec:alpha}, the Tera-$Z$ potential in measuring branching ratios and $CP$ asymmetries of relevant $B\to\pi\pi$ modes and the impact on the CKM global fit are evaluated. The dependence of the measurement precision of these four channels on two key detector performances, $b$-tagging performance and ECAL energy resolution, is analyzed in section~\ref{sec:Dependence}. A conclusion is drawn in section~\ref{sec:Conclusion}.

\section{Simulation samples and detector modeling}
\label{sec:method}

At Tera-$Z$, most events are Bhabha scattering and \Zqq\ ($q = u,\ d,\ s,\ c,\ b$ quarks at $\sqrt{s}$ = 91.2\,GeV) processes. Thanks to the excellent lepton identification performance expected in the future~\cite{CEPC_LeptonID}, Bhabha scattering and other leptonic events can be easily distinguished from the hadronic $Z$ decays. \Bospio\ and \Boseta\ from \Zbb\ decays can be reconstructed by pairing \pio\ or $\eta$ candidates. The major background comes from the inclusive \Zqq\ events with two \pio\ or $\eta$ candidates paired with the invariant mass around the physical mass of \Bos\ coincidentally. The two \pio\ or $\eta$ can originate either from the combinatorics of unrelated decays/hadronic showers or other decay modes of the $b$-hadron.

Table~\ref{tab:BkgSample} summarizes the expected yields and actual sample sizes of the exclusive \Zqq\ background used in this analysis.
Both background and signal samples are generated by the Whizard~\cite{Whizard} and Pythia~\cite{Pythia} packages. Due to the limited computing resources, only around $10^{8}$ \Zqq\ events are used as our inclusive background. The uncertainty induced by the limited sample size is estimated by using the relative statistical uncertainties of the sample events ($\frac{\sqrt{N_{\rm bkg}^{\rm Sample}}}{N_{\rm bkg}^{\rm Sample}} \times N_{\rm bkg}^{\rm TeraZ}$) and propagated to the final precision. The corresponding yields of \Bopio, \Bspio, \Boeta, and \Bseta\ (with \pio\ and $\eta$ decaying to two photons) at Tera-$Z$ are $1.91\times 10^5$, $9.0\times 10^3$, $1.9 \times 10^3$, and $4.74\times 10^4$, respectively, estimated by
\begin{equation}
\label{eq:nSig_B}
N_{B^{0}_{(s)}} \approx \text{Tera-}Z \times \mathcal{B}(Z \to b\bar{b}) \times 2 \times f(b \to B^{0}_{(s)})~,
\end{equation}
\begin{equation}
\label{eq:nSig_Pi0}
N_{B^{0}_{(s)} \to \pi^{0}\pi^{0}(\eta\eta) \to 4\gamma} \approx N_{B^{0}_{(s)}} \times \mathcal{B}(B^{0}_{(s)} \to \pi^{0}\pi^{0}(\eta\eta)) \times \mathcal{B}(\pi^{0}(\eta)\to\gamma\gamma)^{2}~,
\end{equation}
where $f(b \to B^{0}_{(s)})$ is the chance of a $b$ quark producing $B^{0}_{(s)}$ at the $Z$-pole~\cite{HFLAV2021}. The other input values are listed in table~\ref{tab:BrValues}. 
\begin{table}[thp]
	\centering
	\normalsize
	\begin{tabular}[t]{|c|c|c|c|c|}
		\hline
		Process & $\mathcal{B}$ & Tera-$Z$ yield & Sample size \\
		\hline
		$Z\to u\bar{u}$ & 11.17\% & 1.117$\times 10^{11}$ & 4.247$\times 10^{7}$ \\
		$Z\to d\bar{d}$ & 15.84\% & 1.584$\times 10^{11}$ & 5.432$\times 10^{7}$ \\
		$Z\to s\bar{s}$ & 15.84\% & 1.584$\times 10^{11}$ & 5.432$\times 10^{7}$ \\
		$Z\to c\bar{c}$ & 12.03\% & 1.203$\times 10^{11}$ & 2.940$\times 10^{8}$ \\
		$Z\to b\bar{b}$ & 15.12\% & 1.512$\times 10^{11}$ & 3.766$\times 10^{8}$ \\
		\hline
	\end{tabular}
	\caption{Branching ratios ($\mathcal{B}$), expected yields at Tera-$Z$, and sample sizes actually used for exclusive \Zqq\ backgrounds.}
	\label{tab:BkgSample}
\end{table}
\begin{table}[thp]
	\centering
	\normalsize
	\begin{tabular}[t]{|l|l|}
		\hline
		$\mathcal{B}(Z \to b\bar{b})$ & $(15.12\pm0.05)\%$~\cite{PDG2020} \\
		$f(b \to B^{0})$ & $0.407\pm0.007$~\cite{HFLAV2021} \\
		$f(b \to B^{0}_{s})$ & $0.101\pm0.008$~\cite{HFLAV2021} \\
		\hline
		$\mathcal{B}(B^{0} \to \pi^{0}\pi^{0})$ & $(1.59\pm0.26) \times 10^{-6}$~\cite{PDG2020} \\
		$\mathcal{B}(B^{0}_{s} \to \pi^{0}\pi^{0})$ & $3 \times 10^{-7}$~\cite{BsBRIHEP} \\
		$\mathcal{B}(\pi^{0} \to \gamma\gamma)$ & $(98.823\pm0.034)\%$~\cite{PDG2020} \\
		\hline
		$\mathcal{B}(B^{0} \to \eta\eta)$ & $1 \times 10^{-7}$~\cite{B0To2Eta_QCDF_2009, B0To2Eta_QCDF_2003, B0To2Eta_pQCD} \\
		$\mathcal{B}(B^{0}_{s} \to \eta\eta)$ & $1 \times 10^{-5}$~\cite{BsTo2Eta_QCDF, BsTo2Eta_pQCD} \\
		$\mathcal{B}(\eta \to \gamma\gamma)$ & $(39.41\pm0.20)\%$~\cite{PDG2020} \\
		\hline
	\end{tabular}
	\caption{Numerical values used to estimate the yields of \Bospio\ and \Boseta\ at Tera-$Z$. The notation $f(b \to B^{0}_{(s)})$ represents the fraction of \Bos\ produced in the fragmentation of $b$ quarks, and the symbol $\mathcal{B}$ denotes the decay branching ratio.}
	\label{tab:BrValues}
\end{table}

\begin{figure}[htbp]
    \centering
    \subfigure[]{
        \label{fig:btag_Eff}
        \includegraphics[width=0.45\textwidth]{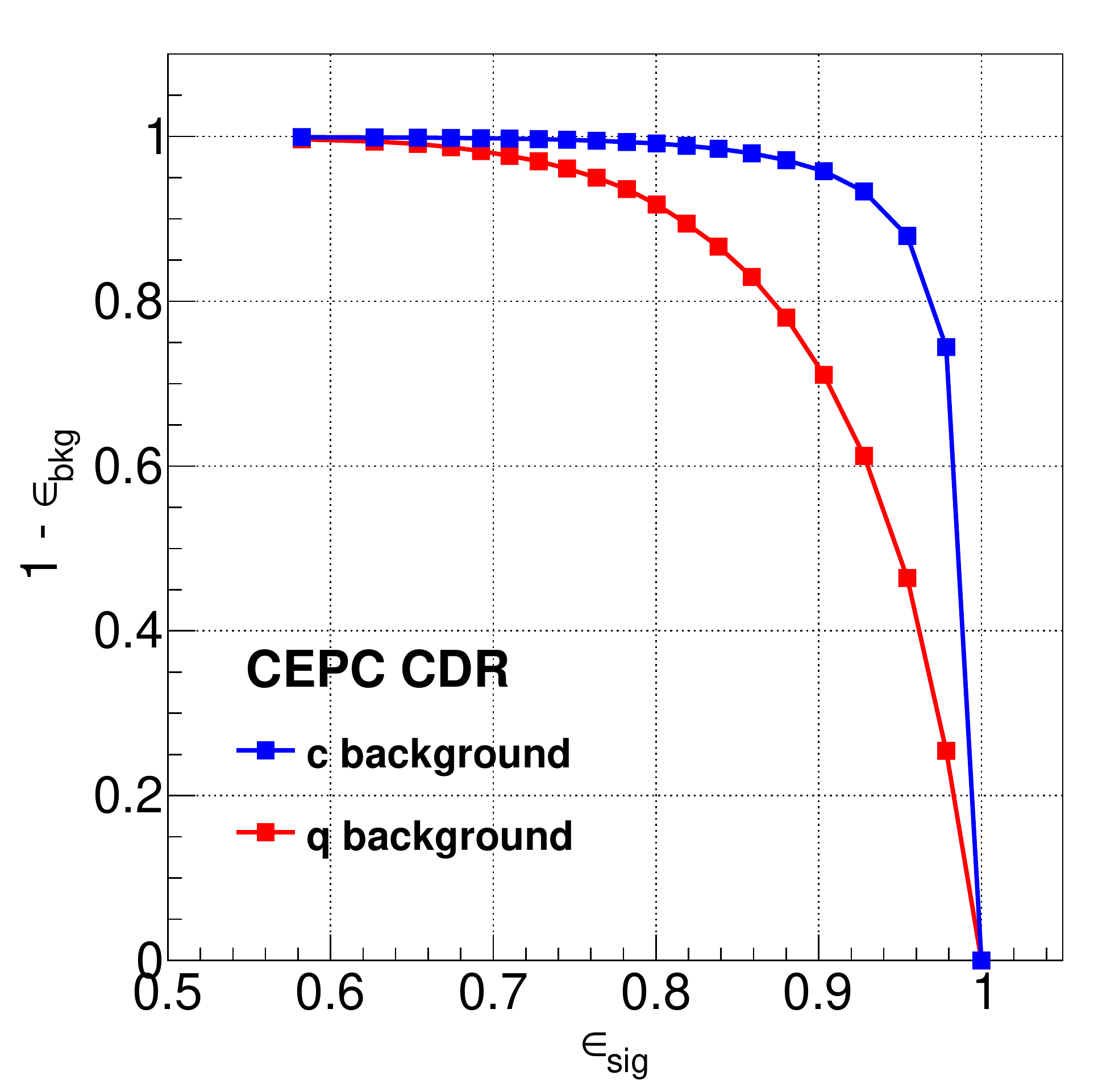}
    }
    \subfigure[]{
        \label{fig:btag_Pur}
        \includegraphics[width=0.45\textwidth]{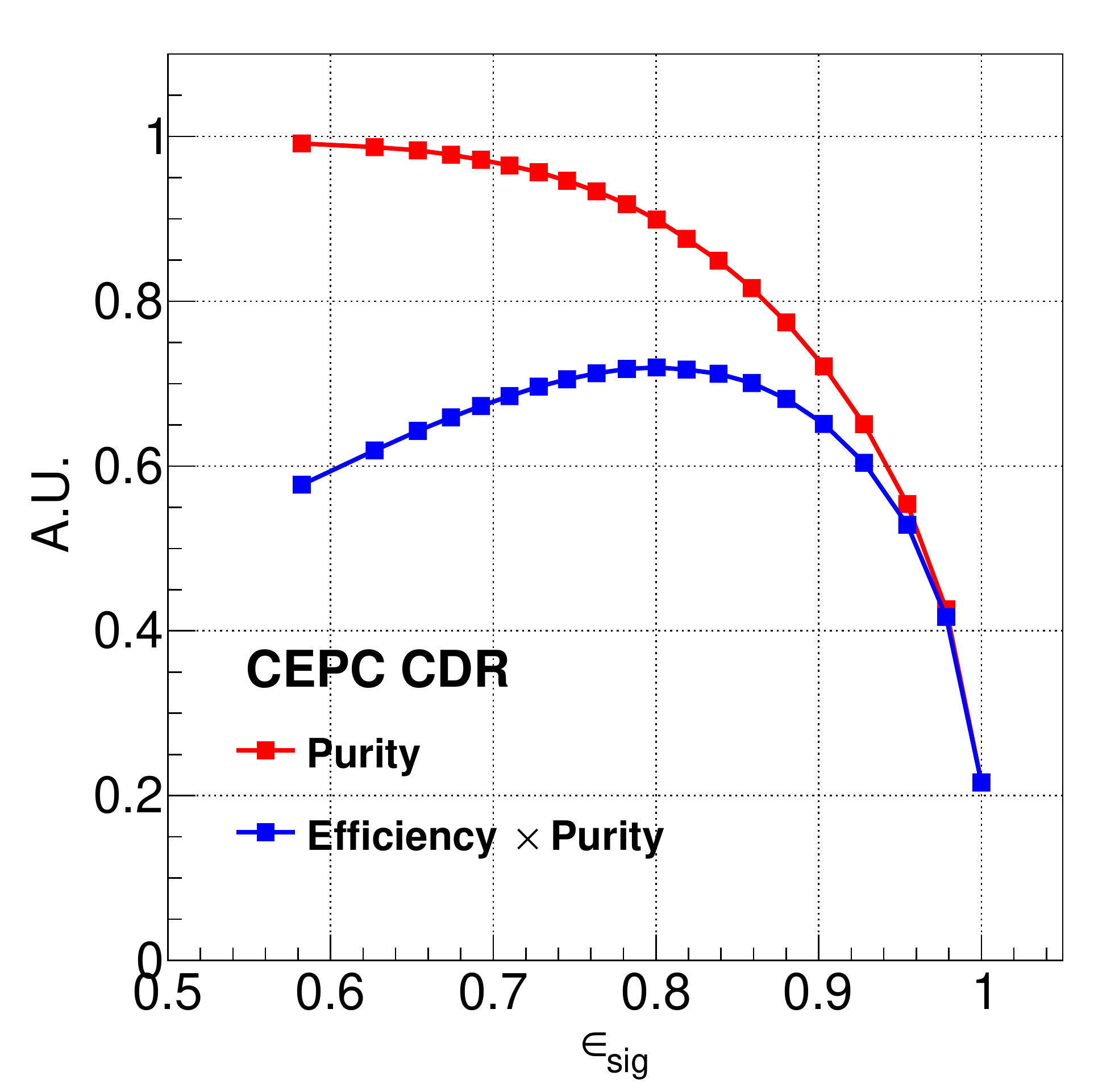}
    }
    \caption{ The $b$-jet tagging performance of the inclusive $Z\to q\bar{q}$ ($\sqrt{s}$ = 91.2\,GeV) sample derived from the full simulation of the CEPC baseline detector. (a) The $b$-jet tagging efficiency ($\epsilon_{\rm sig}$) versus background-jet rejection rate ($1-\epsilon_{\rm bkg}$). The background-jets are classified into the light-flavor jets ($u$, $d$, and $s$, simply denoted by $q$ in the legend) and the $c$-jet in the blue and red lines, respectively. (b) The $b$-jet tagging purity (red line) and efficiency $\times$ purity (blue line) versus $b$-jet tagging efficiency. The optimal performance is defined as the maximum efficiency $\times$ purity and is achieved with an efficiency of 80\% and a purity of 90\%.}
    \label{fig:btag_Perf}
\end{figure}

The $b$-tagging and the ECAL energy resolution are two key detector performances in the reconstruction of these four channels with $B$ mesons as the initial state produced in \bb\ events and fully neutral electromagnetic (EM) objects \piogamma\ and \etagamma\ in the final state.
In this work, we model the $b$-tagging based on the full simulation results described in ref.~\cite{CEPC_CDR_Phy}.
The corresponding $b$-tagging performance of the inclusive $Z\to q\bar{q}$ sample at $\sqrt{s}$ = 91.2\,GeV is shown in figure~\ref{fig:btag_Perf}, which is characterized by the $b$-jet tagging efficiency and the background-jet rejection rates.
The optimal performance can be achieved with an efficiency of 80\% and a purity of about 90\% with corresponding rejection rates of 91.74\% and 99.15\% for c-jet and light-flavor jets, respectively.
This optimal working point can also be denoted by a set of efficiencies ($\epsilon_{b \to b}, \epsilon_{c \to b}, \epsilon_{uds \to b}$) = (80\%, 8.26\%, 0.85\%), where $\epsilon_{udsc \to b}$ represents the probability of background-jets tagged as b-jet, namely the mis-tagging rate of background-jets.
The final number of $Z\to q\bar{q}$ events with different quark flavors is simply rescaled\footnote{Due to the incoherent production of $b\bar{b}$ at the $Z$-pole, the $b$-tagging performance of the signal event (with the $B$ meson decaying to fully neutral mode and only the $b$-jet on the other side can be used to perform the $b$-tagging) is expected to be approximate to that of the inclusive $Z\to q\bar{q}$ background.} using ($\epsilon_{b \to b}, \epsilon_{c \to b}, \epsilon_{uds \to b}$).

Both \pio\ and $\eta$ are reconstructed via their di-photon decays measured by the ECAL.
The energy resolution, which is the key performance of the ECAL, is generally parameterized as $\frac{\sigma_{E}}{E} = \frac{A}{\sqrt{E}} \oplus C$, where $E$ is the photon energy in the unit of GeV, $\frac{A}{\sqrt{E}}$ and $C$ represent the stochastic term and the constant term\footnote{For a realistic calorimeter, there is also a term contributed by the noise which is proportional to $1/E$ and drops fast with increasing energy.
It mainly comes from the readout electronics, the photodetector (e.g. the dark current in the photomultiplier), as well as the detector material.
Depending on the choice of the hardware, the noise term differs a lot between experiments. For simplicity, we only consider the stochastic and constant terms in this paper.}, respectively. Here the symbol $\oplus$ indicates a quadratic sum.
We model the ECAL energy resolution by smearing the energy of Monte Carlo truth-level (MCTruth) photons with errors following a Gaussian distribution with standard deviation of $\sigma_{E}$. The ECAL energy resolution determines the final $B$-meson separation by mass. Since $m_{B_s^0}-m_{B^0}$ is only $\sim$87\,MeV\footnote{Natural units with $\hbar=c=1$ are used throughout.}, the $B$-meson mass resolution needs to be at least 30\,MeV to separate \Bo\ and \Bs\ with 3\,$\sigma$ separation power\footnote{The separation power is defined as $\frac{|m_{B^0_s} - m_{B^0}|}{\sigma_{m_B}}$.}.
We thus propose a reference ECAL energy resolution of $\frac{3\%}{\sqrt{E}} \oplus 0.3\%$,\footnote{There are also other combinations of the stochastic and constant terms (realistically, the noise term should also be included) corresponding to the same $B$-meson mass resolution. The proposed one here is only an instance to performance the analysis. It is more proper to characterize the whole ECAL performance using the $B$-meson mass resolution as we will do in section \ref{sec:BMassReso}} to achieve this ambitious goal in this paper.
This ECAL energy resolution is almost six times as high as the currently typical performance of $\frac{17\%}{\sqrt{E}} \oplus 1\%$ at Tera-$Z$~\cite{CEPC_CDR_Phy,Aleksa:2021ztd} and about four times the high precision of a dual-readout calorimeter concept~\cite{Antonello:2018sna}.
It is also a factor of 1.5--2 better than the performance of the state-of-the-art homogeneous crystal ECALs in CMS~\cite{CMSECAL2013} and Belle~\cite{Belle-II:2010dht} experiments.
Even though the superb resolution we proposed may be challenging for the current technology, we believe that such an improvement is possible through the development and innovation in the future ECAL technology, and we use it as the reference ECAL performance in this paper.

\begin{figure}[htbp]
	\centering
	\subfigure[]{
		\label{fig:Pi0Mass}
		\includegraphics[width=0.45\textwidth]{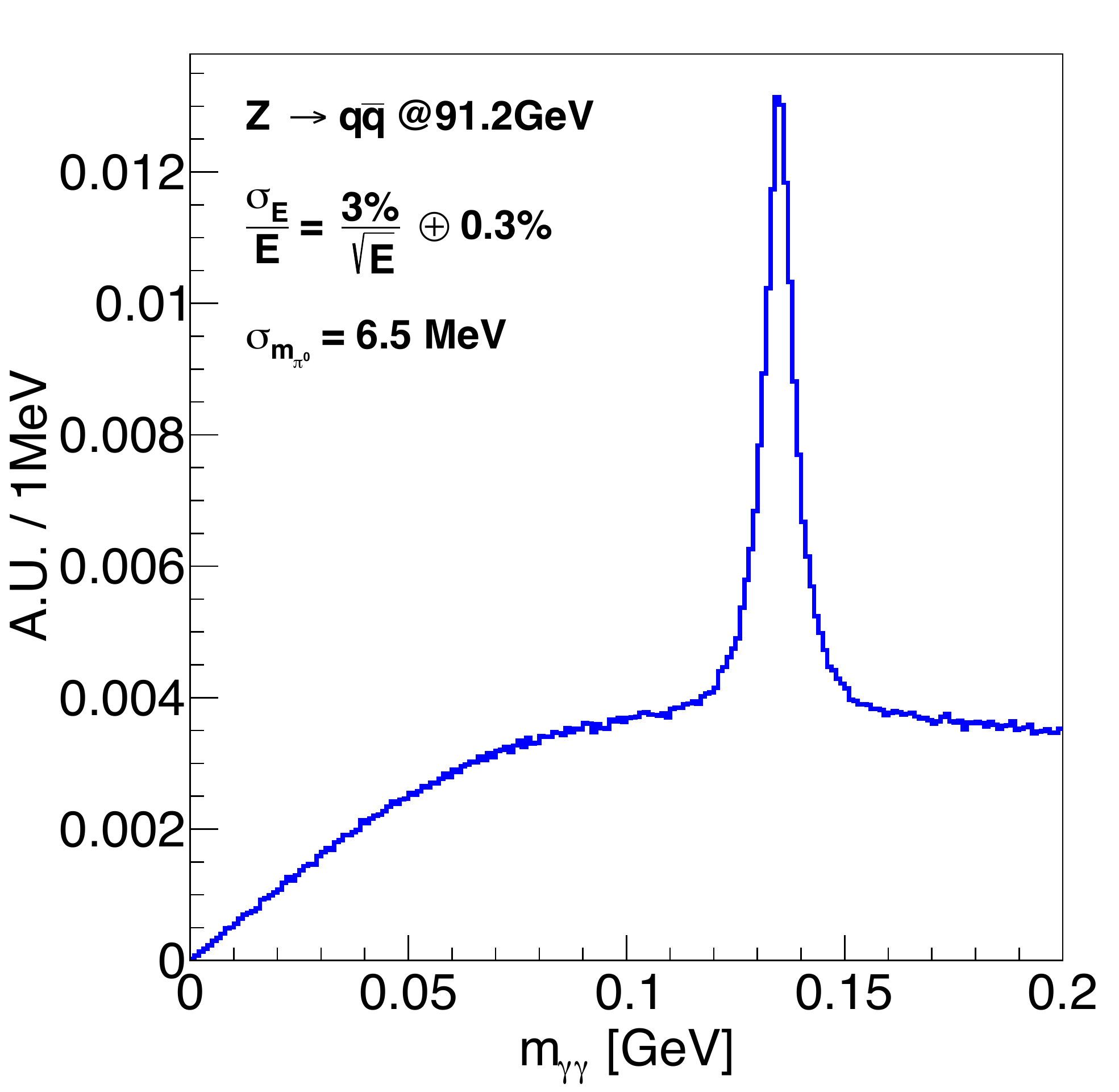}
	}
	\subfigure[]{
		\label{fig:Pi0EffPur}
		\includegraphics[width=0.45\textwidth]{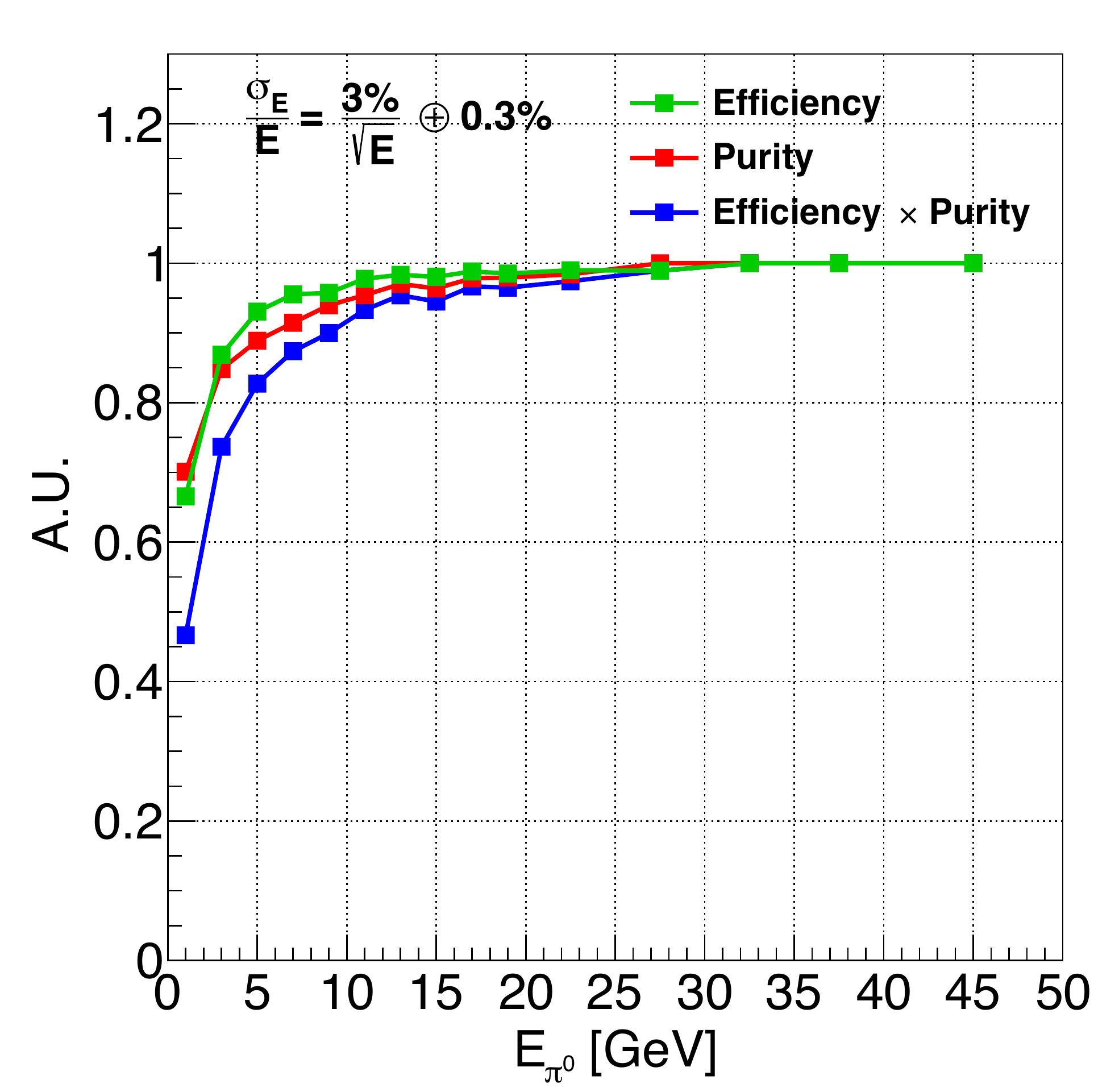}
	}
	\caption{ \pio\ reconstruction performance of the inclusive \Zqq\ ($\sqrt{s}$ = 91.2\,GeV) sample when the ECAL energy resolution is $\frac{3\%}{\sqrt{E}} \oplus 0.3\%$. (a) Invariant mass spectrum of photon pairs with \pio\ mass peak around 135\,MeV and the combinatorial background underneath. (b) Energy differential \pio\ reconstruction efficiency and purity.}
	\label{fig:Pi0Reco}
\end{figure}

\begin{figure}[htbp]
	\centering
	\subfigure[]{
		\label{fig:EtaMass}
		\includegraphics[width=0.45\textwidth]{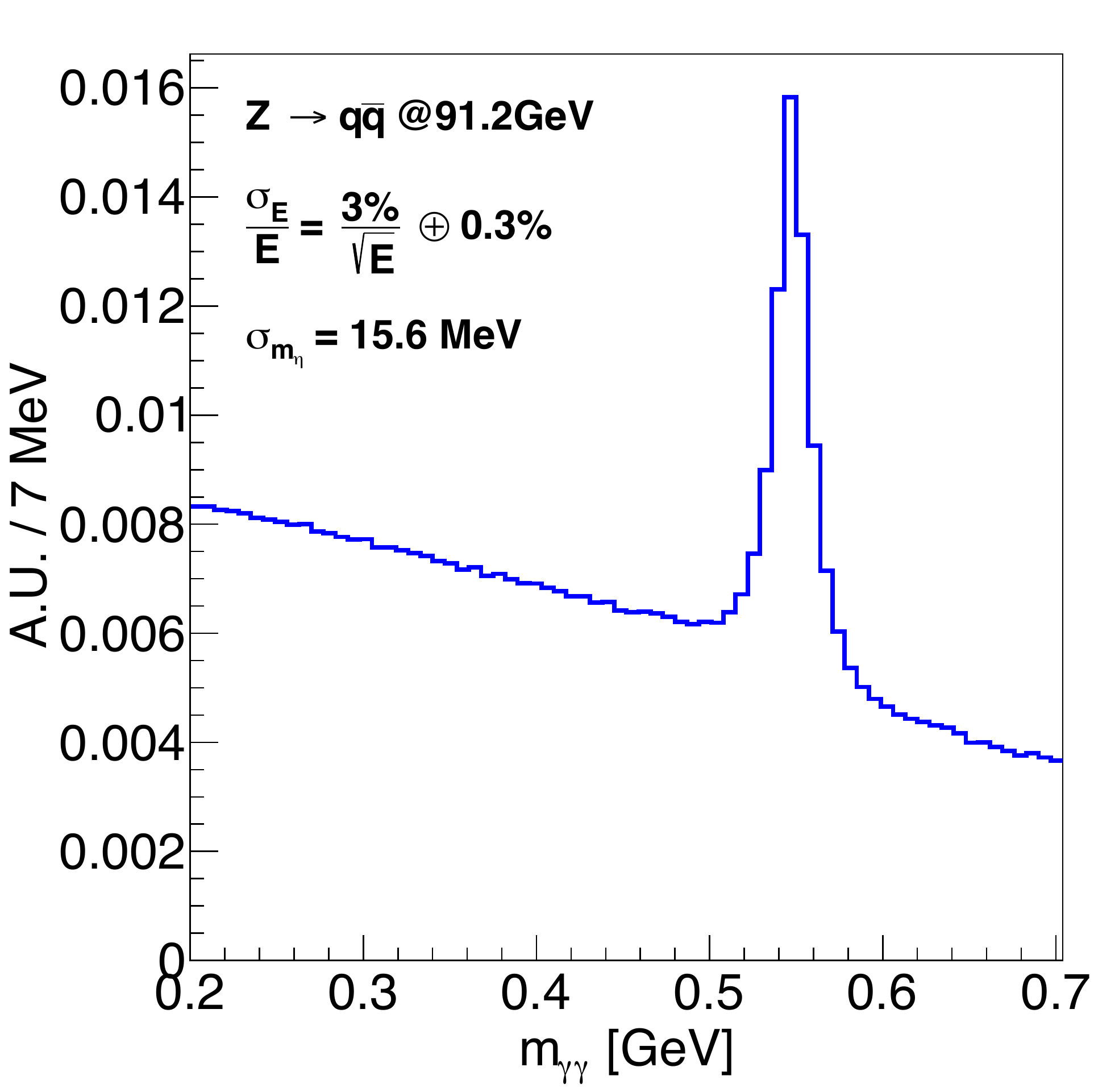}
	}
	\subfigure[]{
		\label{fig:EtaEffPur}
		\includegraphics[width=0.45\textwidth]{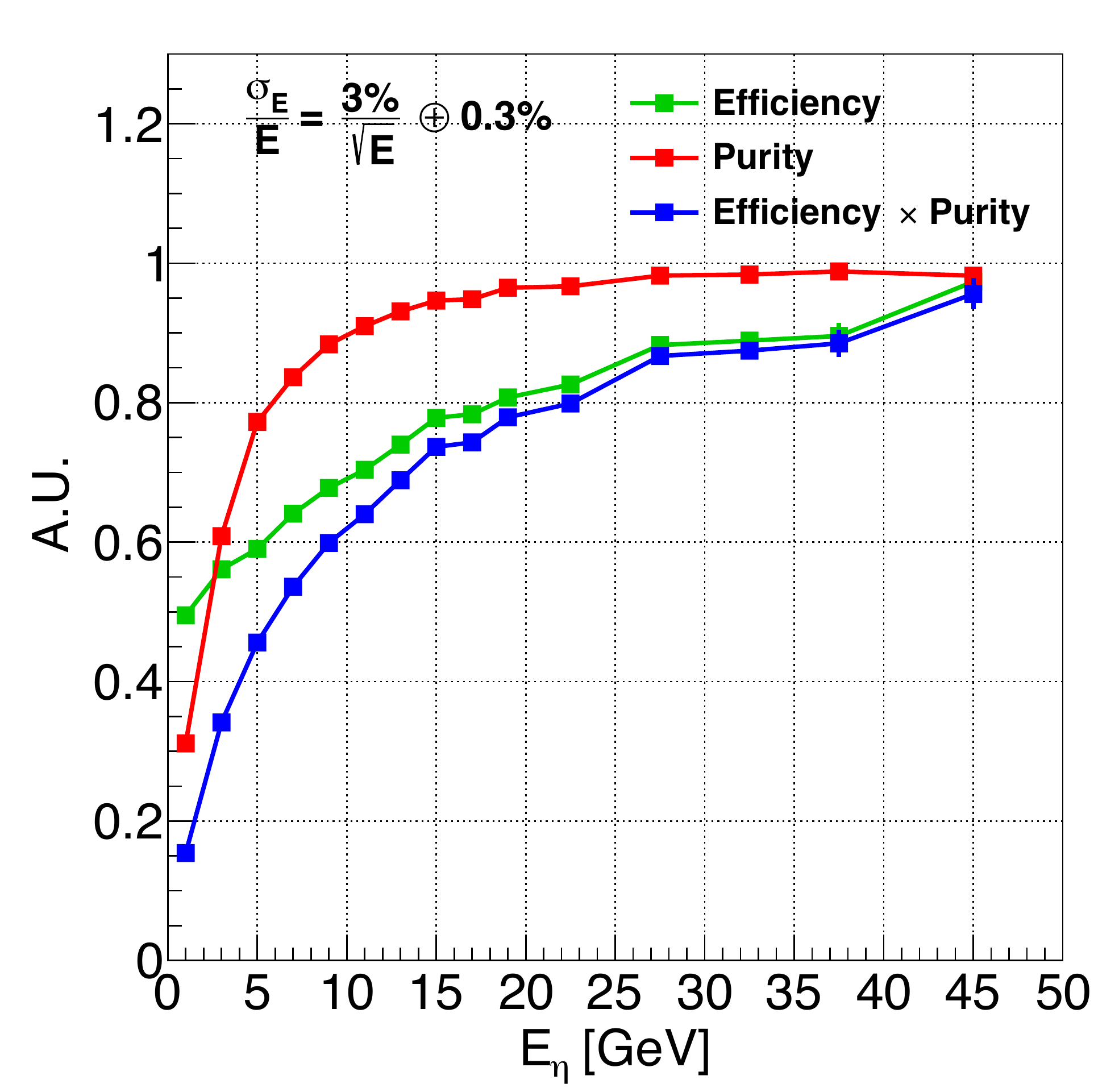}
	}
	\caption{ $\eta$\ reconstruction performance of the inclusive \Zqq\ ($\sqrt{s}$ = 91.2\,GeV) sample when the ECAL energy resolution is $\frac{3\%}{\sqrt{E}} \oplus 0.3\%$. (a) Invariant mass spectrum of photon pairs with $\eta$\ mass peak around 550\,MeV and the combinatorial background underneath. (b) Energy differential $\eta$\ reconstruction efficiency and purity.}
	\label{fig:EtaReco}
\end{figure}

Because about 10 \pio\ are produced on average in each \Zqq\ event, while the average number of $\eta$ is less than 1 ($\sim$0.85), we prioritize the \pio\ reconstruction and use the remaining photons to reconstruct $\eta$. Figure \ref{fig:Pi0Mass} shows the invariant mass spectrum of photon pairs $\in$ [0, 0.2]\,GeV with all possible combinations. The peak around 135\,MeV is the \pio\ resonance, with the extracted \pio\ mass resolution of about 6.5\,MeV. In order to reduce the combinatorial background and avoid photon double counting, we adopt an energy sorting strategy described in~\cite{Pi0Note} to reconstruct \pio. The optimal \pio\ reconstruction efficiency and purity as functions of $E_{\pi^{0}}$ are shown in figure~\ref{fig:Pi0EffPur}. For \pio\ with an energy larger than 10\,GeV, the reconstruction efficiency $\times$ purity can be higher than 90\%. Similar plots of $\eta$ are shown in figure~\ref{fig:EtaReco}. The $\eta$ resonance peaks at around 550\,MeV with the mass resolution of 15.6\,MeV. The combinatorial background in figure~\ref{fig:EtaMass} does not contain the background with photons already used when reconstructing \pio\ candidates. This procedure correlates the $\eta$ reconstruction performance with the \pio's but significantly improves the $\eta$ reconstruction purity at the cost of an acceptable efficiency loss. The optimal reconstruction efficiency $\times$ purity can be better than 60\% for $\eta$ with an energy larger than 10\,GeV, as shown in figure~\ref{fig:EtaEffPur}.


\section{Analyses and results with the reference detector}
\label{sec:CutChain}

This section introduces the event selection and final precisions of \Bospio\ and \Boseta\ under the reference detector setup with the baseline $b$-tagging performance introduced above and the reference ECAL energy resolution of $\frac{3\%}{\sqrt{E}} \oplus 0.3\%$.

\subsection{\Bospio}
\label{sec:CutChain_Pi0}

\begin{figure}[htbp]
	\centering
	\subfigure[2D energy spectrum of \pio\ pairs in \Bopio\ (left), \Bspio\ (middle), and \Zqq\ background (right) events.]{
		\includegraphics[width=0.33\textwidth]{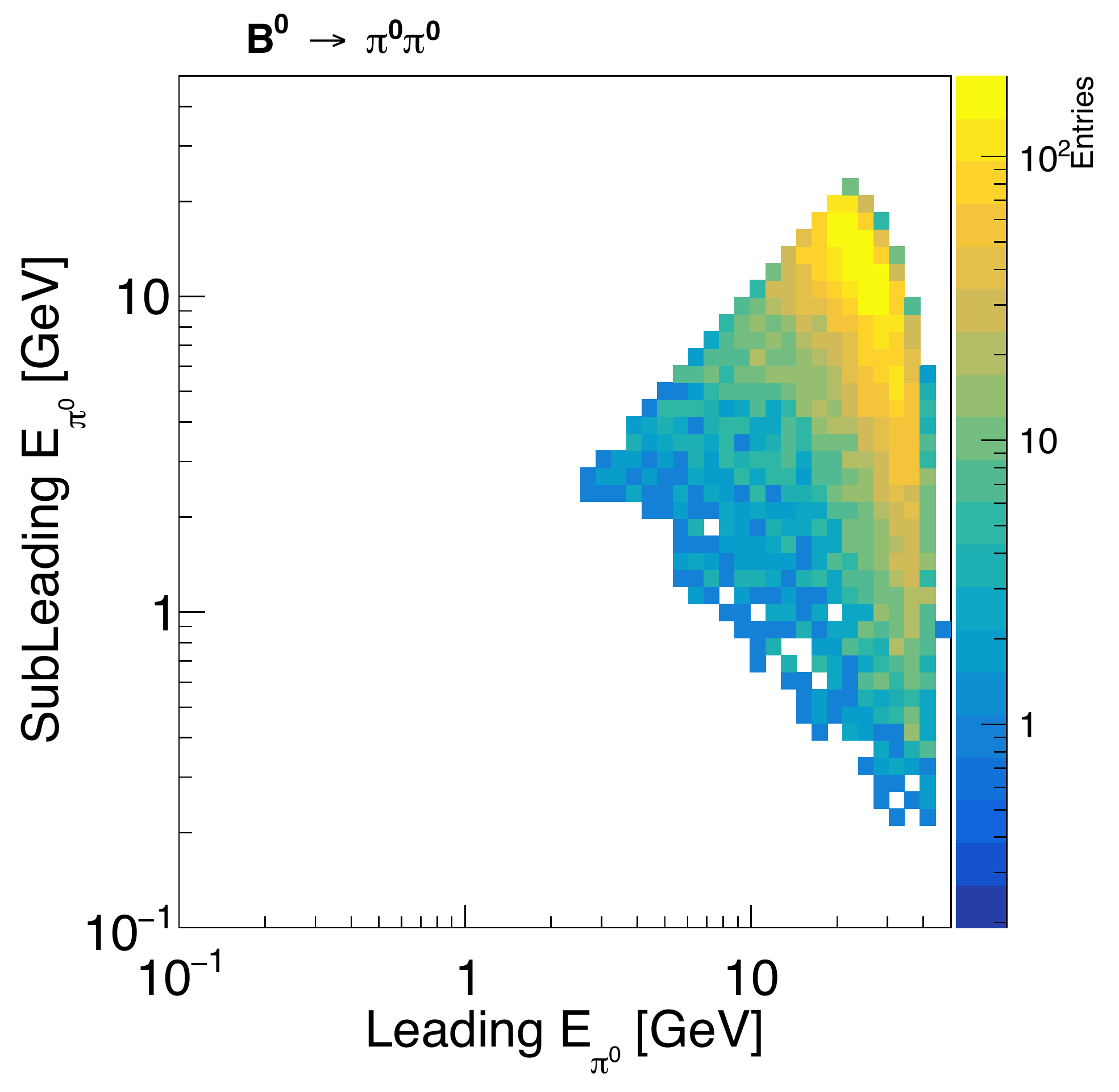}
		\includegraphics[width=0.33\textwidth]{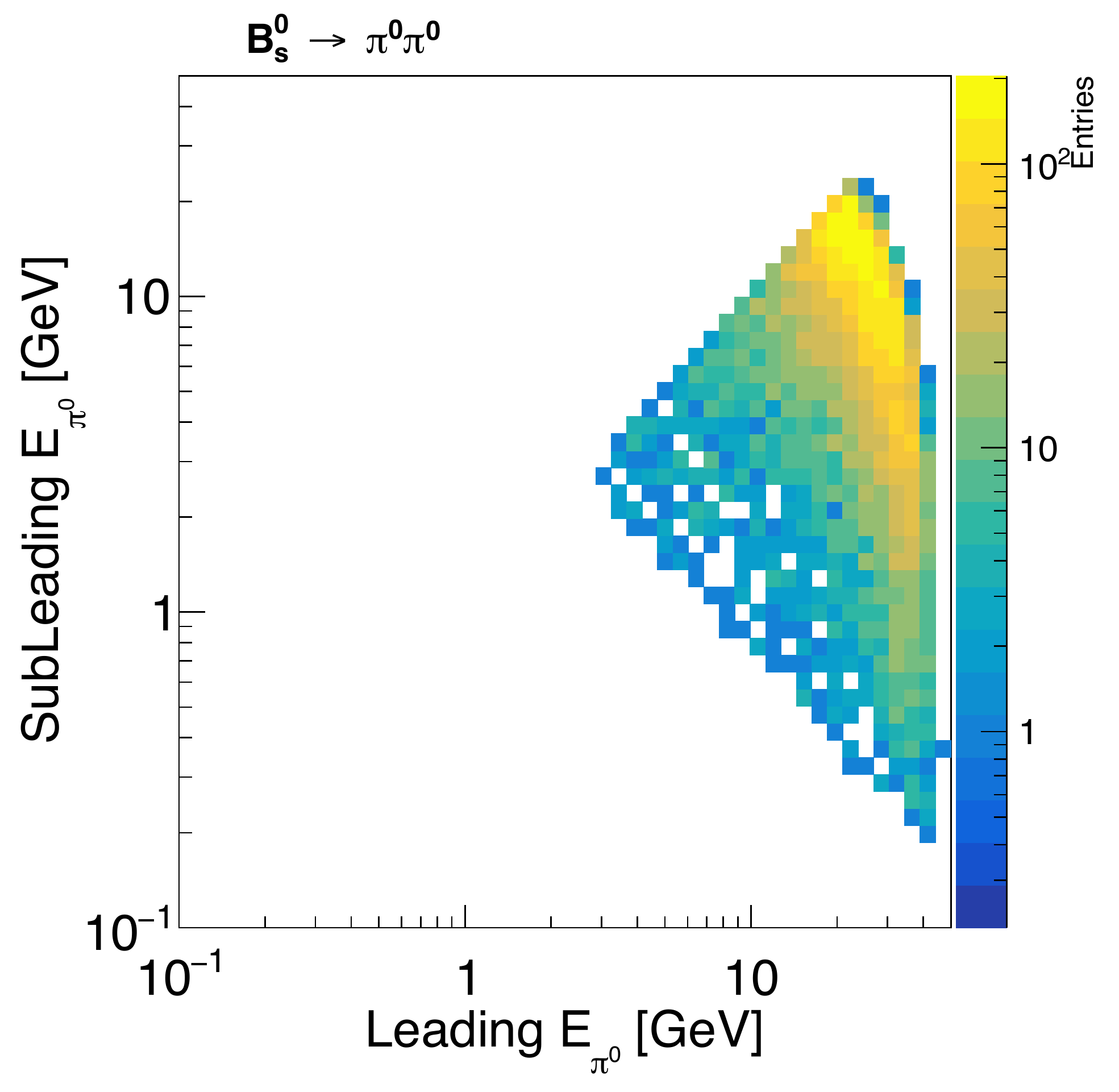}
		\includegraphics[width=0.33\textwidth]{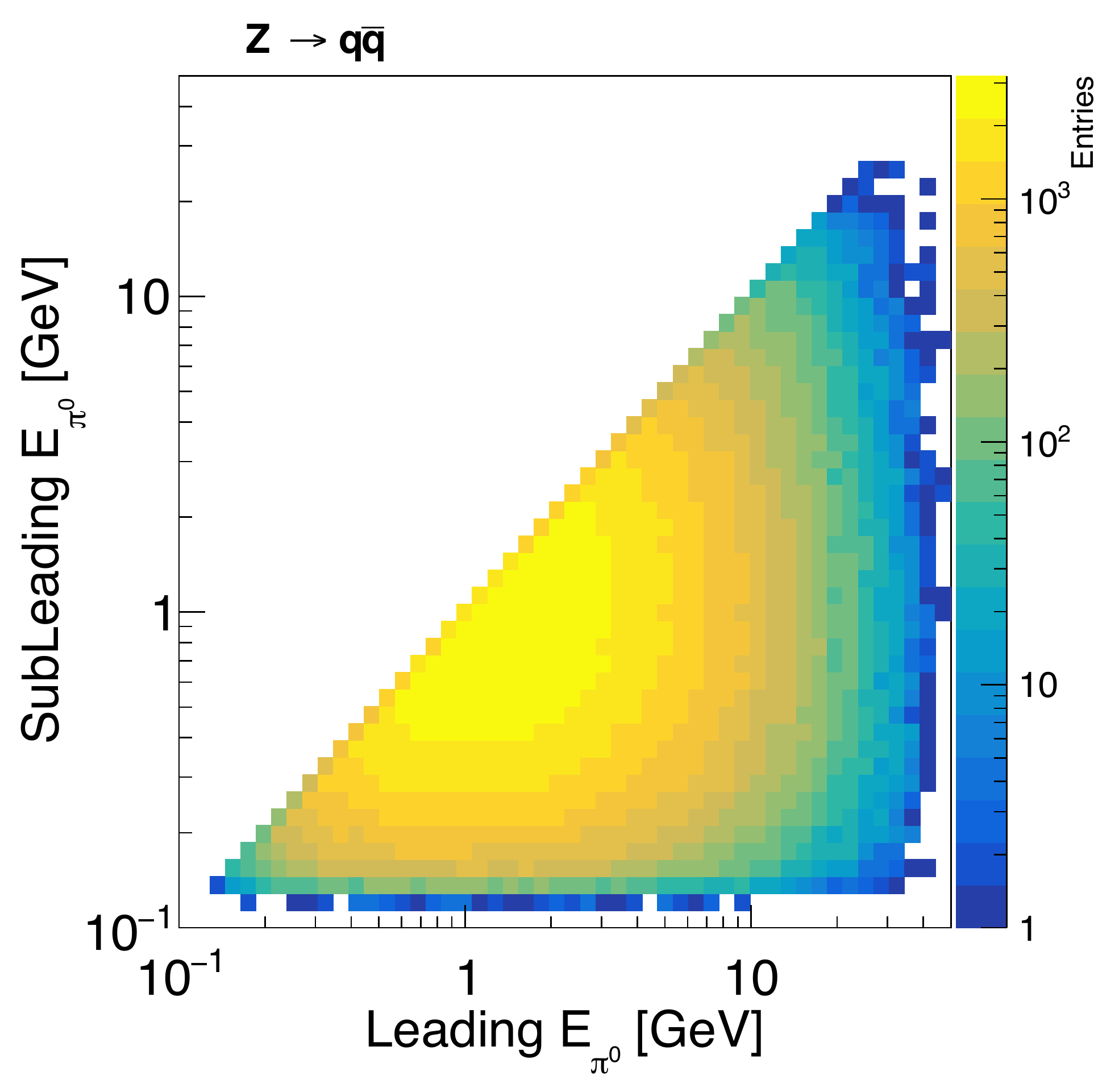}
	}
	\hfill
	\subfigure[Correlation between $E_{\pi^0\pi^0}$ and $\theta_{\pi^0\pi^0}$ in \Bopio\ (left), \Bspio\ (middle), and \Zqq\ background (right) events.]{
		\includegraphics[width=0.33\textwidth]{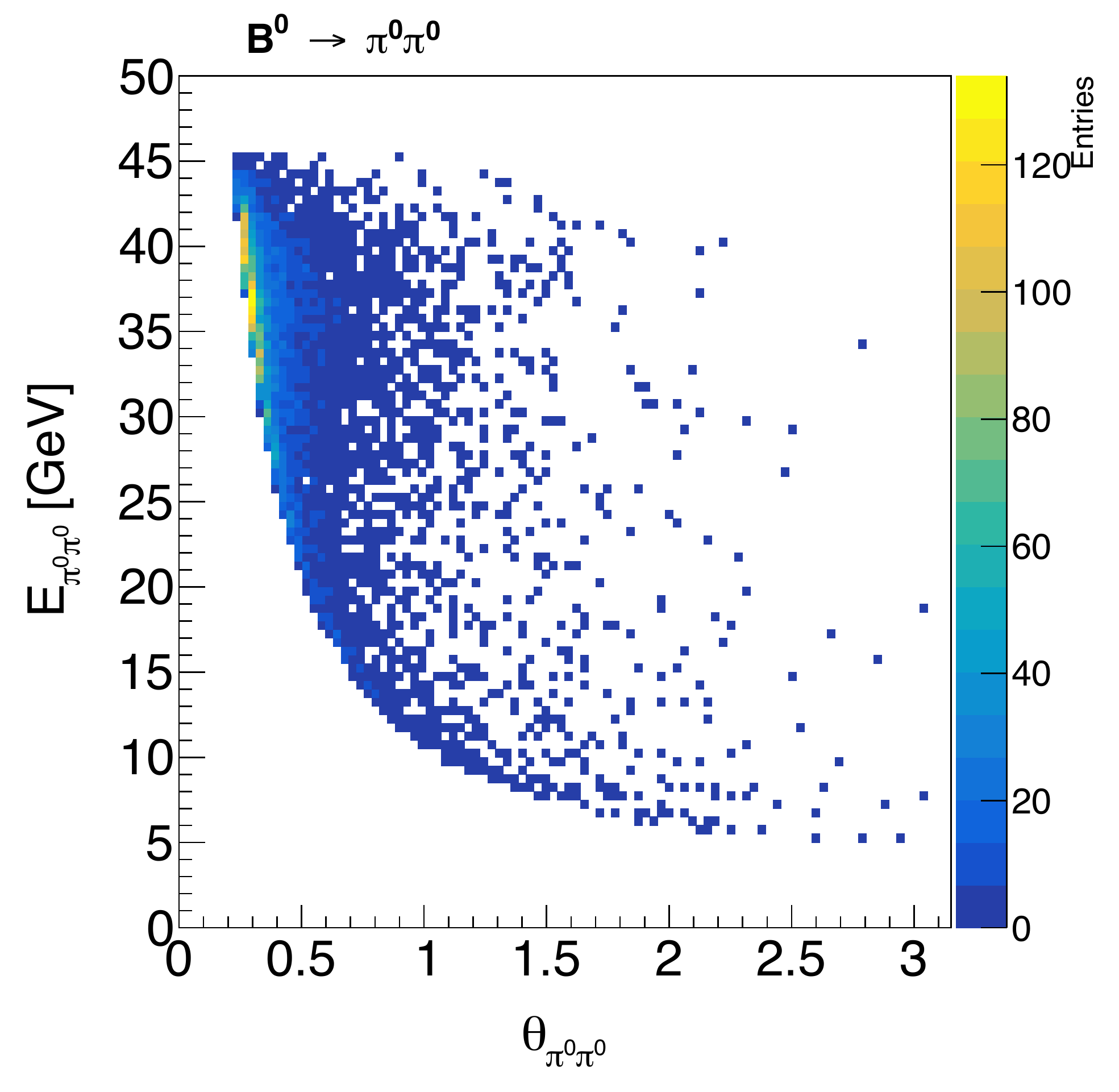}
		\includegraphics[width=0.33\textwidth]{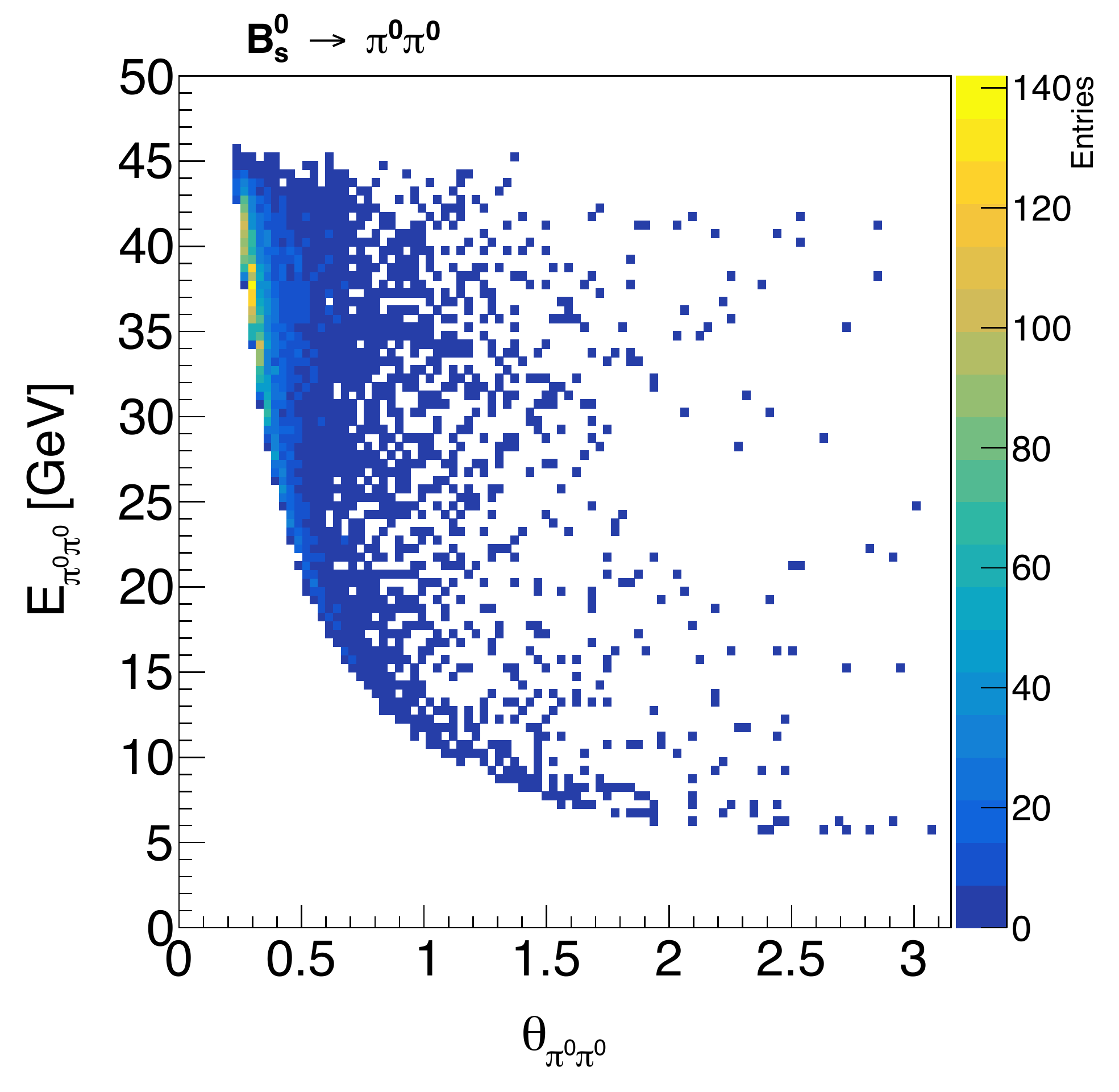}
		\includegraphics[width=0.33\textwidth]{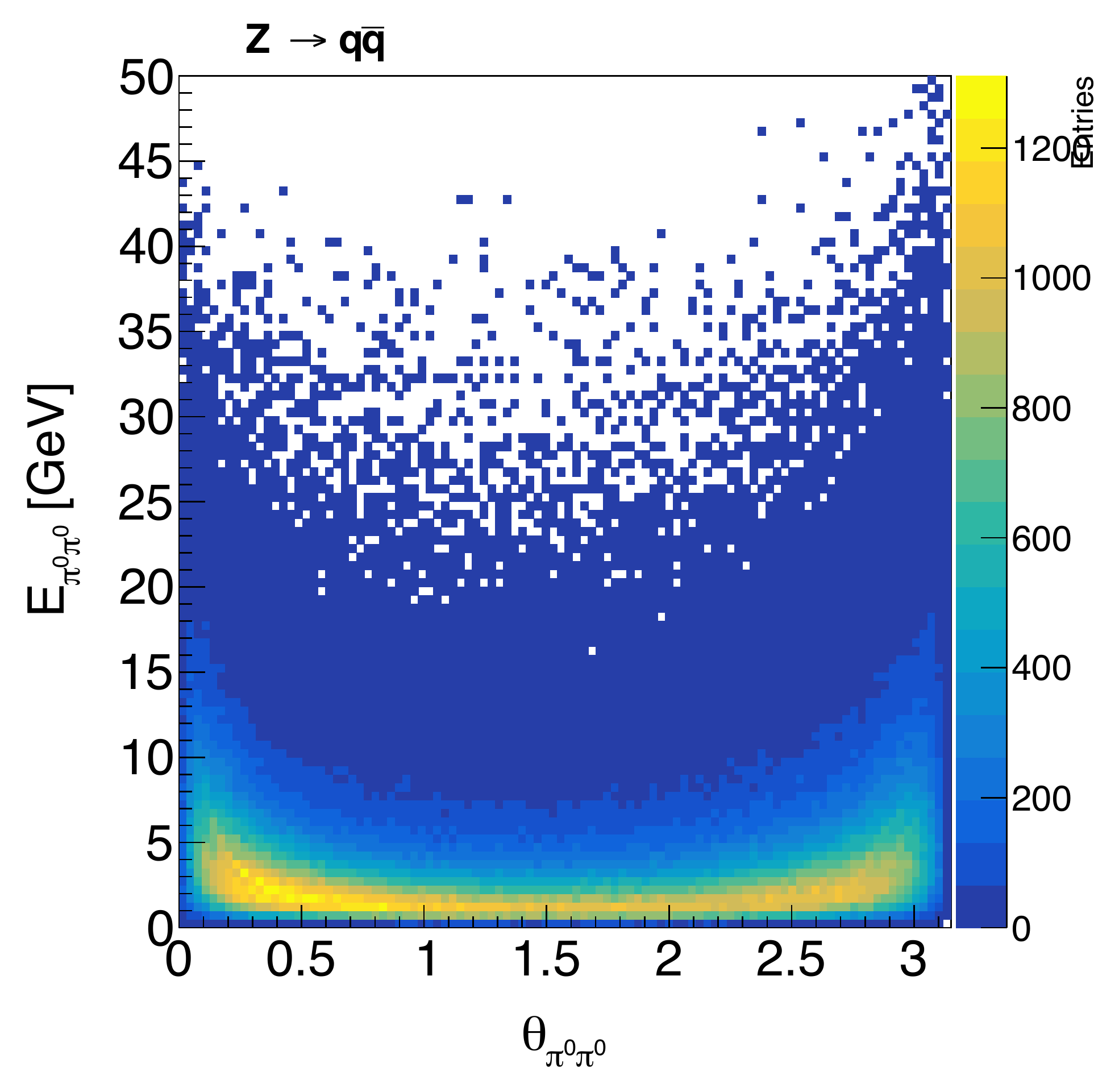}
	}
	\caption{Simulated energy and angle distributions of \pio\ pair in \Bopio, \Bspio, and \Zqq\ background events.}
	\label{fig:Pi0PairEnergyAngle}
\end{figure}

Table~\ref{tab:Pi0CutChain} shows the yields of \Bopio\ and \Bspio\ at each step of the event selection. After applying the baseline $b$-tagging, \bb\ events dominate the background, with less than 10\% of contribution from non-\bb\ events. Since the hard fragmentation of the $b$ quark gives 70\% of the beam energy to \Bos\ on average~\cite{L3}, the \pio\ pairs from \Bos\ are likely to have high energies and small opening angles, as shown in figure \ref{fig:Pi0PairEnergyAngle}. The leading \pio\ from \Bos\ has a typical energy ranging from 10 to 20\,GeV. From our method in section~\ref{sec:method}, the \pio\ reconstruction efficiency reaches above 98\%, leading to an overall \Bospio\ reconstruction efficiency greater than 96\%. On the contrary, the combinatorial background tends to have a low energy and a large spread of opening angle distribution. A series of selection criteria based on the features above are applied, with their details listed in table~\ref{tab:Pi0CutChain}. After these selections (corresponding to the $\theta_{\pi^0\pi^0}<23^\circ$ row), the overall background level is suppressed by nearly two orders of magnitude. It is especially true for the \bb\ background, 99.5\% of which has been removed. The reason is that most \pio\ in light-flavor \qq\ events are generated directly in the hadronization process, while \pio\ in \bb\ and $c\bar{c}$ events are mainly from the long decay chain of $b$- and $c$-hadrons, resulting in the smaller average \pio\ energy and the relatively smaller fraction of $\pi^0$ survived from the high-energy selection.

\begin{table}[thp]
	\centering
	\resizebox{1.\columnwidth}{!}{
	\begin{tabular}[t]{|c|ccccc|c|}
		\hline
		Selection chain
		& $B^0\to \pi^0\pi^0\to 4\gamma$  &  $B^0_s\to \pi^0\pi^0\to 4\gamma$  &  $u\bar{u}$+$d\bar{d}$+$s\bar{s}$  &  $c\bar{c}$  &  $b\bar{b}$ & $\sqrt{S+B}/S$ \\
		\hline
		Yield at Tera-$Z$
		& 1.91$\times 10^{5}$  &  9.0$\times 10^{3}$  &  \tabincell{c}{4.29$\times 10^{11}$\\ (61.21\%)}  &  \tabincell{c}{1.20$\times 10^{11}$\\ (17.19\%)}  &  \tabincell{c}{1.51$\times 10^{11}$\\ (21.60\%)} & \\
		$b$-tagging
		& 1.53$\times 10^{5}$  &  7.2$\times 10^{3}$  &  \tabincell{c}{3.64$\times 10^{9}$\\ (2.70\%)}  &  \tabincell{c}{9.94$\times 10^{9}$\\ (7.38\%)}  &  \tabincell{c}{1.21$\times 10^{11}$\\ (89.92\%)} & \\
		\hline
		\piogamma
		& 1.48$\times 10^{5}$ & 7.0$\times 10^{3}$ & 3.61$\times 10^{9}$ & 9.91$\times 10^{9}$ & 1.21$\times 10^{11}$ & \\
		Lower $E_{\pi^{0}}$ \textgreater\ 6\,GeV
		& 9.24$\times 10^{4}$ & 4.4$\times 10^{3}$ & 8.44$\times 10^{8}$ & 1.60$\times 10^{9}$ & 1.31$\times 10^{10}$ & \\ 
		Higher $E_{\pi^{0}}$ \textgreater\ 14\,GeV
		& 8.74$\times 10^{4}$ & 4.1$\times 10^{3}$ & 3.08$\times 10^{8}$ & 3.15$\times 10^{8}$ & 1.91$\times 10^{9}$ & \\
		$E_{\pi^{0}\pi^{0}}$ \textgreater\ 22\,GeV
		& 8.71$\times 10^{4}$ & 4.1$\times 10^{3}$ & 2.90$\times 10^{8}$ & 2.82$\times 10^{8}$ & 1.66$\times 10^{9}$ & \\ 
		$\theta_{\pi^{0}\pi^{0}}$ \textless\ 23\degree
		& 7.80$\times 10^{4}$ & 3.6$\times 10^{3}$ & 1.19$\times 10^{8}$ & 1.02$\times 10^{8}$ & 6.04$\times 10^{8}$ & \\ 
		\hline
		$m_{\pi^{0}\pi^{0}} \in$ (5.212, 5.347)\,GeV
		& 7.59$\times 10^{4}$ & 9$\times 10^{2}$ & 5.5$\times 10^{3}$ & 1.6$\times 10^{3}$ & 8.7$\times 10^{3}$ & \tabincell{c}{0.40\%\\ $\pm$ 0.01\%} \\
		\hline
		$m_{\pi^{0}\pi^{0}} \in$ (5.336, 5.397)\,GeV
		& 2.8$\times 10^{3}$ & 2.5$\times 10^{3}$ & 2.4$\times 10^{3}$ & 5$\times 10^{2}$ & 2.2$\times 10^{3}$ & \tabincell{c}{4.0\%\\ $\pm$ 0.6\%} \\
		\hline
	\end{tabular}
	}
	\caption{\Bopio\ and \Bspio\ yields at each step of the selection chain and their final precision ($\sqrt{S+B}/S$) when using the baseline $b$-tagging and ECAL energy resolution of $\frac{3\%}{\sqrt{E}} \oplus 0.3\%$. In the last two rows, selections on \MPioPio\ are applied, the first one is for \Bopio, and the other one is for \Bspio. }
	\label{tab:Pi0CutChain}
\end{table}

\begin{figure}[htbp]
		\centering
		\includegraphics[width=0.45\textwidth]{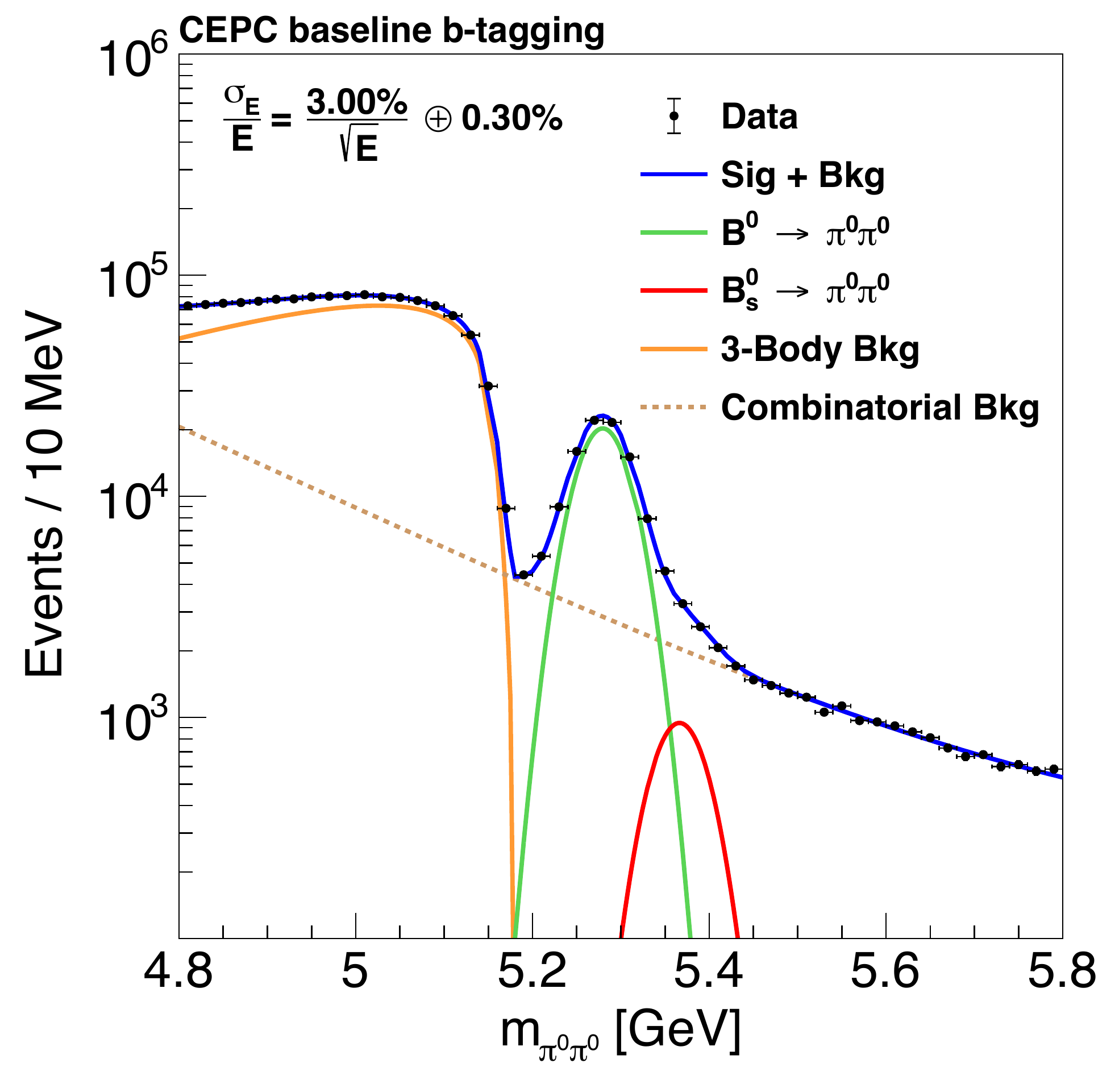}
		\caption{The reconstructed \MPioPio\ distributions of \Bopio, \Bspio, and \Zqq\ background after applying the baseline $b$-tagging and selections on energy and opening angle of \pio\ pairs when the ECAL energy resolution is $\frac{3\%}{\sqrt{E}} \oplus 0.3\%$.}
		\label{fig:Pi0SigBkgCombine}
\end{figure}

\begin{figure}[htbp]
	\centering
	\subfigure[]{
		\includegraphics[width=0.45\textwidth]{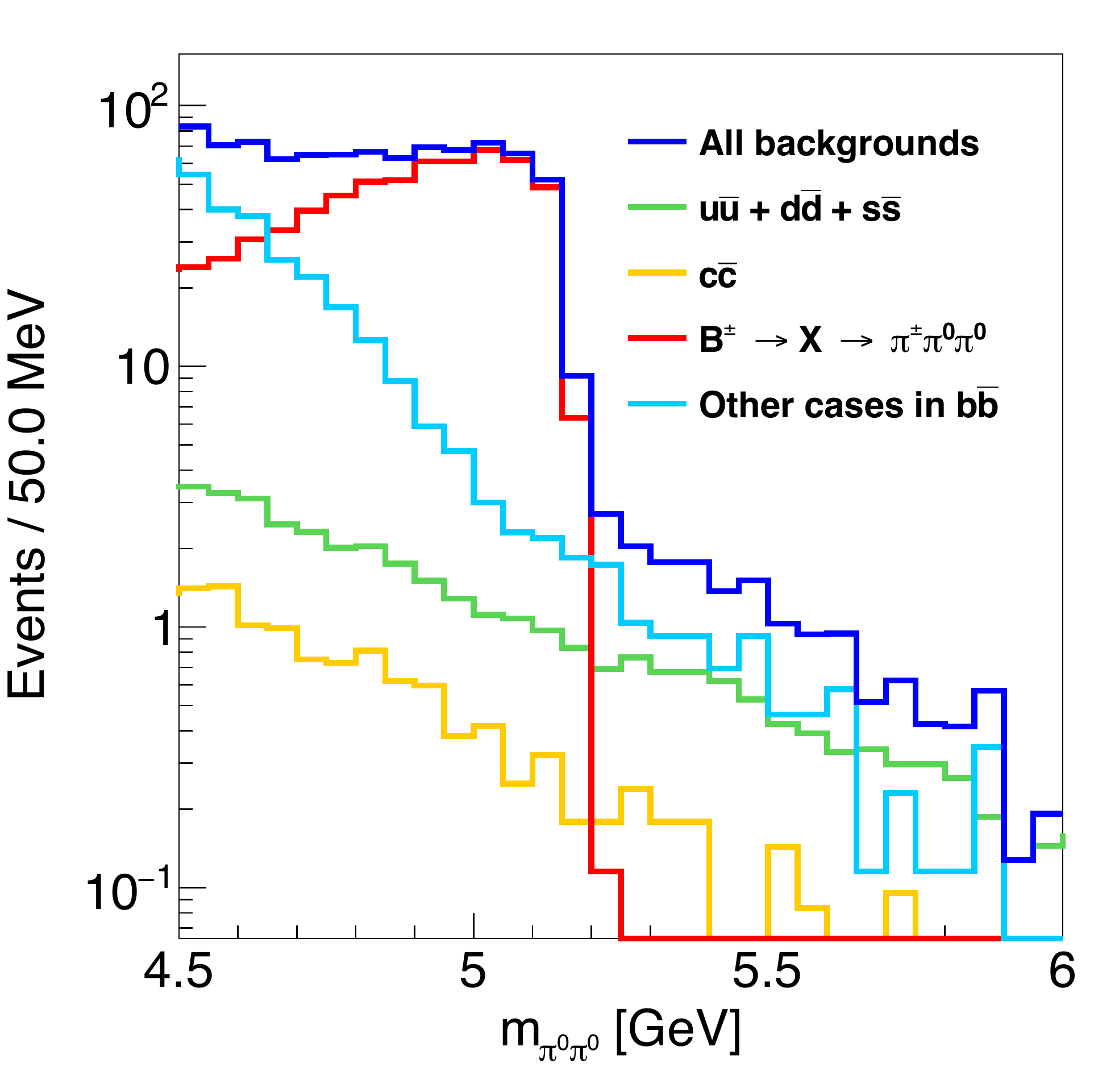}
		\label{fig:BkgComponent}
	}
	\subfigure[]{
		\includegraphics[width=0.45\textwidth]{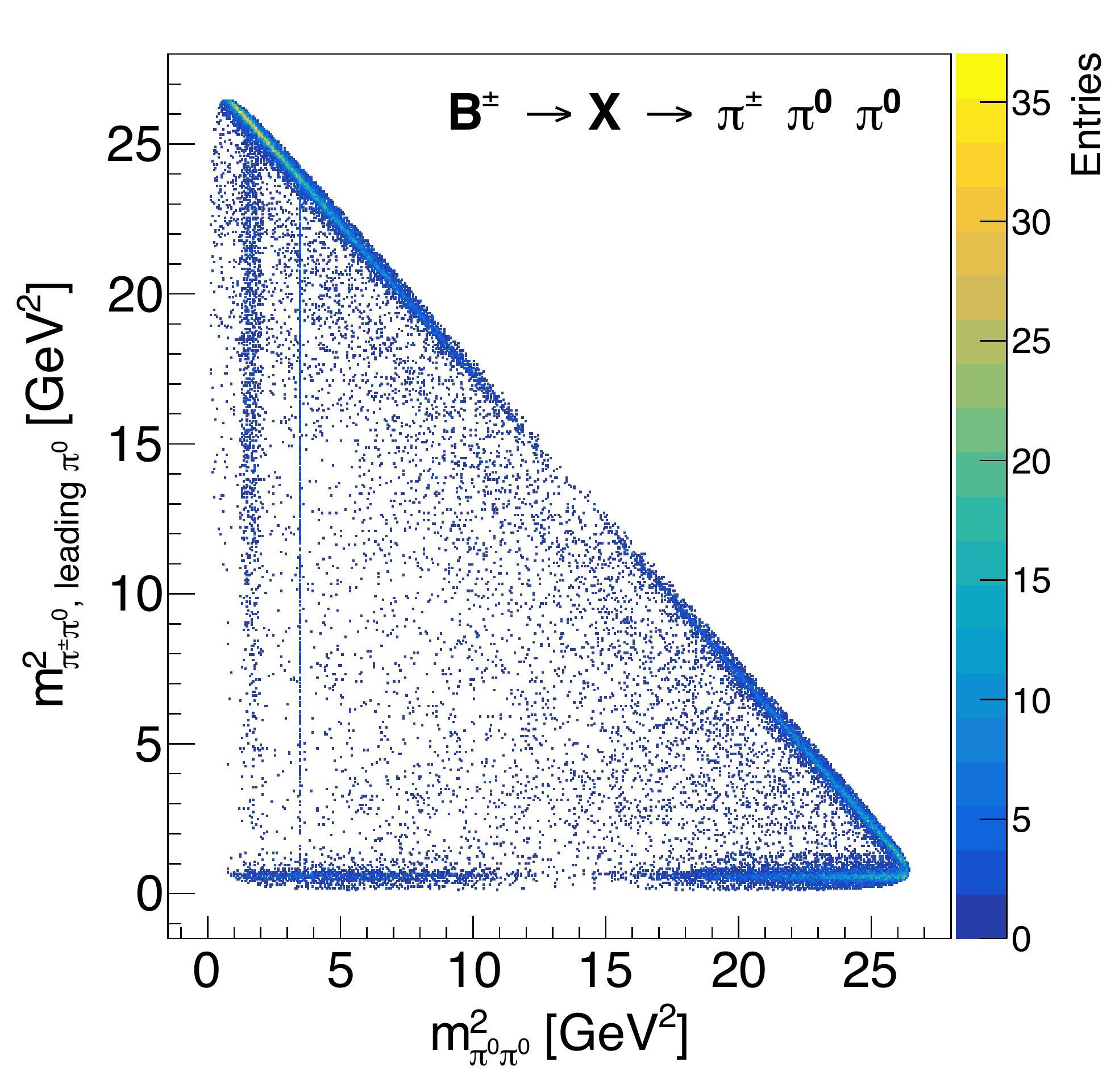}
		\label{fig:BkgDalitz}
	}
	\caption{(a) Background components for \Bospio. (b) Dalitz plot of direct or indirect decays of $B^{\pm}$ to $\pi^{\pm}\pi^{0}\pi^{0}$. The two resonances with $m^{2}_{\pi^{0}\pi^{0}} \approx$ 1.6 and 3.48\,GeV$^{2}$ are $f_{2}(1270)$ and $D^{0}$, respectively.}
\end{figure}

When more than one pair of \pio\ candidates remain, we only select the pair with invariant mass closest to $m_{B^0_{(s)}}$. The consequent \MPioPio\ distributions of \Bopio, \Bspio, and background events are shown in figure~\ref{fig:Pi0SigBkgCombine}. In the plot, one can recognize two types of background. One is the combinatorial background, which can be described by an exponential distribution. In figure~\ref{fig:BkgComponent} we show the contributions of different flavored \qq\ events to the combinatorial background. Another type is the partially reconstructed three- (or more) body $b$-hadron decays, with the dominant contribution from the process of $B^{\pm} \to \rho(770)^{\pm} (\to \pi^{\pm}\pi^{0}) \pi^{0}$. This kind of three-body background can be described by the ARGUS function~\cite{ARGUS:1994rms}. Figure~\ref{fig:BkgDalitz} shows an example of the Dalitz plot of the (in)direct decays of $B^{\pm}$ to $\pi^{\pm}\pi^{0}\pi^{0}$. The kinematic constraint of this kind of three-body $b$-hadron decays results in a cut-off on \MPioPio\ around 5.15\,GeV. The background decreases sharply in the mass range higher than this cut-off, and the $B$-meson mass resolution of 30\,MeV enables the separation of the \Bopio\ peak from the three-body background. Finally, the fitted yields (relative uncertainties) of \Bopio\ and \Bspio\ are 77464 $\pm$ 316 (0.4\%) and 3691 $\pm$ 139 (4\%), respectively.


We also extract the relative uncertainty of the signal yield by counting the numbers of the signal (S) and the background (B) in the mass window centered at $m_{B^0_{(s)}}$ and optimizing its width to minimize $\frac{\sqrt{S+B}}{S}$. The quantity $\frac{\sqrt{S+B}}{S}$ is known as the relative precision of the signal strength, also equivalent to the relative uncertainty of the signal yield. The corresponding results are shown in the last two rows of table~\ref{tab:Pi0CutChain}. The mode \Bopio\ can be measured with an efficiency of 40\%, a purity of 80\%, and relative precision of 0.4\%. Meanwhile, \Bspio\ can only be measured with the relative precision of 4\% (efficiency of 28\%, purity of 24\%) due to its low yield ($\sim$20 times smaller than that of \Bopio).
The results from this optimal mass window method are very similar to the ones obtained by fitting the two signal peaks and backgrounds, provided that we have complete knowledge and control on the signal and background models.

\subsection{\Boseta}

\begin{figure}[htbp]
	\centering
	\includegraphics[width=0.45\textwidth]{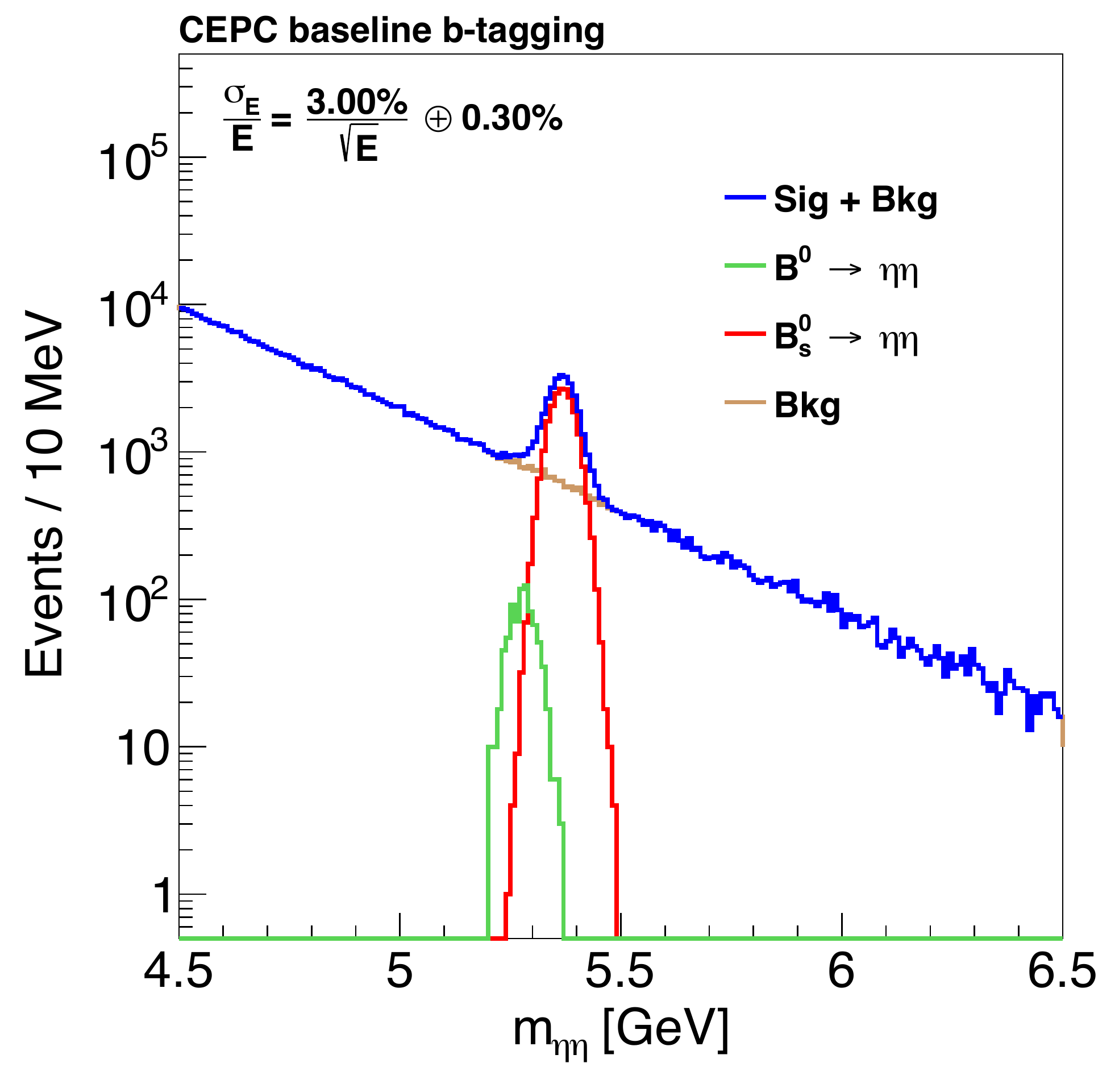}
	\caption{The reconstructed \MEtaEta\ distributions of \Boeta, \Bseta, and \Zqq\ background after applying the baseline $b$-tagging and selections on energy and opening angle of $\eta$\ pairs when the ECAL energy resolution is $\frac{3\%}{\sqrt{E}} \oplus 0.3\%$.}
	\label{fig:EtaSigBkgCombine}
\end{figure}

A similar analysis strategy is applied to \Boseta. Table~\ref{tab:EtaCutChain} summarizes the yields at each step of the event selection and the final precisions of \Boeta\ and \Bseta. The reconstructed \MEtaEta\ distributions of \Boseta\ signals and \Zqq\ background before invariant mass selection are shown in figure~\ref{fig:EtaSigBkgCombine}.
The background decreases exponentially with dominant contributions from the fake $\eta$ (70\%) and combinatorics (22\%).
For \Boeta\ with the SM predicted branching ratio of $\mathcal{O}(10^{-7})$, the relative precision is 17\% $\pm$ 2\%, corresponding to a significance (defined as $\frac{S}{\sqrt{B}}$) of about 6 standard deviations. \Bseta, whose SM predicted branching ratio is two orders of magnitude larger than that of \Boeta, can be measured with a relative precision of 0.90\% $\pm$ 0.05\%.

\begin{table}[thp]
	\centering
	\resizebox{1.\columnwidth}{!}{
		\begin{tabular}[t]{|c|ccccc|c|}
			\hline
			Selection chain
			& $B^0\to \eta\eta\to 4\gamma$  &  $B^0_s\to \eta\eta\to 4\gamma$  &  $u\bar{u}$+$d\bar{d}$+$s\bar{s}$  &  $c\bar{c}$  &  $b\bar{b}$ & $\sqrt{S+B}/S$ \\
			\hline
			Yield at Tera-$Z$
			& 1.9$\times 10^{3}$  &  4.74$\times 10^{4}$  &  4.29$\times 10^{11}$  &  1.20$\times 10^{11}$ & 1.51$\times 10^{11}$ & \\
			$b$-tagging
			& 1.5$\times 10^{3}$  &  3.80$\times 10^{4}$  & 3.64$\times 10^{9}$ &  9.94$\times 10^{9}$ & 1.21$\times 10^{11}$ & \\
			\hline
			\etagamma
			& 1.0$\times 10^{3}$ & 2.58$\times 10^{4}$ & 2.13$\times 10^{8}$ & 5.60$\times 10^{8}$ & 9.41$\times 10^{9}$ & \\
			$E_{\eta\eta}$ \textgreater\ 20\,GeV
			& 9$\times 10^{2}$ & 2.42$\times 10^{4}$ & 1.39$\times 10^{7}$ & 1.09$\times 10^{7}$ & 9.46$\times 10^{7}$ & \\ 
			$\theta_{\eta\eta}$ \textless\ 30\degree
			& 8$\times 10^{2}$ & 2.11$\times 10^{4}$ & 6.76$\times 10^{6}$ & 5.68$\times 10^{6}$ & 5.17$\times 10^{7}$ & \\ 
			\hline
			$m_{\eta\eta} \in$ (5.233, 5.326)\,GeV 
			& 7$\times 10^{2}$ & 2.1$\times 10^{3}$ & 2.3$\times 10^{3}$ & 7$\times 10^{2}$ & 8.0$\times 10^{3}$ & \tabincell{c}{17\%\\ $\pm$ 2\%} \\
			\hline
			$m_{\eta\eta} \in$ (5.310, 5.423)\,GeV
			& 2$\times 10^{2}$ & 1.92$\times 10^{4}$ & 2.2$\times 10^{3}$ & 1.0$\times 10^{3}$ & 7.4$\times 10^{3}$ & \tabincell{c}{0.90\%\\ $\pm$ 0.05\%} \\
			\hline
		\end{tabular}
	}
	\caption{\Boeta\ and \Bseta\ yields at each step of the selection chain and final precisions ($\sqrt{S+B}/S$) when using the baseline $b$-tagging and ECAL energy resolution of $\frac{3\%}{\sqrt{E}} \oplus 0.3\%$. In the last two rows, selections on \MEtaEta\ are applied, the first one is for \Boeta, and the other one is for \Bseta. }
	\label{tab:EtaCutChain}
\end{table}

\subsection{Estimation of other effects}
\label{sec:othereffect}
Besides the $b$-tagging performance and the ECAL energy resolution, the following effects can further deteriorate the measurement precisions of these four channels.

\begin{figure}[htbp]
	\centering
	\subfigure[]{
		\label{fig:PhotonAngle}
		\includegraphics[width=0.45\textwidth]{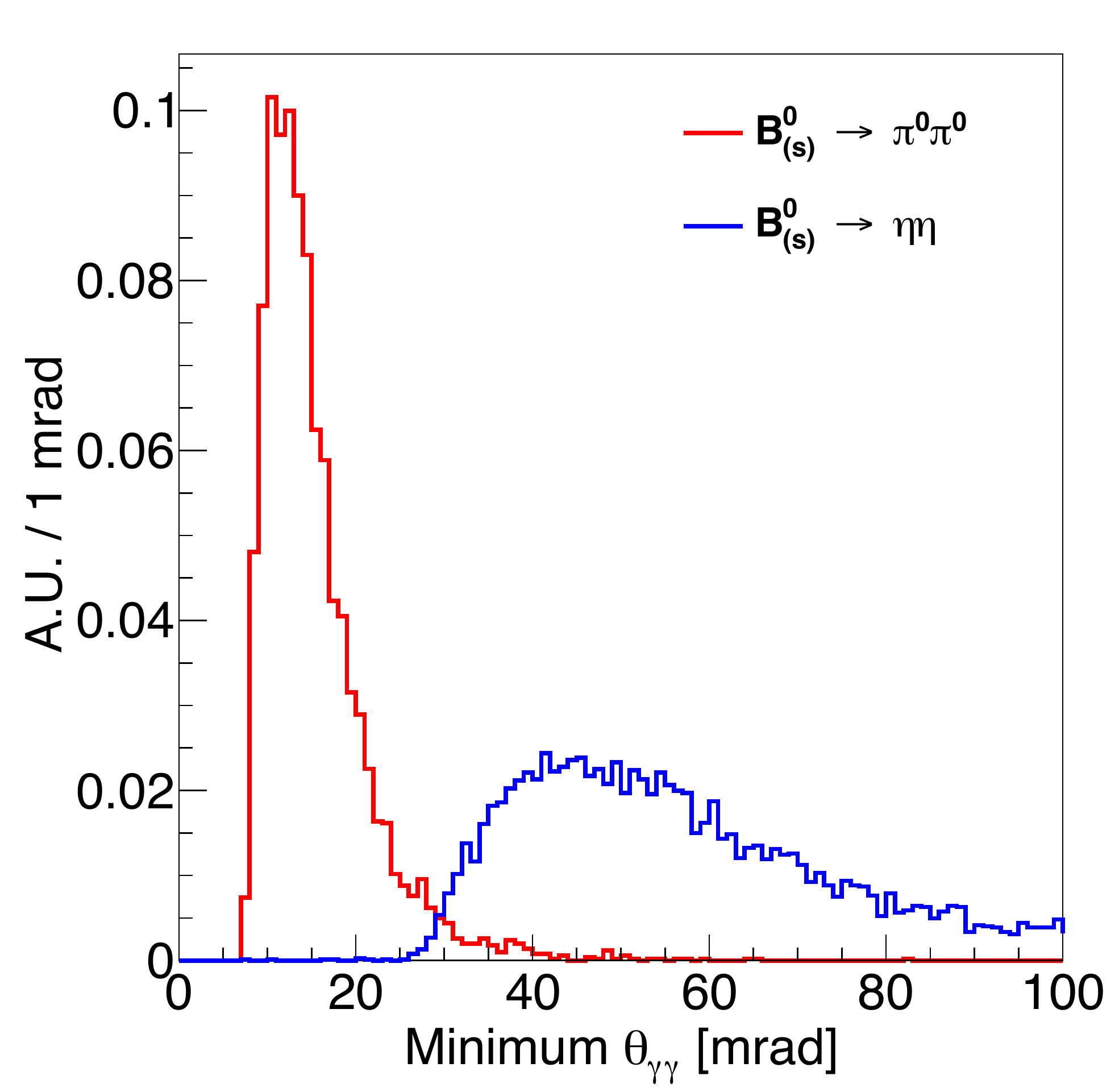}
	}
	\subfigure[]{
		\label{fig:PhotonSeparation}
		\includegraphics[width=0.42\textwidth]{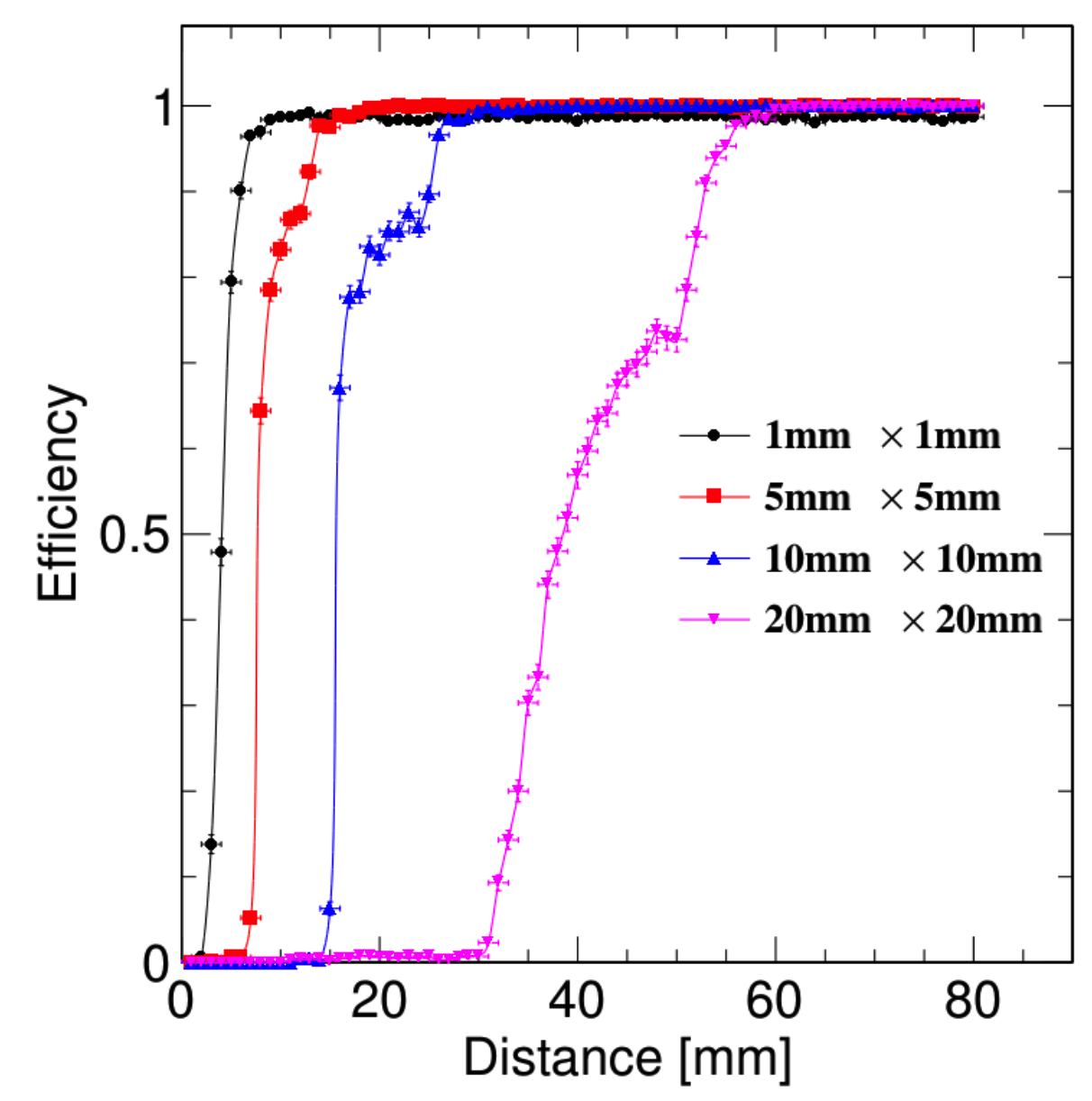}
	}
	\caption{ (a) Distribution of the minimum opening angle among four photons from \Bospio\ and \Boseta. (b) Reconstruction efficiency of two parallel 5\,GeV photons as a function of their separation distance with different ECAL sensor sizes. The blue curve corresponds to the CEPC baseline ECAL~\cite{CEPC_CDR_Phy}.}
\end{figure}

\begin{enumerate}
\item Photon conversion:
About 5--10\% of photons in the central region and nearly 25\% of photons in the forward region may convert into $e^{+}e^{-}$ pairs due to the interaction with the materials in front of the ECAL. Referring to the current study, nearly 80\% of the converted photons can be recovered~\cite{CEPC_CDR_Phy}. Accordingly, we estimate a 3\% effective conversion loss from each photon and a corresponding $\sim$12\% extra efficiency loss for \Bospio\ and \Boseta\ measurements.

\item Photon separation:
In the case of \Bospio\ and \Boseta, photons from high energy \pio\ or $\eta$ are likely to be merged into a single EM cluster. Figure~\ref{fig:PhotonAngle} shows the distribution of the minimum opening angle $\theta_{\gamma\gamma}$ among four photons from \Bospio\ and \Boseta\ after applying the energy and angle selections on \pio\ and $\eta$ pairs described above. Due to the larger $m_\eta$, the opening angle between two photons from $\eta$\ decay is larger than that from \pio\ decay with the same energy.
As for the separation power of the ECAL, it not only depends on the granularity, but also on the intrinsic lateral development of the EM shower which is determined by the Moli\`ere radius ($R_{\rm M}$) of the ECAL material.
According to the full simulation study of di-photon separation power in figure~\ref{fig:PhotonSeparation}, the CEPC baseline sampling silicon-tungsten (Si-W) ECAL can separate two 5\,GeV photons 2\,cm apart with $\sim$80\% efficiency~\cite{CEPC_CDR_Phy}. Assuming an ECAL with an inner radius of 2\,m, 2\,cm separation distance corresponds to an angular separation of 10 mrad. The separation power of 10 mrad well satisfies the photon separation requirement of \Boseta, while the loss from \Bospio\ events with the minimum $\theta_{\gamma\gamma}$ smaller than 10\,mrad is about 10\%.
But for a crystal ECAL, which has larger $R_{\rm M}$ than that of tungsten\footnote{$R_{\rm M} \approx$ 2\,cm for crystals such as PbWO$_4$, BGO, and LYSO. $R_{\rm M} \approx$ 1\,cm for tungsten.}, the shower lateral profile is wider.
If consider at least 3\,$R_{\rm M}$ ($\sim$6\,cm) separation distance of two photons in the crystal ECAL, the corresponding angular separation is about 30\,mrad, which will lead to about 97\% efficiency loss of \Bospio\ signal.
So the successful separation of two photons with opening angle larger than 10\,mrad is critical to the efficient reconstruction of \Bospio, and is thus chosen as the reference point in this paper.
To satisfy this separation requirement with an ECAL which also has high energy resolution indeed needs dedicated studies and innovative designs of the ECAL configuration and the clustering algorithm.
We leave it for the future exploration.

\item Photon angular resolution:
The photon angular resolution $\sigma_{\theta, \phi}$ is expected to reach the level of $\frac{1\sim2\ \text{mrad}}{\sqrt{E (\text{GeV})}}$ for the future high granular ECAL. When the ECAL energy resolution is $\frac{3\%}{\sqrt{E}} \oplus 0.3\%$ or worse, the contribution of the photon angular resolution of $\sigma_{\theta} < \frac{2\ \text{mrad}}{\sqrt{E (\text{GeV})}}$ to the $B$-meson mass resolution is no more than 2\,MeV. Compared to the ultimate $B$-meson mass resolution of $\mathcal{O}(50)$\,MeV, the effect of the angular resolution can be safely ignored.

\item Unknown photon vertex:
All photons are reconstructed assuming they originate from the primary vertex (PV) of the collision events. Therefore, a photon's momentum direction inferred by ECAL hits could be shifted if it is from a secondary vertex (SV) away from the PV. For the four photons from the $B$ meson whose typical flight length is $\mathcal{O}$(mm), photon vertices assigned to the PV will shift the reconstructed $B$-meson mass spectrum to the lower mass. However, the maximum mass shift does not exceed 50\,MeV and will not further degrade the separation between \Bo\ and \Bs. In particular, the full width at half maximum (FWHM) of the mass difference caused by the unknown SV position is less than 3\,MeV and can also be safely ignored.

\item Fake photons:
In a realistic reconstruction procedure, there could be many fake photons from hadronic interactions, fluctuations of electromagnetic showers, or the overlapping of neighboring showers. However, the typical energies of fake photons are below 1--2\,GeV~\cite{ALEPH_FakePhoton}, which is not the energy range concerned in this analysis. The expected high granularity and the ultra-fast timing measurement of future detectors are also promising to suppress fake photons. Henceforth, we anticipate the impact of the fake photons on \Bospio\ and \Boseta\ measurements to be negligible.
\end{enumerate}

Considering all the above, the estimated uncertainties of \Bospio\ modes further deteriorate by about 12\% due to the efficiency loss caused by the photon conversion and unresolved photon pairs. The precisions of \Boseta\ mainly suffer from the photon conversion and only drop by about 6\%. The resulting precisions of \Bopio, \Bspio, \Boeta, and \Bseta\ are projected to be 0.45\%, 4.5\%, 18\%, and 0.95\%, respectively, under the reference detector setup. 

\section{Impact on the CKM angle $\alpha$ extraction and the CKM global fit}
\label{sec:alpha}
Amongst the four $B^0_{(s)}$ rare decay modes considered here, \Bopio\ is particularly relevant as it provides a way to extract the CKM angle $\alpha$ via an isospin analysis~\cite{Gronau:1990ka,IsospinAna,Descotes-Genon:2017thz}. Following the reference detector setup, $\mathcal{B}$(\Bopio) will be measured with an unprecedented precision. The high reconstruction efficiency and signal purity indicate that the Tera-$Z$ will be able to contribute to the measurement of $CP$ violation parameters needed to constrain $\alpha$.

\subsection{Sensitivity to $CP$ observables in $B\to \pi\pi$ modes}
\label{sec:alpha_input}
The extraction of $\alpha$ through the isospin analysis of the $B \to \pi\pi$ system involves at least six observables, as the fit has six unknown parameters to be constrained~\cite{Gronau:1990ka,IsospinAna,Descotes-Genon:2017thz}.  Note that the isospin analysis of the $B\to\pi\pi$ system requires some inputs from the other two isospin-related modes, i.e. $B^+ \to \pi^+\pi^0$ and $B^0 \to \pi^+\pi^-$. In the following, we will adopt the shorthand notation of $00$, $+0$, and $+-$ in the superscript to stand for the three relevant modes. The current inputs to the fit are summarized in table~\ref{tab:SixParameterWA}, together with projections from Belle II and LHCb. They include three branching ratios (BR), denoted by $\mathcal{B}$, and three $CP$ violation parameters.

\begin{table}[thp]
	\centering
	\resizebox{1.\columnwidth}{!}{
		\begin{tabular}[t]{|c|cccc|}
			\hline
			Parameters & World average \cite{PDG2020} & Belle (0.8 ab$^{-1}$) & Belle II (50 ab$^{-1}$) \cite{BelleII2019} & LHCb \\
			\hline
			$\mathcal{B}^{00}$ ($\times 10^{-6}$) & \makecell{1.59\\ $\pm$ 0.26 (16\%)} & \makecell{1.31 \cite{Belle:2017lyb} \\ $\pm$ 0.19 (14.5\%) $\pm$ 0.19} & \makecell{1.31\\ $\pm$ 0.03 (2.3\%) $\pm$ 0.03} & - \\
			\hline
			$\mathcal{B}^{+0}$ ($\times 10^{-6}$) & \makecell{5.5\\ $\pm$ 0.4 (7.3\%)} & \makecell{5.86 \cite{Belle:2012dmz} \\ $\pm$ 0.26 (4.4\%) $\pm$ 0.38} & \makecell{5.86\\ $\pm$ 0.03 (0.6\%) $\pm$ 0.09} & - \\
			\hline
			$\mathcal{B}^{+-}$ ($\times 10^{-6}$) & \makecell{5.12\\ $\pm$ 0.19 (3.7\%)} & \makecell{5.04 \cite{Belle:2012dmz} \\ $\pm$ 0.21 (4.2\%) $\pm$ 0.18} & \makecell{5.04\\ $\pm$ 0.03 (0.6\%) $\pm$ 0.08} & - \\
			\hline
			$C_{CP}^{00}$ & \makecell{-0.33\\ $\pm$ 0.22} & \makecell{-0.14 \cite{Belle:2017lyb} \\ $\pm$ 0.36 $\pm$ 0.10} & \makecell{-0.14\\ $\pm$ 0.03 $\pm$ 0.01} & - \\
			\hline
			$C_{CP}^{+-}$ & \makecell{-0.314\\ $\pm$ 0.030} & \makecell{-0.33 \cite{Belle:2013epq} \\ $\pm$ 0.06 $\pm$ 0.03} & \makecell{-0.33\\ $\pm$ 0.01 $\pm$ 0.03} & \makecell{-0.34 $\pm$ 0.06 $\pm$ 0.01 \cite{LHCb:2018pff} \\ (7 \& 8 TeV, 3.0 fb$^{-1}$)\\ -0.311 $\pm$ 0.045 $\pm$ 0.015 \cite{LHCb:2020byh} \\ (13 TeV, 1.9 fb$^{-1}$)\\ $\pm$ 0.004 (stat. only) \cite{LHCb:2018roe}\\ (Run 1--6, 300\,fb$^{-1}$)}  \\
			\hline
			$S_{CP}^{+-}$ & \makecell{-0.670\\ $\pm$ 0.030} & \makecell{-0.64 \cite{Belle:2013epq} \\ $\pm$ 0.08 $\pm$ 0.03} & \makecell{-0.64\\ $\pm$ 0.01 $\pm$ 0.01} & \makecell{-0.63 $\pm$ 0.05 $\pm$ 0.01 \cite{LHCb:2018pff} \\ (7 \& 8 TeV, 3.0 fb$^{-1}$)\\ -0.706 $\pm$ 0.042 $\pm$ 0.013 \cite{LHCb:2020byh} \\ (13 TeV, 1.9 fb$^{-1}$)\\ $\pm$ 0.004 (stat. only) \cite{LHCb:2018roe}\\ (Run 1--6, 300\,fb$^{-1}$)} \\
			\hline
		\end{tabular}
	}
	\caption{ Six input parameters currently used to determine the CKM angle $\alpha$ via $B\to\pi\pi$ modes. The percentages in parentheses represent the relative statistical uncertainties of the three BRs.}
	\label{tab:SixParameterWA}
\end{table}

The relative statistical uncertainties of $\mathcal{B}$($B\to \pi\pi$) are estimated through a simplified approach using the relation
\begin{equation}
    \frac{\sigma_{\mathcal{B}}}{\mathcal{B}} \simeq \frac{1}{\sqrt{N_{\rm eff}}}\equiv\frac{1}{\sqrt{ \text{Yield}\times \epsilon \times p }}~,
\end{equation}
where $\epsilon$ and $p$ stand for the signal efficiency and purity $\in [0,1]$. The anticipated relative precision of the signal strength of $B^0 \to \pi^0\pi^0$ at Tera-$Z$ is evaluated to be 0.45\% this way, according to the discussion at the end of section~\ref{sec:othereffect}. Meanwhile, given the high $\epsilon \times p$ and statistics of the $4\gamma$ final state achieved in section~\ref{sec:CutChain}, it is reasonable to expect the performance of $B^+\to\pi^+\pi^0$ and $B^0\to\pi^+\pi^-$ channels at Tera-$Z$ will also be excellent. Both decay channels feature charged final state particles allowing tracking with a momentum resolution almost one order of magnitude better than that of the calorimetry. Their BRs are also a few times larger than \Bopio, leading to higher signal efficiency, purity, and more precise determinations of $B$-meson lifetimes.

\begin{figure}[htbp]
	\centering
	\includegraphics[width=0.45\textwidth]{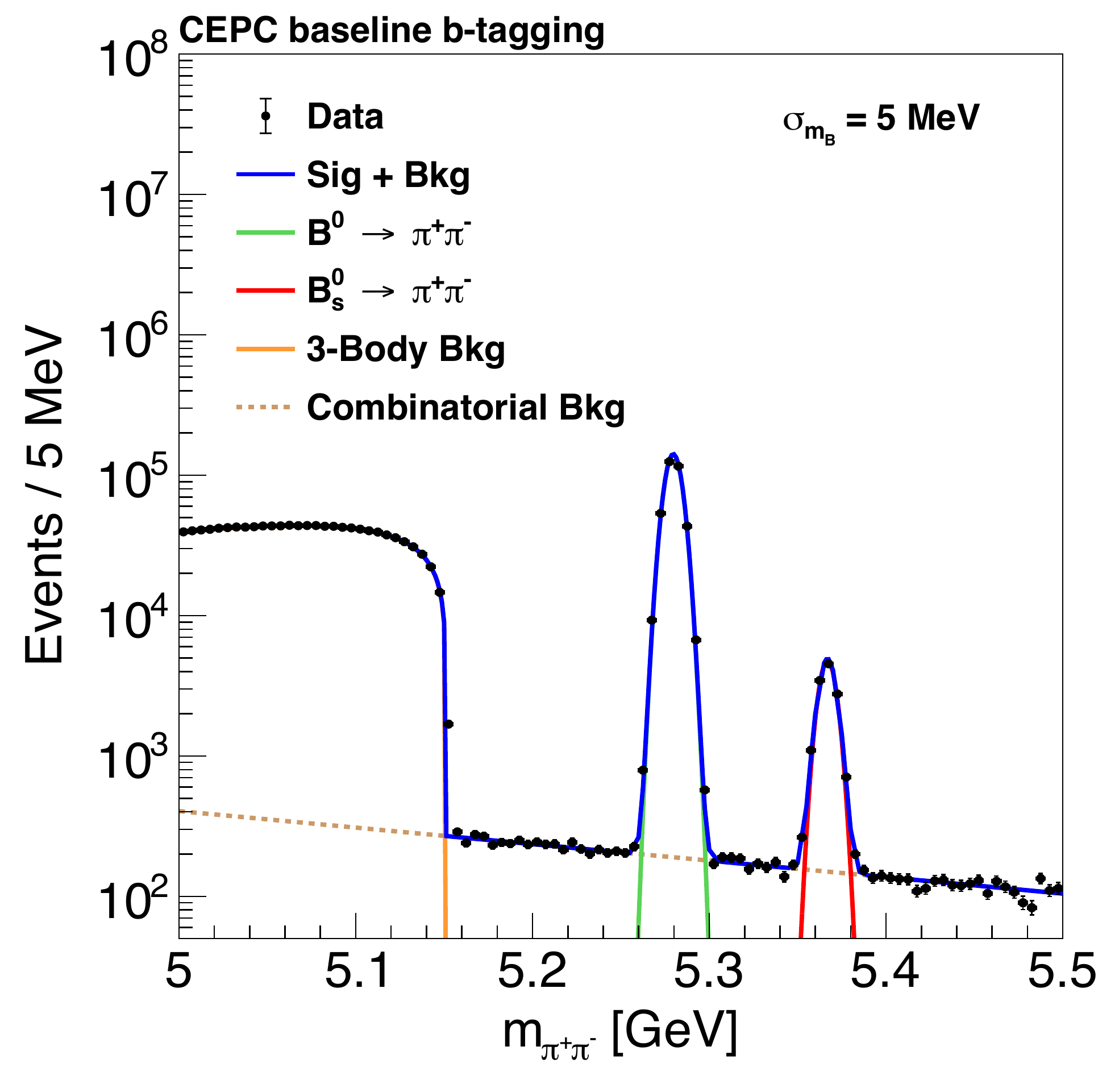}
	\caption{The anticipated $m_{\pi^+\pi^-}$ distribution in the branching ratio measurement of $B^0\to\pi^+\pi^-$ at Tera-$Z$.}
	\label{fig:PiPiSigBkg}
\end{figure}

The relatively reasonable guess at the efficiency and purity of $B^0\to\pi^+\pi^-$ is obtained from a simple analysis at the truth level. In particular, events with hard $\pi^+\pi^-$ pairs, each with energy $>$ 1\,GeV and their total energy $>$ 20\,GeV, are selected. The $\pi^+\pi^-$ pair also needs to share the same vertex and a separation angle $<30^\circ$. The combinatorial background originating from the PV is further suppressed by applying the criterion that the displacement of the $\pi^+\pi^-$ vertex is larger than 100\,$\upmu$m. We assume the $\pi^\pm$ track reconstruction efficiencies are close to 100\%, with a typical track momentum resolution of 0.1\%, corresponding to resonance peak widths about 5\,MeV. Finally, since all events must be tagged by the inclusive $b$-tagging algorithm, the signal efficiency and purity are rescaled by corresponding factors of 0.8 and 0.9, respectively. The $B^0_{(s)}\to\pi^+\pi^-$ events and backgrounds, including the combinatorial ones and the partially reconstructed three-body $b$-hadron decays, are shown in figure~\ref{fig:PiPiSigBkg}. The signal efficiency for $B^0\to\pi^+\pi^-$ is $\gtrsim$ 55\% with a purity of 99\% (in 3\,$\sigma$ mass window)\footnote{The effect from an imperfect particle identification (PID) performance is also assumed to be negligible. Even if the charged kaons are misidentified as pions, the small combinatorial background is insensitive to the small kaon population~\cite{An:2018jtk}. The mass difference between the peaks of the $B^0\to\pi^+\pi^-$ signal and the misidentified $B^0\to K^+\pi^-$ background will be about 90\,MeV at the $Z$-pole, which is enough to separate these two modes when the $B$-meson mass resolution is 5\,MeV.}. We finally take the modest efficiency of 55\%, purity of 95\%, and the corresponding relative precision of 0.18\% as the baseline performance of the $B^0\to\pi^+\pi^-$ measurement at Tera-$Z$, leaving room for more optimistic assumptions. The performance of $B^+\to\pi^+\pi^0$ is anticipated to be between $B^0\to\pi^0\pi^0$ and $B^0\to\pi^+\pi^-$. As the crude estimation, we assign the mode's efficiency $\simeq 50\%$ and purity $ \simeq 85\%$, leading to a relative precision of 0.19\%. The estimates above are summarized in table~\ref{tab:B2Pi_EffPur}.

\begin{table}[thp]
	\centering
	\resizebox{1.\columnwidth}{!}{
		\begin{tabular}[t]{|c|ccccc|c|}
			\hline
			Channel
			& \makecell{Branching ratio \\ $\mathcal{B}$ ($\times 10^{-6}$)} & \makecell{Yield \\ at Tera-$Z$} & \makecell{Efficiency \\ $\epsilon$} & \makecell{Purity \\ $p$} & $\epsilon \times p$ & \makecell{$\sigma_\mathcal{B} / \mathcal{B}$} \\
			\hline
			$B^0 \to \pi^0\pi^0$
			& 1.59 & $1.9 \times 10^{5}$ & 32\% & 80\% & 25\% & 0.45\% \\
			$B^+ \to \pi^+\pi^0$
			& 5.50 & $6.6 \times 10^{5}$ & 50\% & $\gtrsim$ 85\% & 43\% & 0.19\% \\
			$B^0 \to \pi^+\pi^-$
			& 5.12 & $6.1 \times 10^{5}$ & 55\% & $\gtrsim$ 95\% & 52\% & 0.18\% \\
			\hline
		\end{tabular}
	}
	\caption{ Estimation of $B \to \pi\pi$ relative precision (signal yield, efficiency, and purity) at Tera-$Z$. }
	\label{tab:B2Pi_EffPur}
\end{table}

In addition to the three branching ratios above, the $CP$ violation parameters of $B\to\pi\pi$ decays are crucial for the determination of $\alpha$. Besides the three $CP$ asymmetries listed in table~\ref{tab:SixParameterWA} used to reconstruct the $B$- and $\bar{B}$-meson isospin triangles separately, the mixing-induced $CP$ asymmetry of \Bopio, denoted by $S_{CP}^{00}$, carries the information on the relative phase between the two isospin triangles. It proves useful to recall the interpretation of these asymmetries in order to discuss their status in the Tera-$Z$ context. In decays of $B$ mesons to a $CP$ eigenstate final state $f_{CP}$, the $CP$ violation originates from the interference between the decay and $B$--$\bar{B}$ mixing. Firstly we define the relevant $CP$-violating parameters as
\begin{equation}
\label{eq:Acp_CS}
C_{f_{CP}} \equiv \frac{1-|\lambda_{f_{CP}}|^2}{1+|\lambda_{f_{CP}}|^2}, ~ S_{f_{CP}} \equiv \frac{2{\rm Im}\lambda_{f_{CP}}}{1+|\lambda_{f_{CP}}|^2}, ~ \lambda_{f_{CP}} \equiv \frac{q_d}{p_d}\frac{\bar{\mathcal{A}}_{f_{CP}}}{\mathcal{A}_{f_{CP}}}~,
\end{equation}
where $\mathcal{A}(\bar{\mathcal{A}})_{f_{CP}}\equiv \langle f_{CP} |\mathcal{H}| B(\bar{B})\rangle$ represent the $B(\bar{B})$ decay amplitudes to the $CP$ eigenstate $f_{CP}$. The parameters $q_d$ and $p_d$ are defined by the linear decomposition of the two mass eigenstates $B_{1,2}\equiv q_d| \bar{B}_d\rangle \pm p_d | B_d\rangle$. The time-dependent decay rates of $B^0 \to \pi \pi$ are complicated functions due to the interplay of multiple mixing parameters and time scales like the mass difference of the two mass eigenstates $\Delta m_d$ and the their averaged decay width $\Gamma_d$. However, they get simplified hugely if we ignore the small decay-width difference of the two $B$ mass eigenstates $\Delta \Gamma_d \ll \Gamma_d$~\cite{PDG2020} and if we adopt the approximation that |$q_d/p_d|\simeq 1$~\cite{PDG2020}. The time-dependent decay rates become~\cite{Dunietz:2000cr,IsospinAna,Gersabeck:2011xj}:
\begin{equation}\label{eq:timeevolBBbar}
    \Gamma_{B^0/\bar{B}^0 \to f_{CP}}(t) \propto e^{-\Gamma_d t}\bigg[ 1\pm C_{f_{CP}} \cos(\Delta m_d t) \mp S_{f_{CP}}\sin(\Delta m_d t) \bigg]~.
\end{equation}
This leads to the definition of the measured time-dependent $CP$ asymmetry:
\begin{equation}
\label{eq:Acp_t}
a_{f_{CP}}(t) \equiv \frac{\Gamma_{\bar{B} \to f_{CP}}(t) - \Gamma_{B \to f_{CP}}(t)}{\Gamma_{\bar{B} \to f_{CP}}(t) + \Gamma_{B \to f_{CP}}(t)} \simeq -C_{f_{CP}} \cos(\Delta m_d t) + S_{f_{CP}} \sin(\Delta m_d t)~,
\end{equation}
and the time-integrated $CP$ asymmetry:
\begin{equation}
\label{eq:Acp}
a_{f_{CP}} \equiv \bigg(1+\frac{\Delta m_d^2}{\Gamma_d ^2}\bigg)\frac{\int_0^\infty \Gamma_{\bar{B} \to f_{CP}}(t) - \Gamma_{B \to f_{CP}}(t) dt}{\int_0^\infty \Gamma_{\bar{B} \to f_{CP}}(t)  + \Gamma_{B \to f_{CP}}(t) dt} \simeq   -C_{f_{CP}}+ S_{f_{CP}}\frac{\Delta m_d}{\Gamma_d}~.
\end{equation}
We notice that neutral-meson mixing affects $a_{f_{CP}}$. It was indeed shown to impact time-integrated measurements of decays of $b\bar{b}$ pairs produced incoherently at hadronic machines or $Z$-factories in various modes, in particular non-leptonic two-body $B_{d,s}$ decays~\cite{Descotes-Genon:2011rgs}, $B_s\to\mu\mu$~\cite{DeBruyn:2012wk,DeBruyn:2012wj}, and rare $b\to s\ell\ell$ transitions~\cite{Descotes-Genon:2015hea,Descotes-Genon:2020tnz}. Similar expressions for the time-integrated $CP$ asymmetry also occur in the measurement of $D^0\to \pi^+\pi^-$ and $D^0\to K^+K^-$ decays~\cite{Gersabeck:2011xj,CDF:2011ejf}.

For {\Bopio} measurement through the four-photon final state without tracking, the lifetime of $B$ mesons cannot be precisely measured, and the $CP$ asymmetry takes the time-integrated form. In this case, the $Z$-factory measures a linear combination of $C_{CP}^{00}$ and $S_{CP}^{00}$ instead of $C_{CP}^{00}$. Such a situation differs from the $C_{CP}^{00}$ measurement at $B$-factories, where the contribution from $S_{CP}^{00}$ cancels out after time integration (see the Appendices of refs.~\cite{Abudinen:2018wce,Descotes-Genon:2011rgs} for detailed derivations)\footnote{More specifically, in $B$-factories~\cite{BaBar:1998yfb}, the $B^0\bar{B}^0$ pair is produced coherently, and eq.~\eqref{eq:timeevolBBbar} is modified in two ways: $t$ measures the difference of decay time between the two $B$-mesons and can thus run from $-\infty$ to $\infty$, and the exponential factor becomes $e^{-\Gamma|t|}$. This means in turn that the integration over time will wash away $S_{f_{CP}}$ in eq.~\eqref{eq:Acp} in the case of $B$-factories.}.

The measured time-integrated $CP$ asymmetry eq.~\eqref{eq:Acp} of \Bopio, denoted by $a_{CP}^{00}$, is also diluted by the $B^0$--$\bar{B}^0$ mixing by a factor of $1+(\Delta m_d/\Gamma_d)^2= (1-2\chi_d)^{-1}+\mathcal{O}(\Delta \Gamma_d/\Gamma_d)^2$, where $\chi_d = 0.1858 \pm 0.0011$ is the time-integrated mixing probability~\cite{PDG2020}. In contrast, for the measurement of $C_{CP}^{+-}$ and $S_{CP}^{+-}$, the two charged $\pi^\pm$ tracks in the final state enable high-precision decay-time measurements via the reconstructed vertex location. Both parameters can be obtained by fitting the measured $a_{CP}^{+-}(t)$ as a function of time and remain free from the dilution factor $1-2\chi_d$.

Another key element for $CP$ parameter measurement is the $b$-flavor charge tagging (also known as the jet charge measurement), i.e. the ability to distinguish whether the observed $B\to\pi\pi$ event stems from \Bo\ or $\bar{B}^{0}$ as an initial state from the $Z\to b\bar{b}$ process. The flavor tagging is thus recovering the initial $b$-flavor charge (+1 for $\bar{B}^0$, $-$1 for $B^0$ and 0 for backgrounds) hidden by the $B^0$--$\bar{B}^0$ mixing. At the $Z$-pole, we expect flavor tagging techniques are similar to the ones at LEP~\cite{ALEPH:1998unp} or LHCb~\cite{LHCb:2012dgy,LHCb:2016mtq} by identifying accompanying particles from either hadronization or another $b$-hadron. The performance is often evaluated by the effective tagging efficiency (power)
\begin{equation}
\epsilon_{\rm eff} \equiv \epsilon_{\rm tag} (1-2\omega)^2~,
\end{equation}
where $\epsilon_{\rm tag}$ is the flavor tagging efficiency and $\omega$ is the wrong tagging fraction. The effective signal statistics when measuring $CP$ properties are further corrected by $\epsilon_{\rm eff}$. The straightforward and comprehensive study in~\cite{JetChargeNote} concludes that the overall $\omega$ of $b$-flavor charge at CEPC is no worse than 35\%, corresponding to an $\epsilon_{\rm eff}$ of 9\%. The dedicated $b$-flavor charge identification algorithm developed for the specific study of $B_s \to J/\psi\phi$~\cite{Mingrui} shows the potential to reduce $\omega$ to 22.5\% and improve $\epsilon_{\rm eff}$ to 20\%. The range of the future $b$-flavor charge tagging power at Tera-$Z$ is thus determined as $\epsilon_{\rm eff} \in [15,25]\%$ accordingly, while a systematic exploration of flavor tagging will be left to future work.

Statistical uncertainties of $CP$ violation parameters at Tera-$Z$ are then approximated by the well-known relation\footnote{In time-dependent measurements, there is another factor $e^{-(\Delta m_d \sigma_t)}/2$ corresponding to the decay lifetime resolution. However, the expected time resolution $\sigma_t$ at Tera-$Z$ is of a few fs~\cite{Mingrui}, which is much less than $\Delta m_d^{-1}$, and the factor can be safely ignored.}:
\begin{equation}
\label{eq:AcpErr_00}
\sigma_{a_{CP}^{00}} \simeq  \frac{1}{(1-2\chi_d) \sqrt{N_{\rm eff} \times \epsilon_{\rm eff} }}~\text{(time-integrated)}~,
\end{equation}
\begin{equation}
\label{eq:AcpErr_PN}
\sigma_{S_{CP}^{+-}} \simeq\sigma_{C_{CP}^{+-}} \simeq \frac{1}{\sqrt{N_{\rm eff} \times \epsilon_{\rm eff} }}~\text{(time-dependent)}~.
\end{equation}
Compared to the time-dependent ones, results from time-integrated measurements are diluted by the factor of $(1-2\chi_d)^{-1} \approx 1.59$ as derived in eq.~\eqref{eq:Acp}. To cross-check, eqs.~\eqref{eq:AcpErr_00} and~\eqref{eq:AcpErr_PN} are also applied to studies listed in table~\ref{tab:SixParameterWA}. The approximation matches with the reported statistical uncertainties within 30\%. The estimated results are shown in the last column of table~\ref{tab:SixParameterTZ} where their variations correspond to $\epsilon_{\rm eff}\in [15,25]\%$. The consequent statistical uncertainties of the three $CP$ asymmetries at Tera-$Z$ are estimated to be $\sigma_{a_{CP}^{00}}$ = 0.014--0.018 and $\sigma_{C_{CP}^{+-}} \approx \sigma_{S_{CP}^{+-}}$ = 0.004--0.005.

\begin{table}[thp]
	\centering
	\resizebox{0.5\columnwidth}{!}{
		\begin{tabular}[t]{|c|c|}
			\hline
			Parameters   & Tera-$Z$ Projection \\
			\hline
			$\sigma_{\mathcal{B}^{00}}/\mathcal{B}^{00}$ & 0.45\% \\
			$\sigma_{\mathcal{B}^{+0}}/\mathcal{B}^{+0}$ & 0.19\% \\
			$\sigma_{\mathcal{B}^{+-}}/\mathcal{B}^{+-}$ & 0.18\% \\
 			$\sigma_{a_{CP}^{00}}$  & $\pm$ (0.014--0.018) \\
			$\sigma_{C_{CP}^{+-}}$  & $\pm$ (0.004--0.005) \\
			$\sigma_{S_{CP}^{+-}}$  & $\pm$ (0.004--0.005) \\
			\hline
		\end{tabular}}
	\caption{Statistical uncertainties for relevant input parameters used to determine $\alpha$ at Tera-$Z$. The ranges of estimation correspond to $\epsilon_{\rm eff}\in [15,25]\%$.}
	\label{tab:SixParameterTZ}
\end{table}

In the above, we discussed the time-integrated measurement of the \Bopio\ mode. Its time-dependent measurement, which requires the vertex information of $\pi^0$ decays to deduce $B^0$ lifetimes, is also possible with an improved precision. In principle, the Dalitz decay of $\pi^0\to e^+e^-\gamma$~\cite{BelleII2019,Monteil:2021ith} or the vertex from converted photons~\cite{BelleII2019} can be used to reconstruct the vertex of \Bo. Both methods are challenged by small effective statistics, either from the small $\pi^0$ Dalitz decay rate ($\sim$1\%) or the low percentage of converted photons. In~\cite{BelleII2019}, a statistical uncertainty of $\sim$0.29 on $S_{CP}^{00}$ is extracted. Considering the much larger $B^0$ boost and the excellent vertex reconstruction at Tera-$Z$, $B$ decay lifetime resolution from $\pi^0$ Dalitz decays is estimated to be $\sim$15\,fs, which is much smaller than the time resolution of $\mathcal{O}$(1)\,ps at Belle II~\cite{BelleII2019}. Combining the larger effective statistics, we argue that the time-dependent $S_{CP}^{00}$ constraint at Tera-$Z$ would be stronger than that at Belle II. Intensive studies on this challenging measurement are strongly recommended for future works.

On the other hand, the physics picture of the time-integrated $CP$ asymmetry measurement at the $Z$-factory is quite different from that at a $B$-factory. The measured $a_{CP}^{00}$ is a combination of $C_{CP}^{00}$ and $S_{CP}^{00}$ at Tera-$Z$, while it is a function of $C_{CP}^{00}$ only at Belle II. This provides another feasible method to extract $S_{CP}^{00}$ by combining the results of Tera-$Z$ and Belle II. Given that the statistical uncertainties of $a_{CP}^{00}$ at Tera-$Z$ and $C_{CP}^{00}$ at Belle II are both $\mathcal{O}(10^{-2})$, the precision of $S_{CP}^{00}$ extracted via this method is expected to be of the same order. The sensitivity of such a novel method to $S_{CP}^{00}$ is thus much higher than the ones from the time-dependent measurements.

\subsection{Extraction of the CKM angle $\alpha$ and the global fit}
\label{sec:CKMfit}

\begin{figure}[h!]
    \centering
    \includegraphics[width=7.5cm]{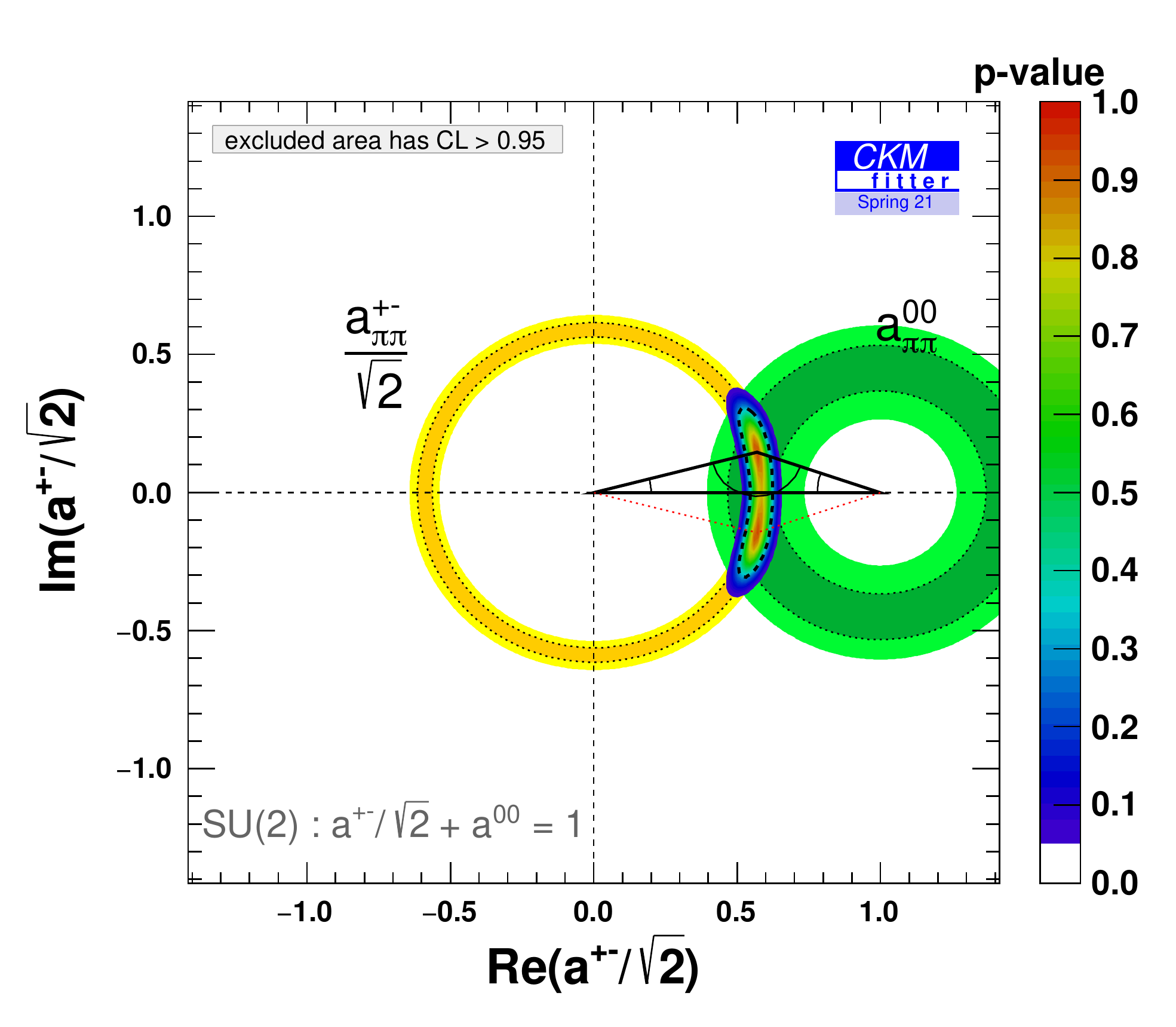}
    \includegraphics[width=7.5cm]{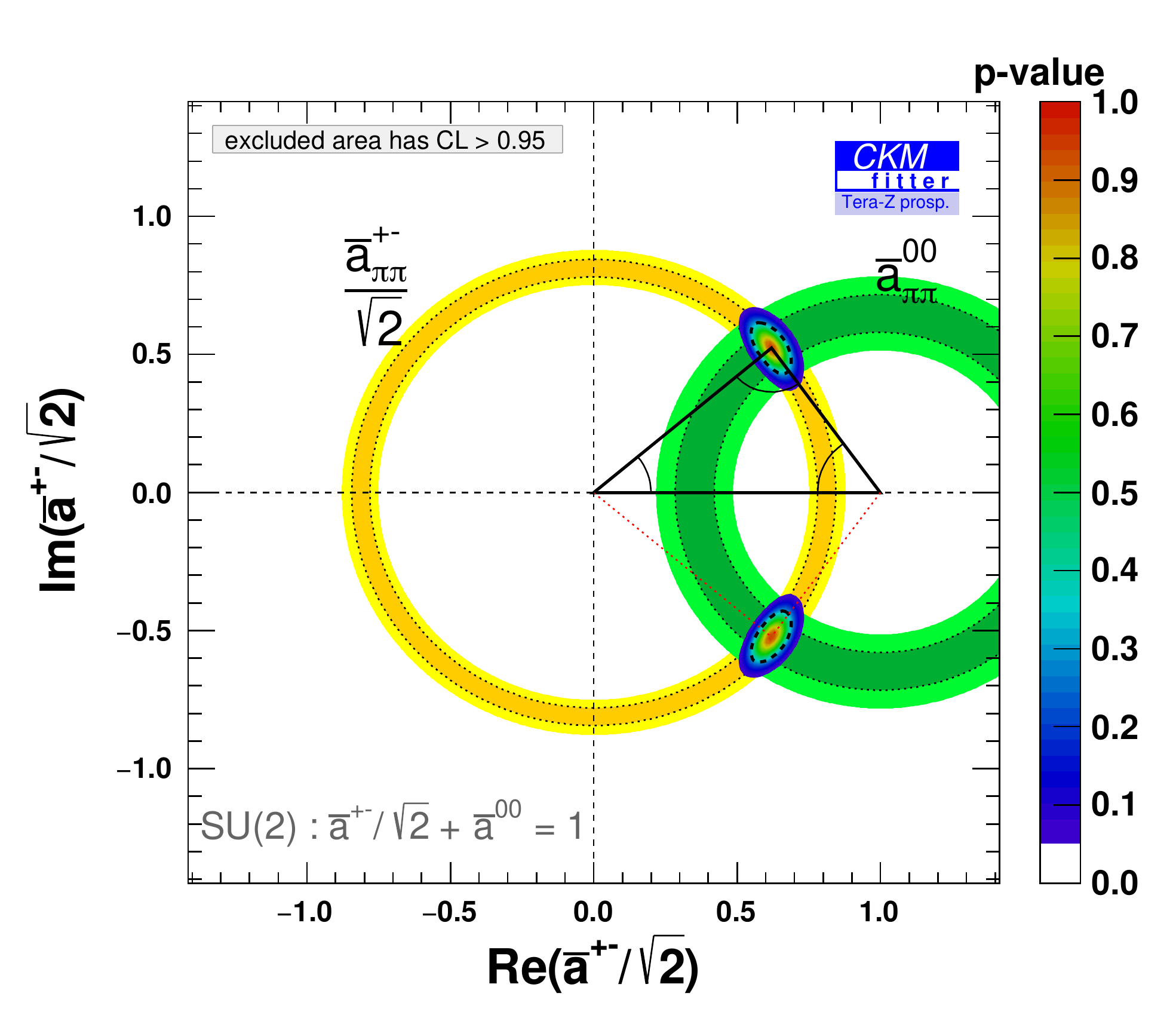}
    \caption{Isospin triangles for $B\to \pi\pi$ ($B^0,B^+$ modes on the left-hand side, $\bar{B}^0,B^-$ modes on the right-hand side) based on the current world average~\cite{Charles:2004jd,IsospinAna}.
    }
    \label{fig:IsospinTrianglesWA}
\end{figure}

\begin{figure}[h!]
    \centering
    \includegraphics[width=8cm]{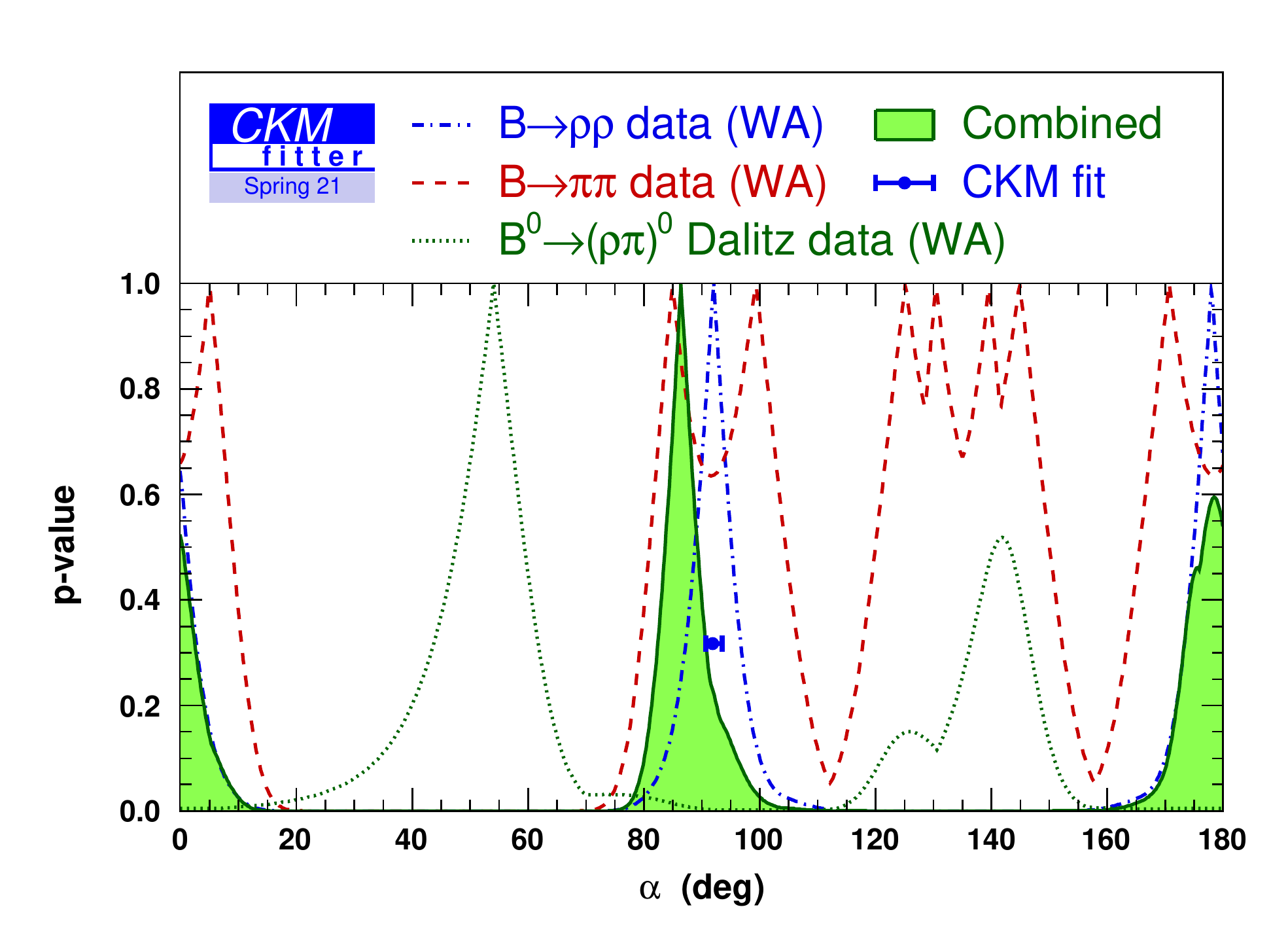}
    \caption{$p$-value for $\alpha$ from various modes ($B\to\pi\pi$ dashed red, $B\to \rho\pi$ dotted green, $B\to \rho\rho$ dotted-dashed blue), from the combination of all modes (solid green) and the global fit prediction (blue interval) based on the current world average~\cite{Charles:2004jd,IsospinAna}.
    }
    \label{fig:alphaWA}
\end{figure}

We can consider the improvements discussed above in the framework of the CKM global fits~\cite{Descotes-Genon:2017thz}. We start from the current determination of $\alpha$ performed by the CKMfitter group~\cite{Charles:2004jd}. As discussed extensively in ref.~\cite{IsospinAna}, the extraction of $\alpha$ from $S_{CP}^{+-}$ is polluted by penguin contributions which can be assessed using isospin symmetry~\cite{Gronau:1990ka,Lipkin:1991st,Charles:1998qx}. Indeed, isospin symmetry enforces triangular relations among hadronic amplitudes, which can be determined from $B\to\pi\pi$ $CP$-averaged branching ratios and the direct $CP$ asymmetries $C_{CP}^{+-}$, $C_{CP}^{00}$ and which can be used to assess the penguin pollution in $S_{CP}^{+-}$ ($B\to \rho\rho$ and $B\to \rho\pi$ modes can be used for similar purposes). The reconstruction of the corresponding triangles from the data suffers from discrete ambiguities that result in (potentially) 8 mirror solutions for $\alpha$. Currently, as shown in figure~\ref{fig:IsospinTrianglesWA}, the $B$-triangle is almost flat, whereas the $\bar{B}$-triangle is not, and the $00$ side has a large uncertainty. This yields the current world average (WA) of $\alpha$ based on $\pi\pi$ alone:
\begin{equation*}
    {\rm WA}: \alpha(\pi\pi)=(93.0\pm 13.6)^\circ~.
\end{equation*}
The resulting determination of $\alpha$ combining $B\to\pi\pi,\rho\pi,\rho\rho$ modes can be compared with the indirect prediction from the rest of the global fit (see also figure~\ref{fig:alphaWA}):
\begin{equation*}
{\rm WA}: \alpha[{\rm combined}]=(86.4^{+4.3}_{-4.0} \cup 178.5^{+3.1}_{-5.2})^\circ~,
\qquad
\alpha[{\rm indirect}]=(91.9^{+1.6}_{-1.2})^\circ~,
\end{equation*}
showing a good agreement (pull of 1.3\,$\sigma$), but an uncertainty significantly larger for the combined (direct) determination.


\begin{figure}[h!]
    \centering
    \includegraphics[width=7cm]{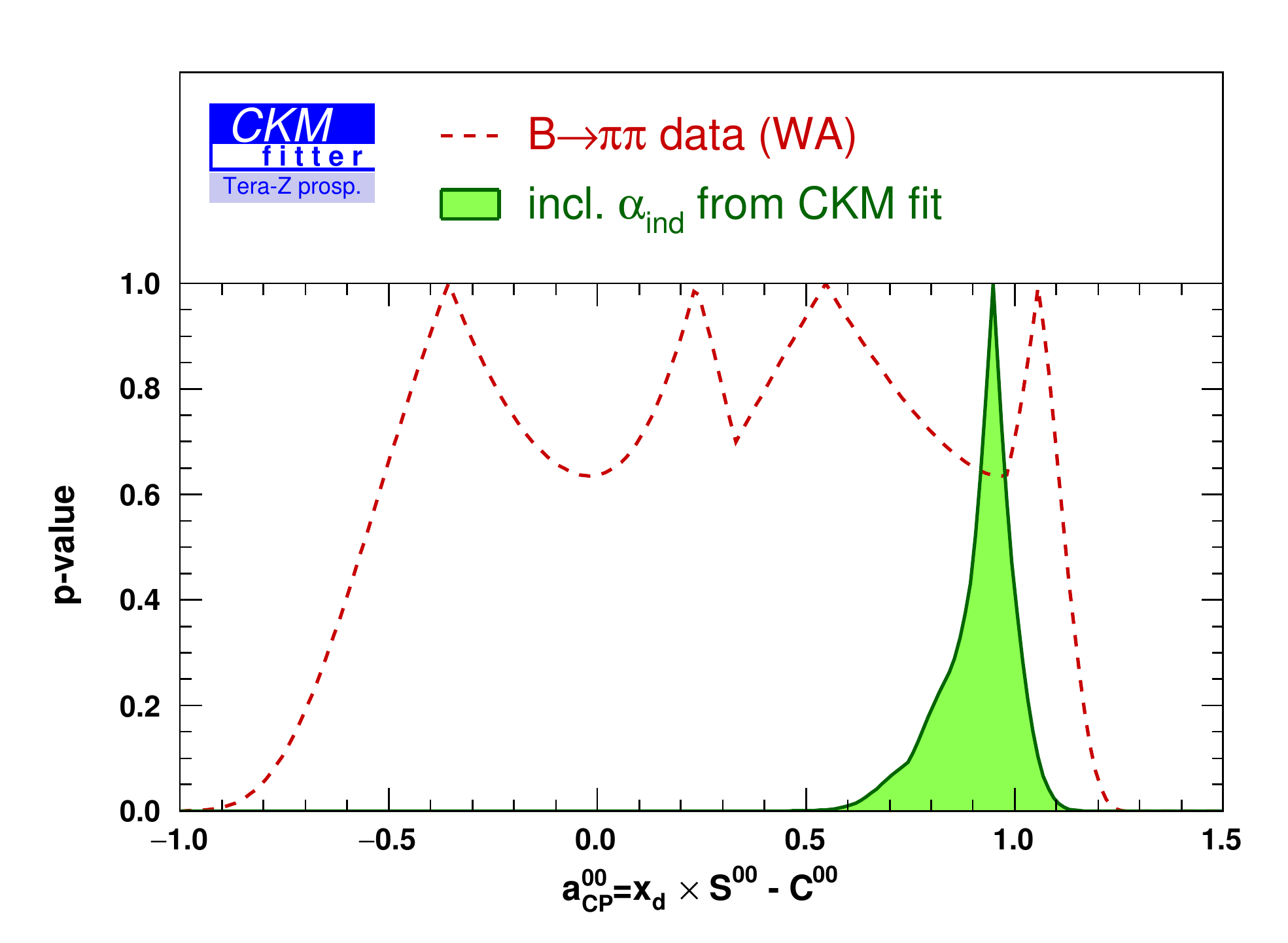}
    \includegraphics[width=7cm]{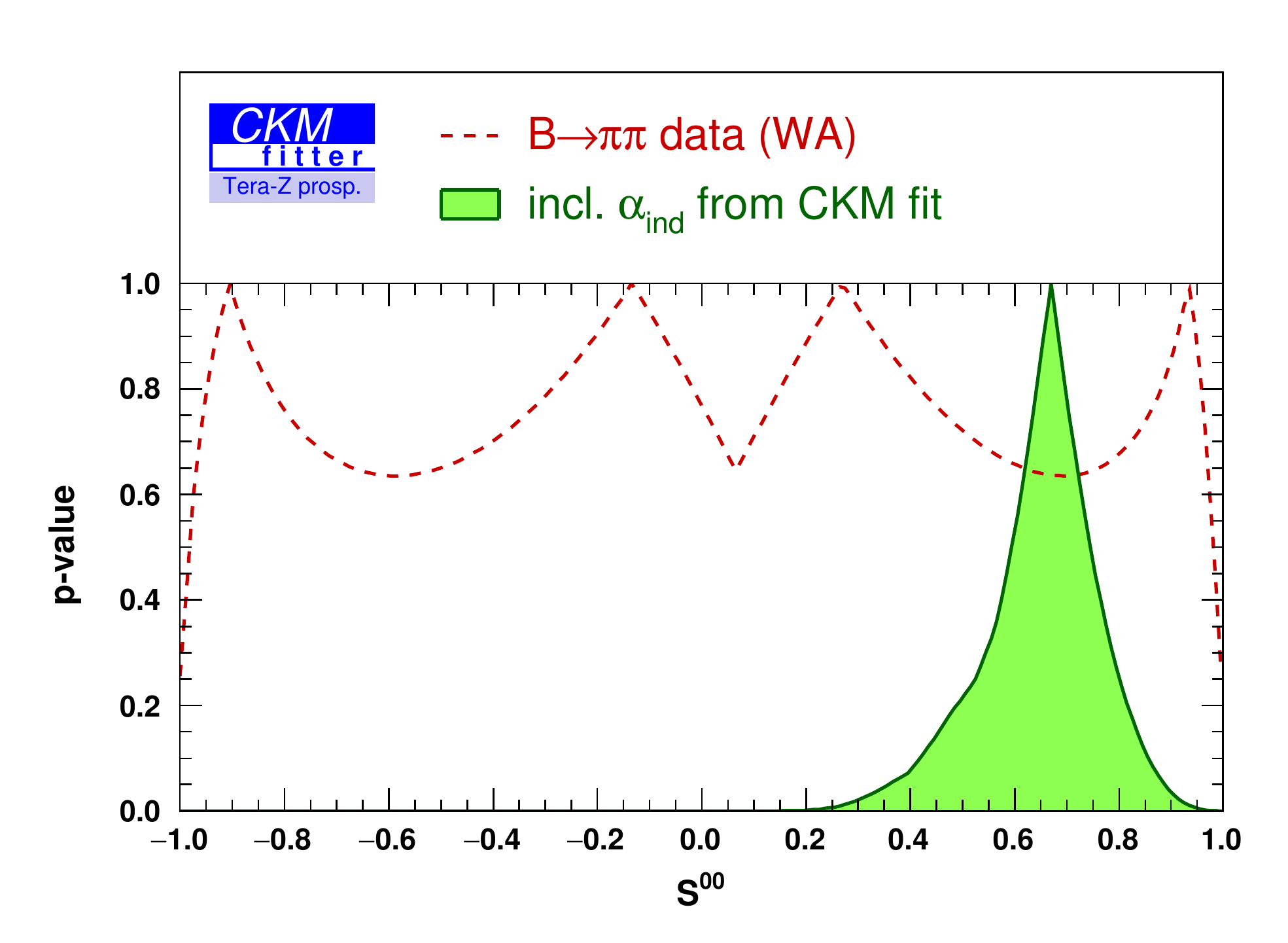}
\caption{$p$-values for $a_{CP}^{00}$ (left) and $S_{CP}^{00}$ (right) from current $B\to \pi\pi$ data alone (dashed red) and the global fit (solid green). The latter yields the predictions $a_{CP}^{00}=0.95^{+0.07}_{-0.08}$ and $S_{CP}^{00}=0.67\pm 0.12$ (as well as $C_{CP}^{00}=-0.43^{+0.12}_{-0.09}$).
}
\label{fig:aCP00S00current}
\end{figure}

\begin{figure}[h!]
    \centering
    \includegraphics[width=7cm]{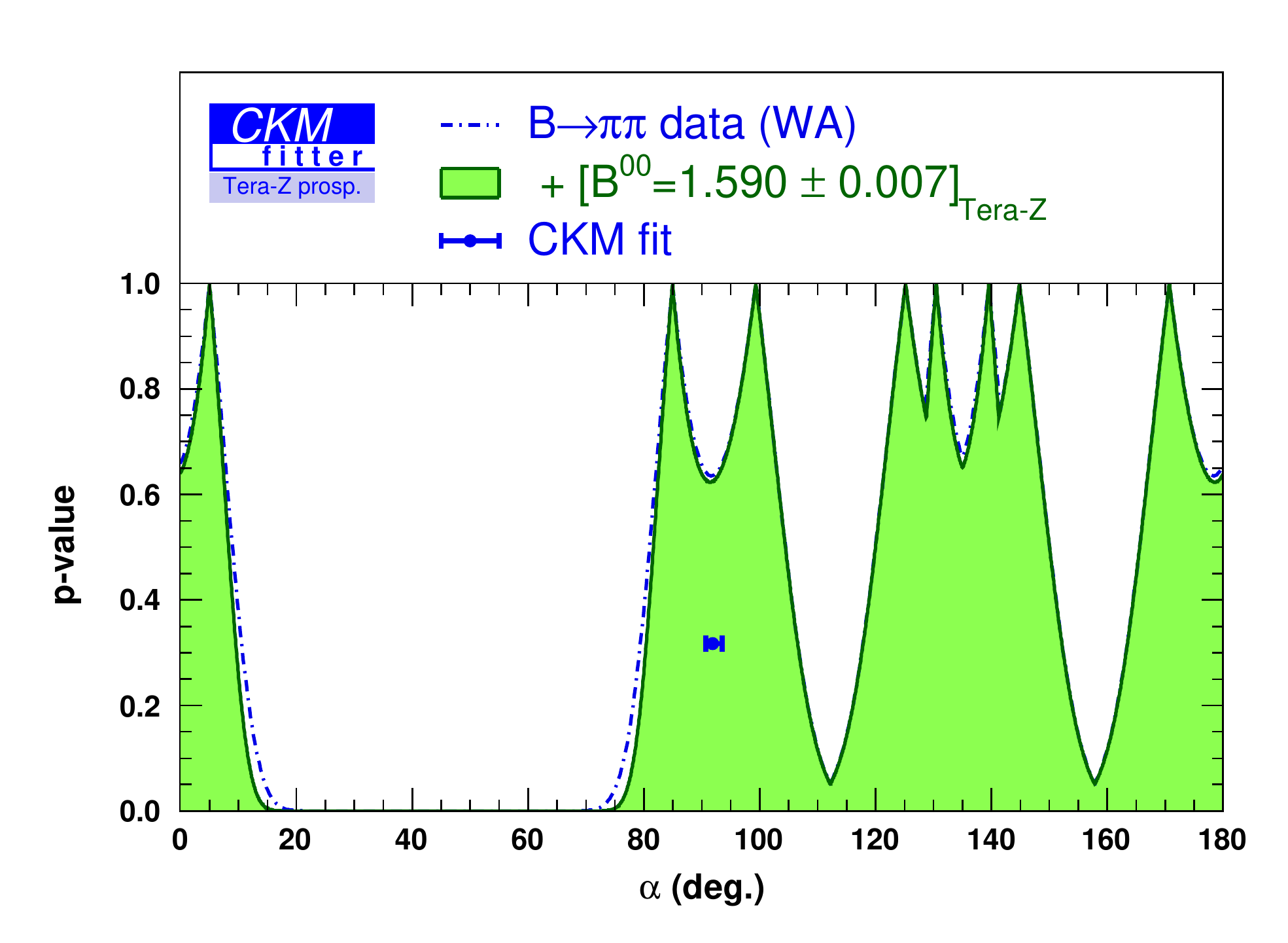}
    \includegraphics[width=7cm]{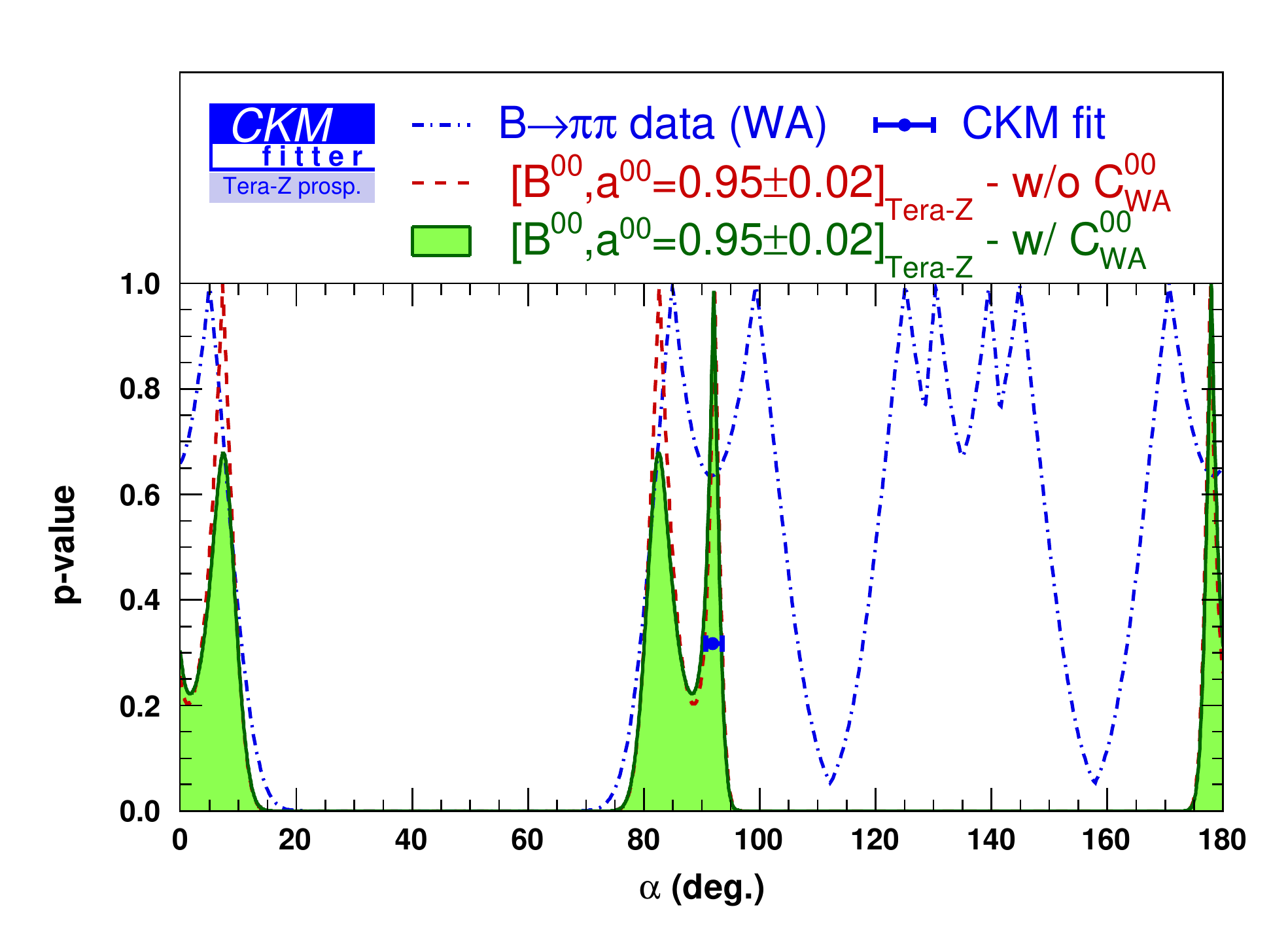}
    \caption{$p$-value for $\alpha$ from $B\to\pi\pi$ measurements.
    On the left: using the current data (dotted-dashed blue) and the improved measurement of $\mathcal{B}(B\to\pi^0\pi^0)$ alone.
    On the right: using the current data (dotted-dashed blue), the improved measurements of $B\to\pi^0\pi^0$ observables but without the current measurement of $C_{CP}^{00}$ (dashed red) and with this measurement (solid green, scenario 1). We take the baseline value $a_{CP}^{00}=0.950\pm 0.018$.
    }
    \label{fig:alphaTera00}
\end{figure}

We are now in a position to study the impact of the Tera-$Z$ measurements on the extraction of $\alpha$. We keep the central values of the various inputs currently used, but rescale the uncertainties according to two scenarios. Moreover, we have to choose a central value for the input of $a_{CP}^{00}$, which combines the two $B^0\to\pi^0\pi^0$  $CP$ asymmetries $C_{CP}^{00}$ (for which there is a determination from $B$-factories) and $S_{CP}^{00}$ (unknown currently)\footnote{In the following, we neglect the (small) uncertainty on $\Delta m_d/\Gamma_d$.}. As discussed in the introduction, these $CP$ asymmetries are expected to have a significant impact on the extraction of $\alpha$, which we will illustrate by considering three different central values for $a_{CP}^{00}$:
\begin{itemize}
\item $a_{CP}^{00}=0.95$, which corresponds to the current prediction of the global fit, as shown on the left-hand side of figure~\ref{fig:aCP00S00current}, and will be our baseline value,
\item $a_{CP}^{00}=0.53$, which matches the peak for $\alpha$ around $100^\circ$ on the left-hand side of figure~\ref{fig:alphaTera00},
\item  $a_{CP}^{00}=0.84$, obtained by taking the central value of the current determination of $C_{CP}^{00}$ from $B$-factories and the prediction for $S_{CP}^{00}$ from the global fit (on the right-hand side of figure~\ref{fig:aCP00S00current}).
\end{itemize}


\begin{figure}[h!]
    \centering
    \includegraphics[width=7.5cm]{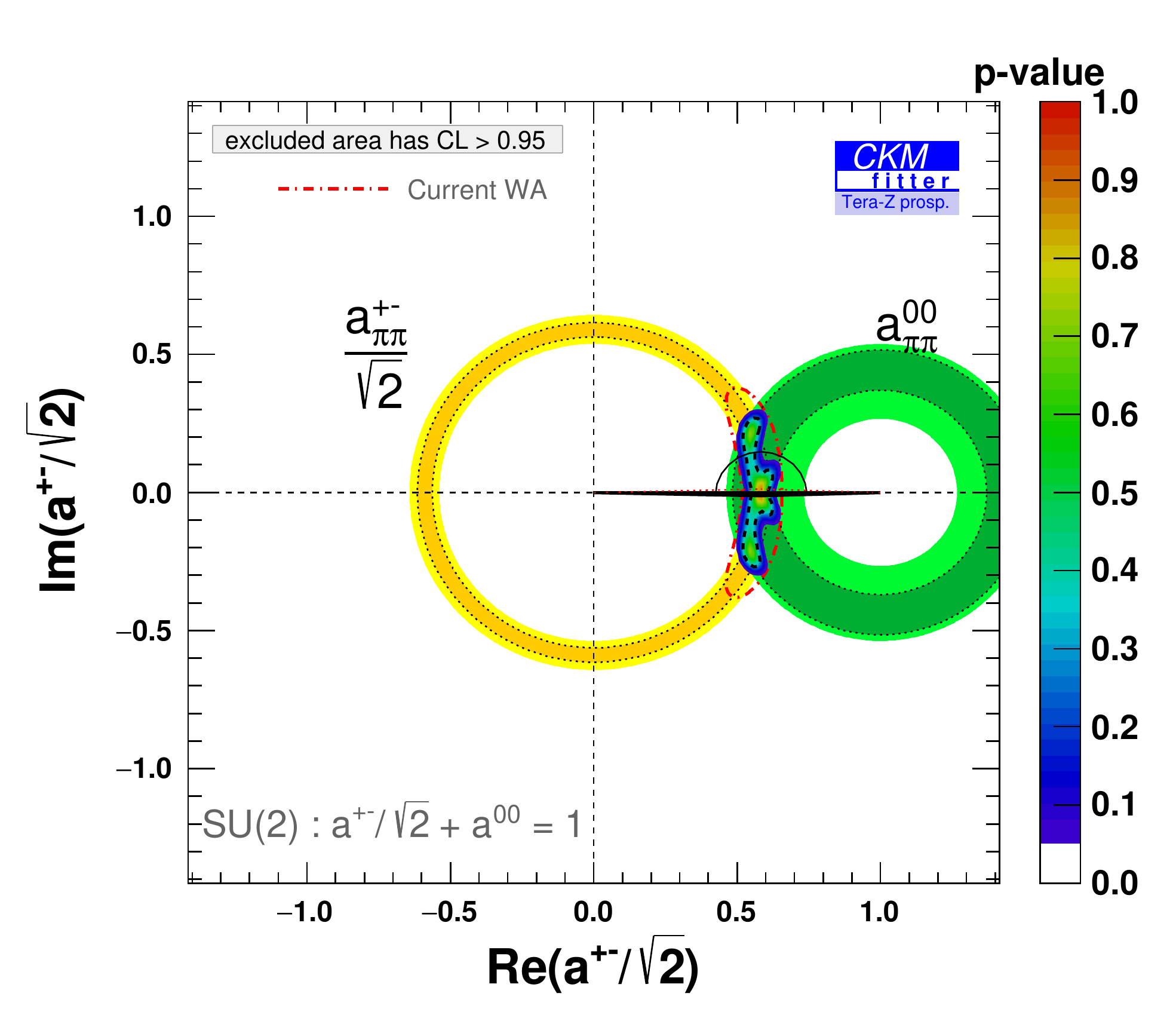}
    \includegraphics[width=7.5cm]{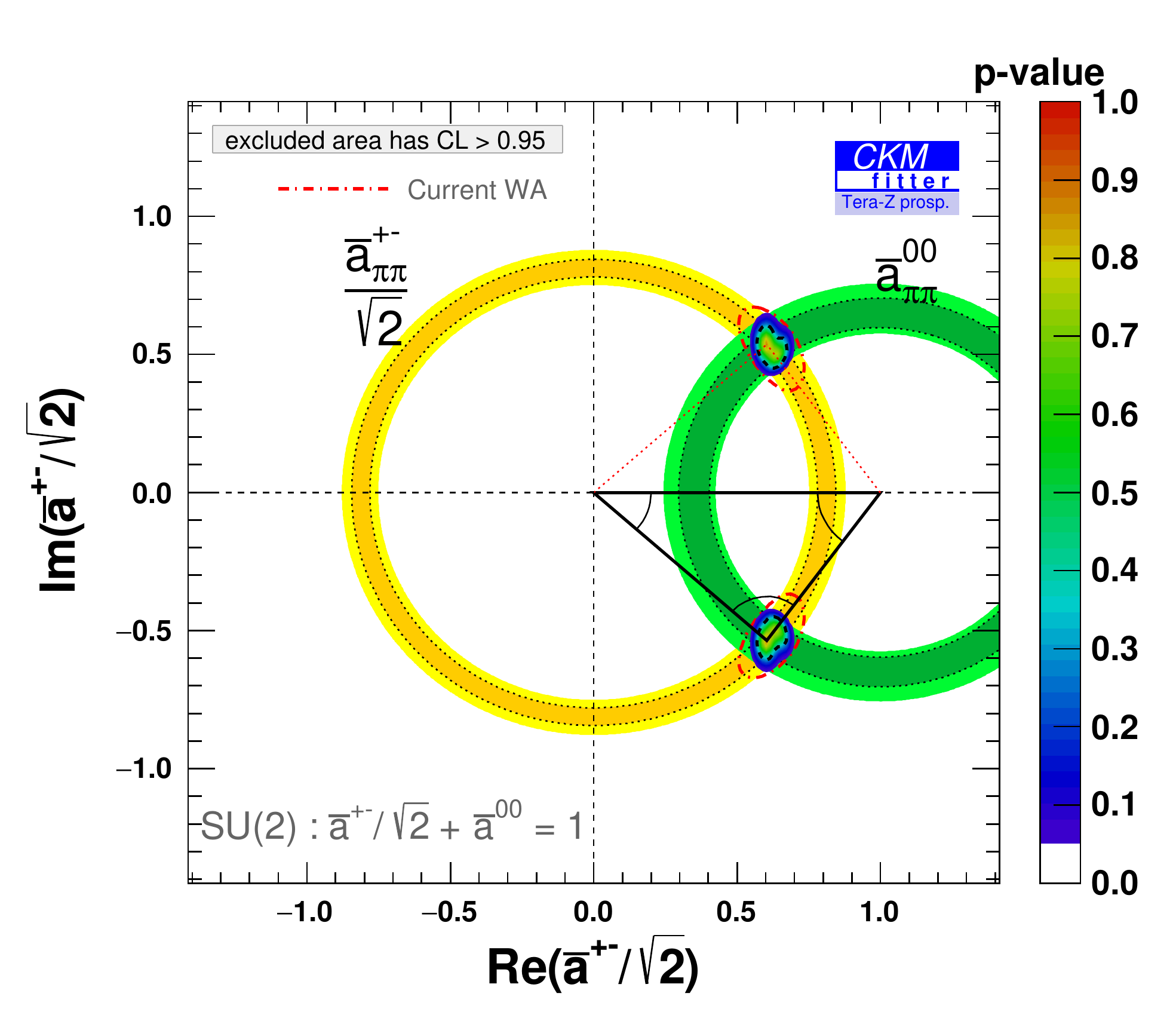}
\caption{Isospin triangles for $B\to \pi\pi$ ($B^0,B^+$ modes on the left, $\bar{B}^0,B^-$ modes on the right) in scenario 1 improving only the neutral modes. We take the baseline value $a_{CP}^{00}=0.950\pm 0.018$.
}
\label{fig:IsospinTrianglesTera00}
\end{figure}

\begin{figure}[h!]
    \centering
    \includegraphics[width=7cm]{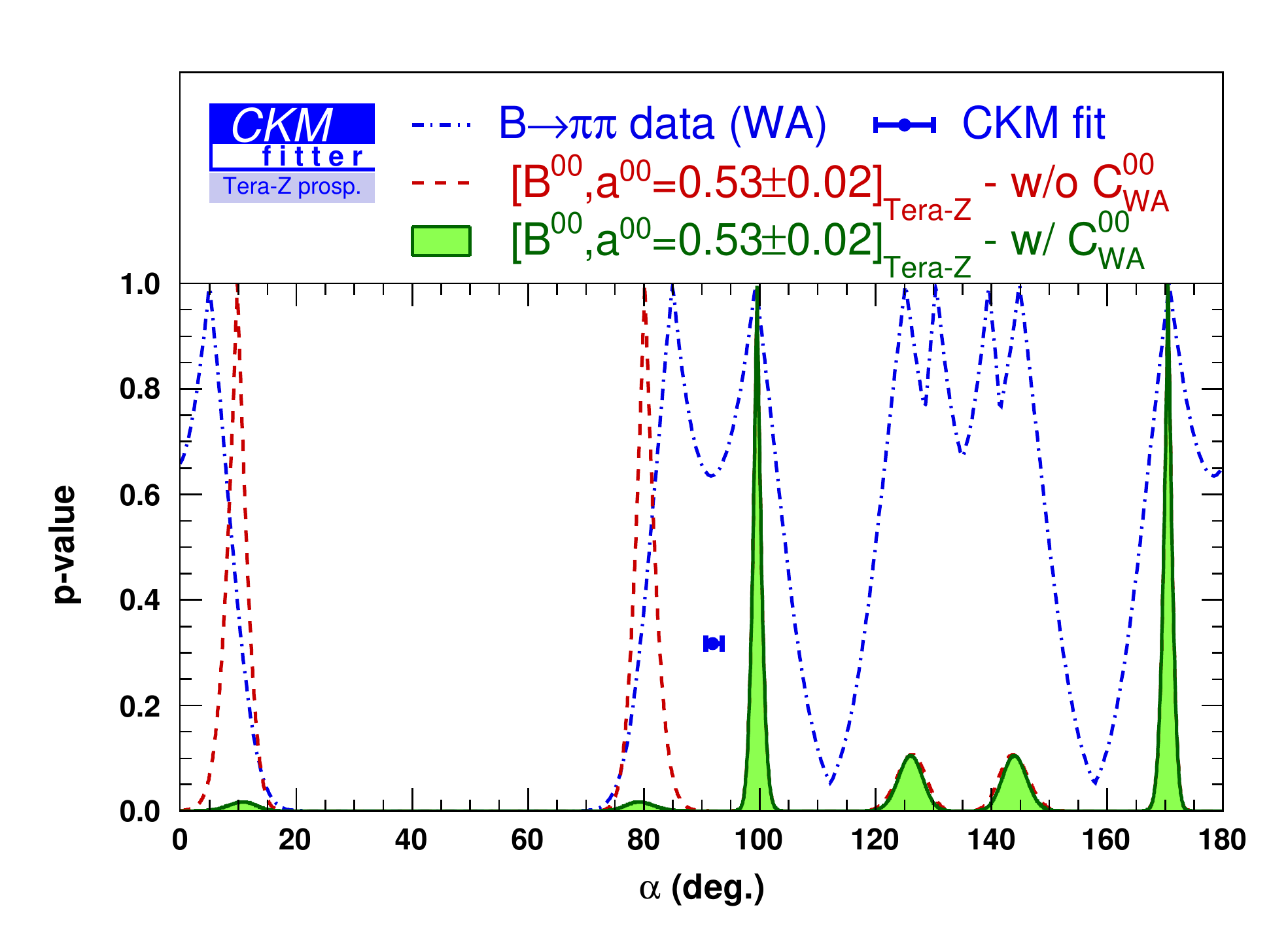}
    \includegraphics[width=7cm]{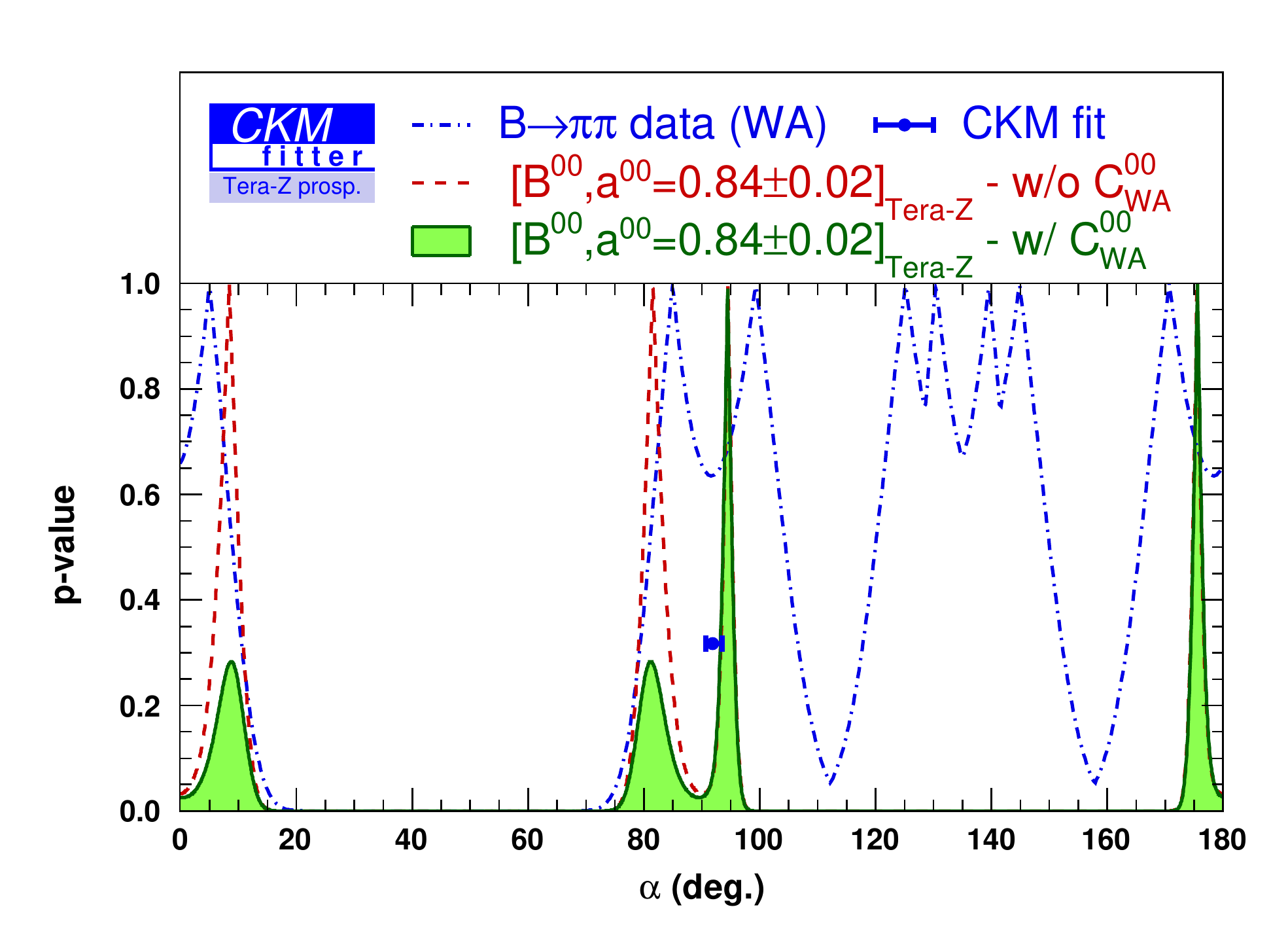}
    \caption{Same plot as the right-hand side of figure~\ref{fig:alphaTera00} but with alternative $a_{CP}^{00}$ central values. 
    In particular, we take $a_{CP}^{00}=0.530\pm 0.018$ on the left-hand side and $a_{CP}^{00}=0.840\pm 0.018$ on the right-hand side.}
    \label{fig:alphaTera00alt}
\end{figure}

We consider two scenarios, the first one focusing on improving the neutral mode only, the second one considering an additional improvement in charged modes. More precisely, in scenario 1, we consider only the improvement expected for $\mathcal{B}^{00}$ and $a^{00}_{CP}$ from table~\ref{tab:SixParameterTZ} (using the upper value for the statistical uncertainties). If we compare figures~\ref{fig:IsospinTrianglesWA} and \ref{fig:IsospinTrianglesTera00}, we observe a reduced uncertainty on one side of the isospin triangles, without lifting the ambiguity between the almost degenerate solutions in the $B$-meson isospin triangle. The corresponding improvement in the determination of $\alpha$ can be considered using $B\to\pi\pi$ data only, using our baseline input for $a_{CP}^{00}=0.950\pm 0.018$ in figure~\ref{fig:alphaTera00}. As can be seen on the left-hand side of figure~\ref{fig:alphaTera00}, the main improvement does not come from the branching ratio but from the $CP$ asymmetries $a_{CP}^{00}$ and $C_{CP}^{00}$, leading to a significant suppression of the mirror solutions affecting the determination of $\alpha$ using $B\to \pi\pi$. On the right-hand side of figure~\ref{fig:alphaTera00}, we notice that the combination of $a_{CP}^{00}$ with the input from $B$-factories on $C_{CP}^{00}$ sharpens further the constraint, leading to:
\begin{equation*}
  {\rm WA}: \alpha(\pi\pi)=(93.0\pm 13.6)^\circ \to \text{Tera-$Z$ scenario 1} : \alpha(\pi\pi)=(82.6^{+3.5}_{-2.5} \cup 92.0^{+1.4}_{-2.0})^\circ~.
\end{equation*}
Let us recall that these conclusions depend not only on the reduction of uncertainties on $B\to \pi^0\pi^0$ observables, but also our choice of central values. We can illustrate this point with our two other choices for $a_{CP}^{00}$, shown in figure~\ref{fig:alphaTera00alt}, leading to $\alpha(\pi\pi)=(99.5\pm 1.1)^\circ$ and $\alpha(\pi\pi)=(94.5\pm 1.2)^\circ$ for the central values $a_{CP}^{00}=0.53$ and 0.84 respectively. The confidence intervals are smaller in both cases compared to our baseline scenario, related to the fact that the input for $a_{CP}^{00}$ is then less compatible with the rest of the inputs than in the baseline case $a_{CP}^{00}=0.950\pm 0.018$.


\begin{figure}[t]
    \centering
    \includegraphics[width=7cm]{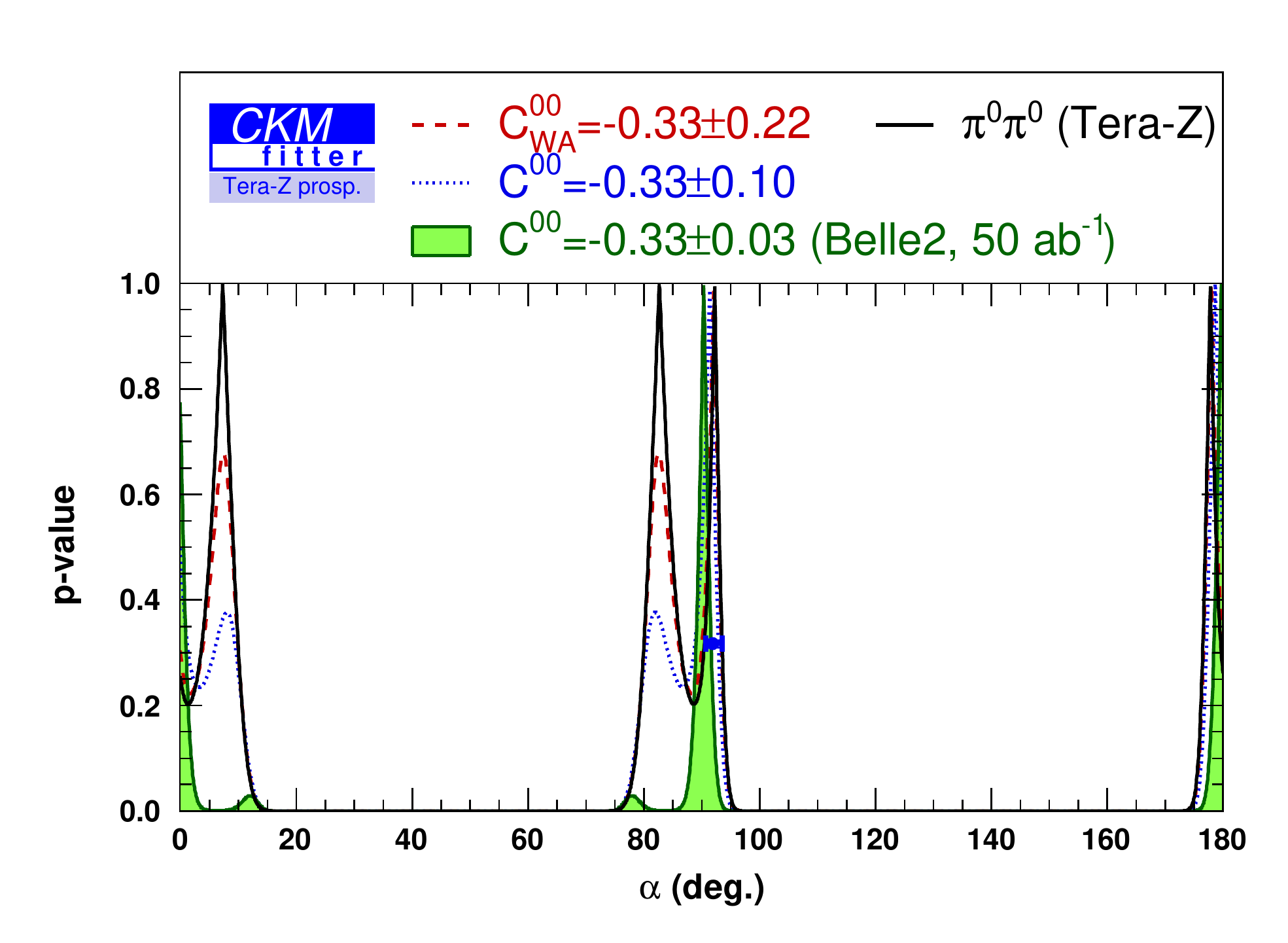}
    \includegraphics[width=7cm]{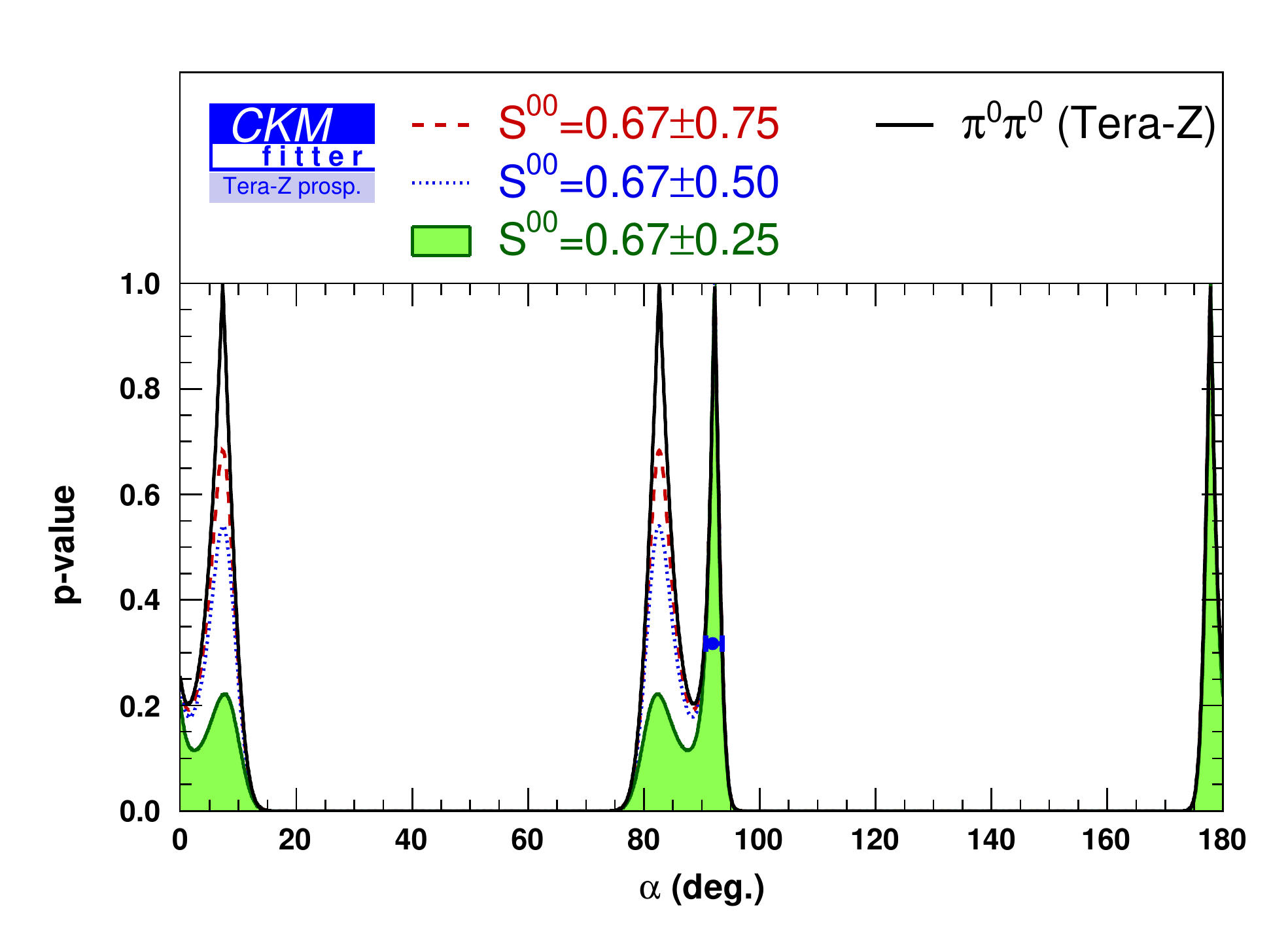}
\caption{$p$-value for $\alpha$ from $B\to\pi\pi$ measurements within the scenario 1 improving only neutral modes without the current measurement of $C_{CP}^{00}$  (black solid line). On the left: with additional information on $C_{CP}^{00}$ with increasing precision up to 50 ab$^{-1}$ Belle II prospects (dashed red, dotted blue, solid green). On the right: with additional information on $S_{CP}^{00}$ with increasing precision (dashed red, dotted blue, solid green).}
\label{fig:alphaS00}
\end{figure}

The impact of the $CP$ asymmetries $a_{CP}^{00}$ and $C_{CP}^{00}$ shows indirectly that another observable can play an important role here, namely the mixing-induced asymmetry $S_{CP}^{00}$. As an illustration of the impact of these $CP$ asymmetries, we consider the same scenario 1 (improved $B^0\to\pi^0\pi^0$ observables) without the current measurement of $C_{CP}^{00}$, to which we add either a measurement of $S_{CP}^{00}$ or $C_{CP}^{00}$. In each case, we take central value given by the result of the global fit, assuming decreasingly small uncertainties. As can be seen in figure~\ref{fig:alphaS00}, the knowledge of either quantity lifts the degeneracy among the mirror solutions, as they carry information either on the relative size or the relative phase between $B$- and $\bar{B}$-meson amplitudes. If we assume $C_{CP}^{00}=-0.33\pm 0.03$ (prospected statistical uncertainty for Belle II at 50\,ab$^{-1}$) together with scenario 1, we obtain $\alpha(\pi\pi)=(90.3\pm 1.3)^\circ$.

\begin{figure}[h!]
    \centering
    \includegraphics[width=7.5cm]{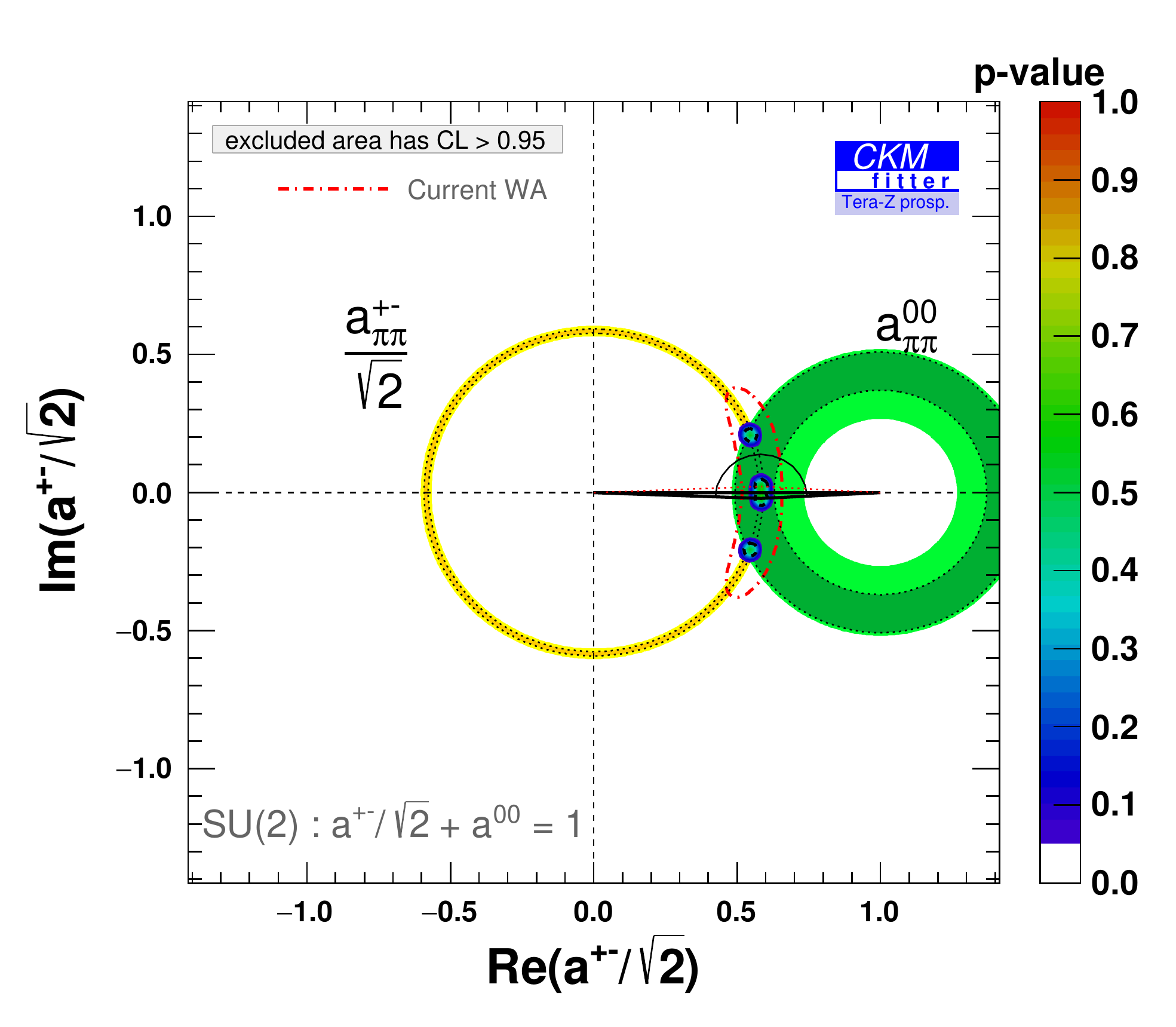}
    \includegraphics[width=7.5cm]{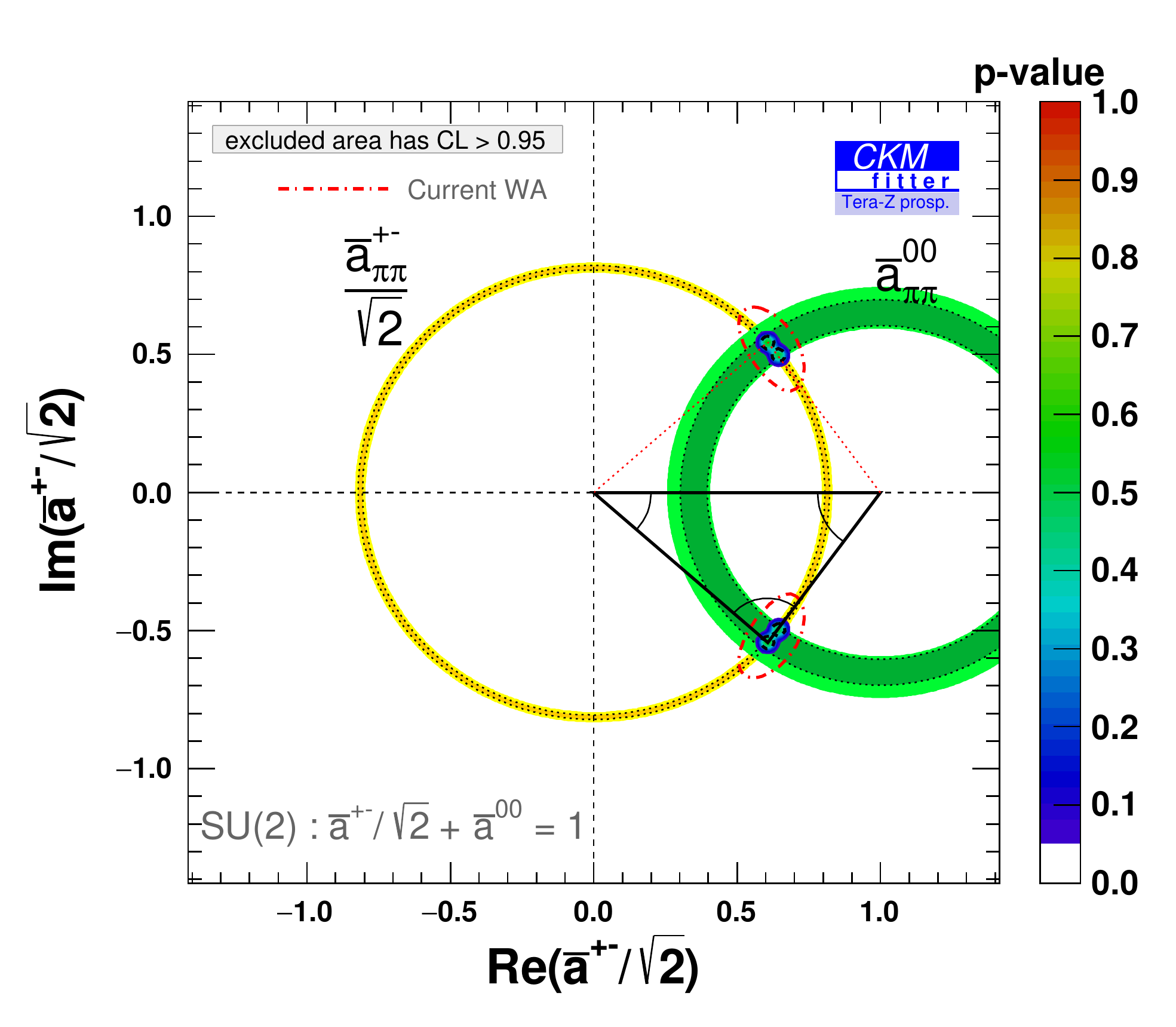}
\caption{Isospin triangles for $B\to \pi\pi$ ($B^0,B^+$ modes on the left, $\bar{B}^0,B^-$ modes on the right) in scenario 2 improving both neutral and charged modes. We take the baseline value $a_{CP}^{00}=0.950\pm 0.018$.
}
\label{fig:IsospinTrianglesTeraFull}
\end{figure}

\begin{figure}[h!]
    \centering
    \includegraphics[width=7cm]{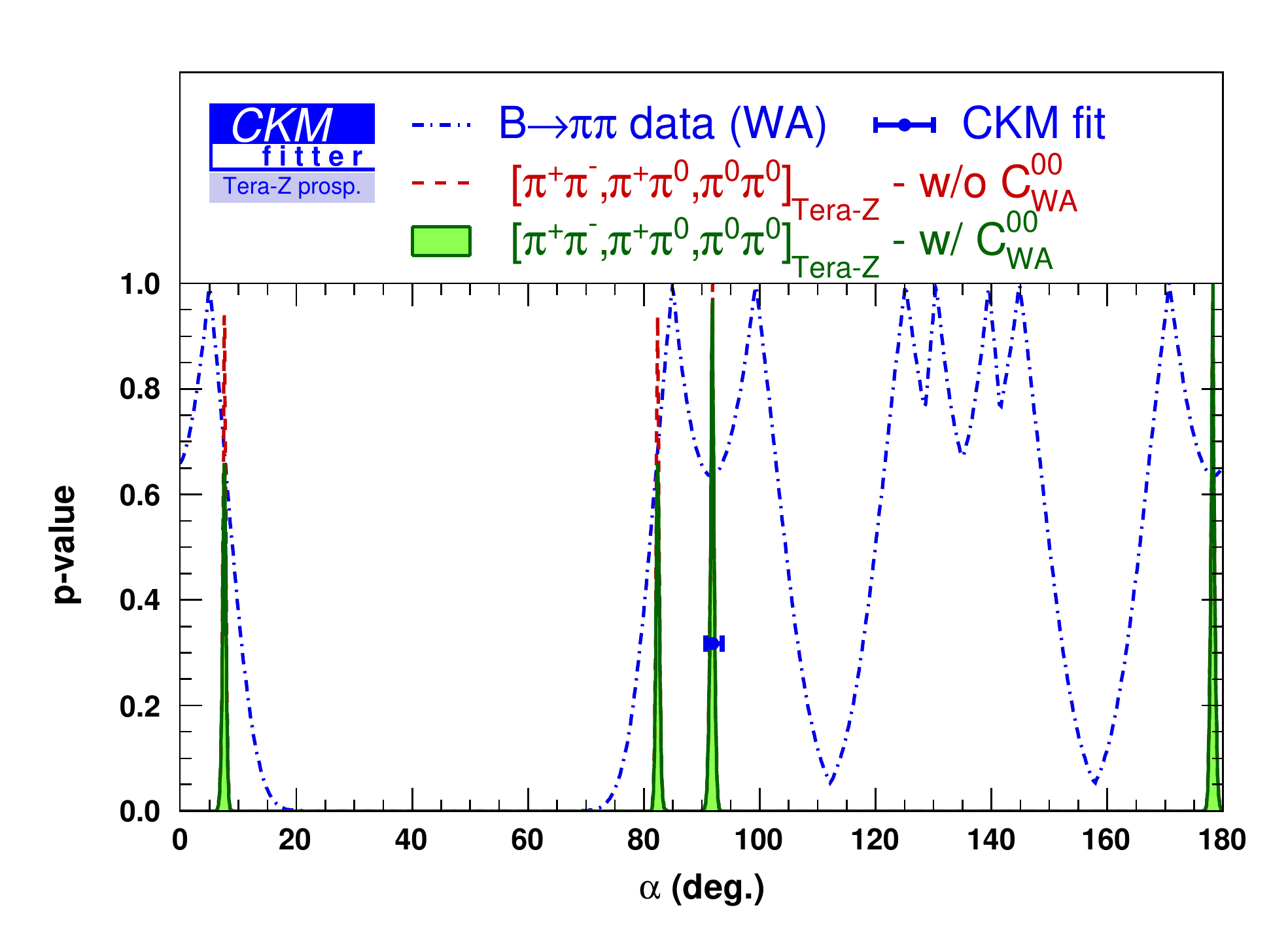}
    \includegraphics[width=7cm]{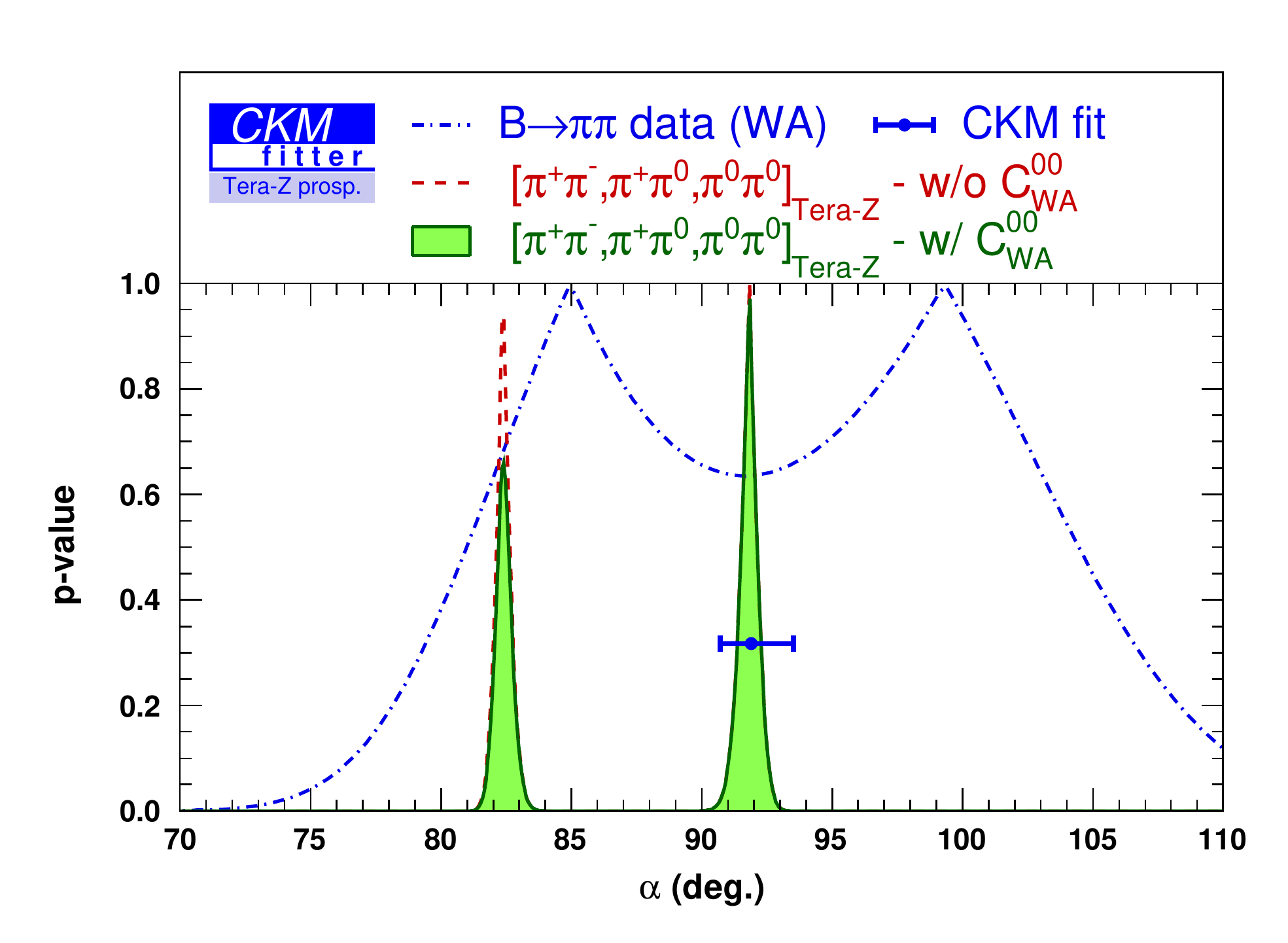}
    \caption{$p$-value for $\alpha$ from $B\to\pi\pi$ measurements
using the current world average (dotted-dashed blue), within scenario 2 improving both neutral and charged modes without (dashed red) and with (solid green) the current measurement of $C_{CP}^{00}$. We show the scan over the whole range of $\alpha$ (on the left) and around the value favored by the global CKM fit (on the right). We take the baseline value $a_{CP}^{00}=0.950\pm 0.018$.
}
\label{fig:alphaTeraFull}
\end{figure}

\begin{figure}[h!]
    \centering
    \includegraphics[width=7cm]{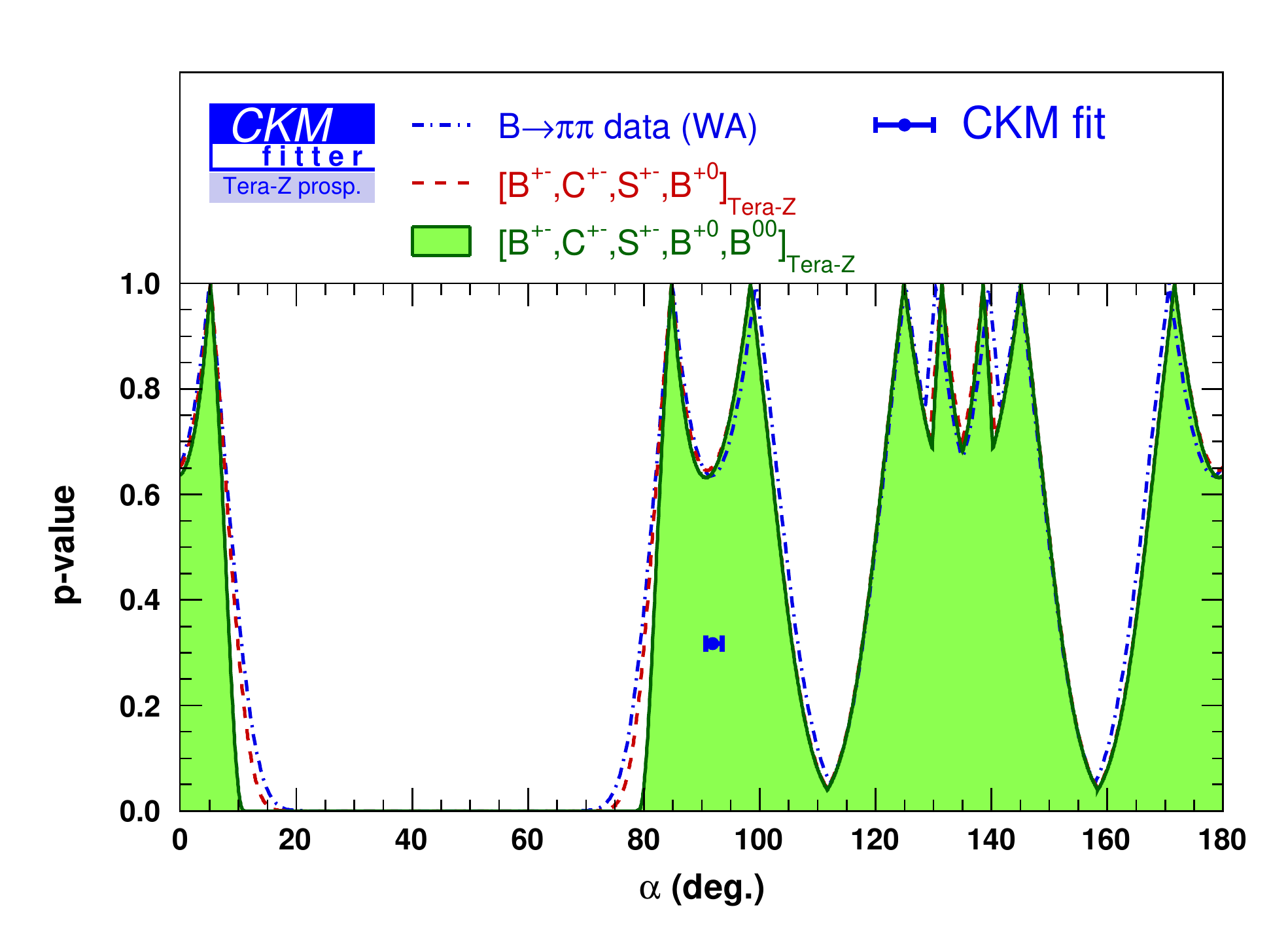} 
\caption{$p$-value for $\alpha$ from $B\to\pi\pi$ measurements
using current data (dotted-dashed blue),
improving only charged modes (dashed red), and improving also neutral modes (only BR, solid green) but without additional information on neither $a_{CP}^{00}$ nor $C_{CP}^{00}$.
}
\label{fig:alphaTeraFull-noCP}
\end{figure}

\begin{figure}[h!]
    \centering
    \includegraphics[width=7cm]{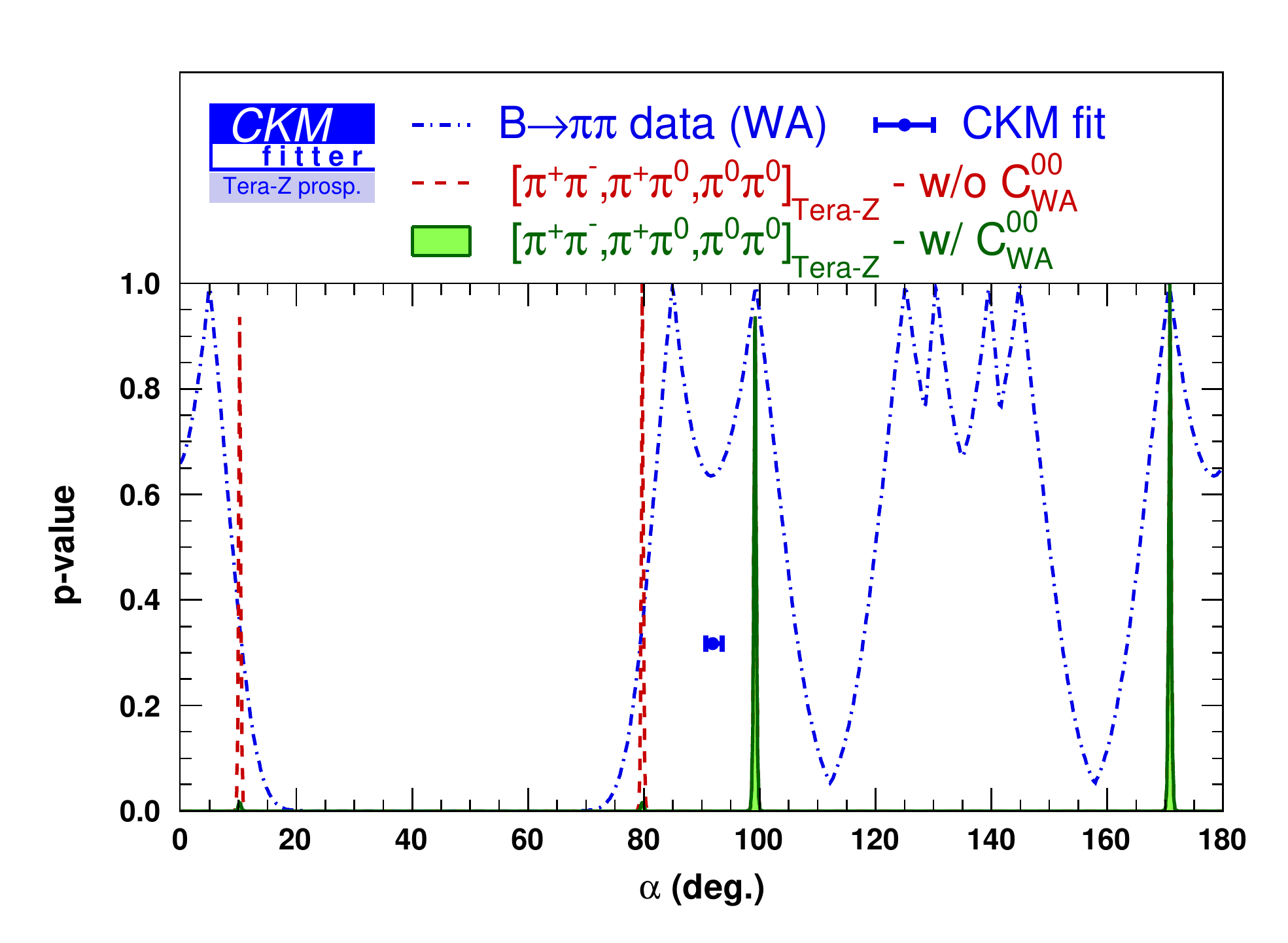}
    \includegraphics[width=7cm]{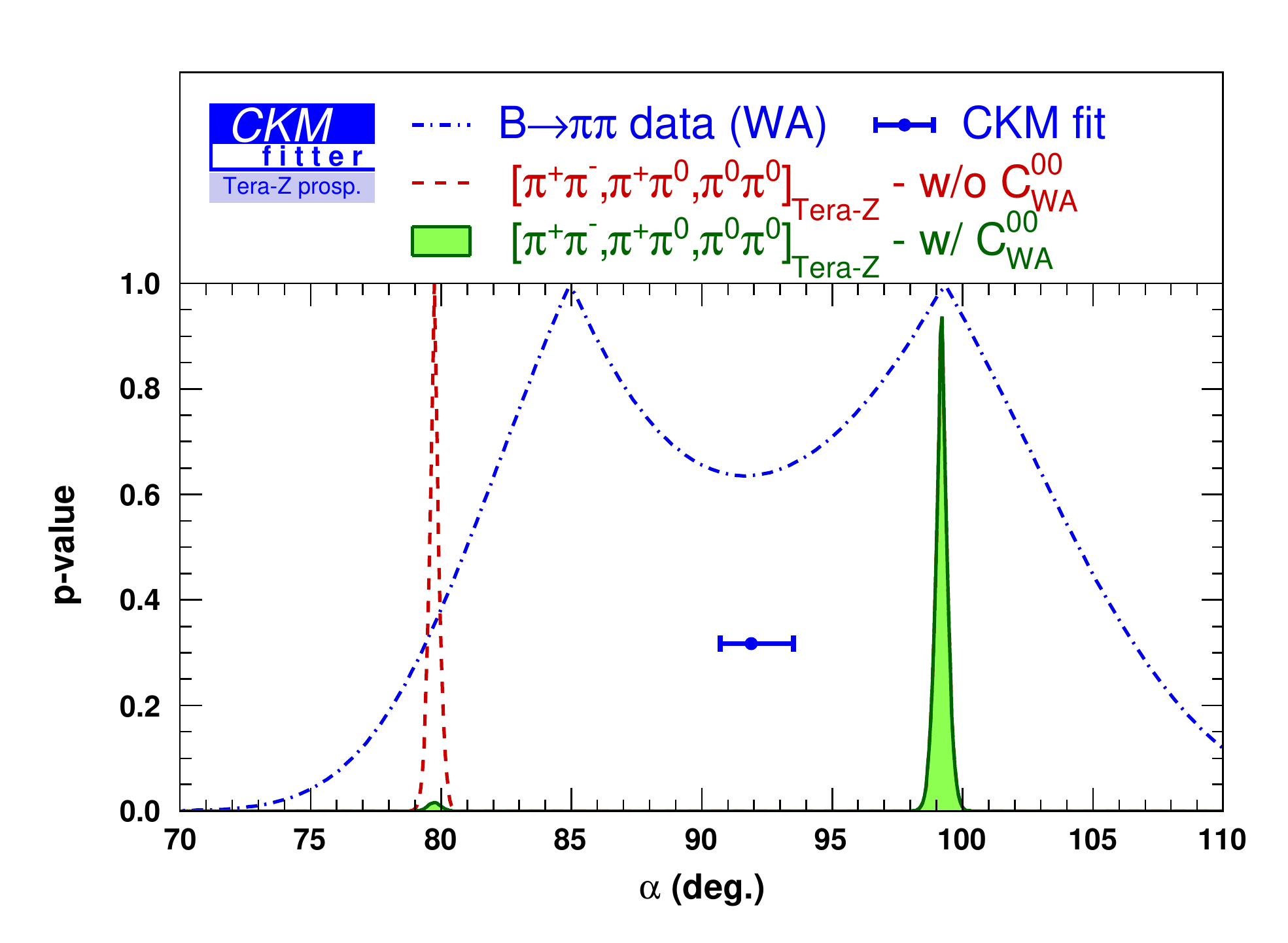}
\caption{Same plots as figure~\ref{fig:alphaTeraFull} for the alternative value $a_{CP}^{00}=0.530\pm 0.018$
}
\label{fig:alphaTeraFull-alt1}
\end{figure}

\begin{figure}[h!]
    \centering
    \includegraphics[width=7cm]{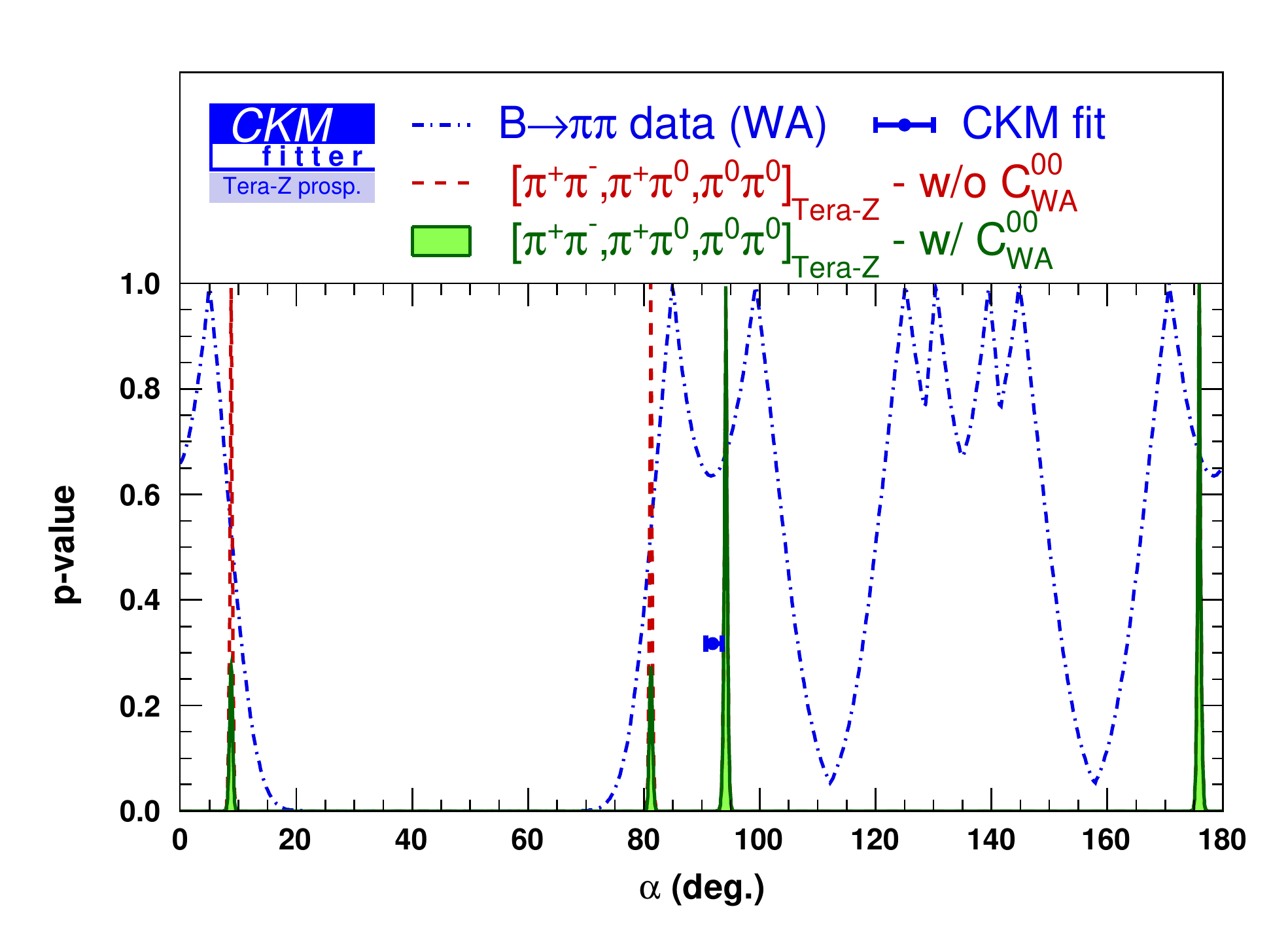}
    \includegraphics[width=7cm]{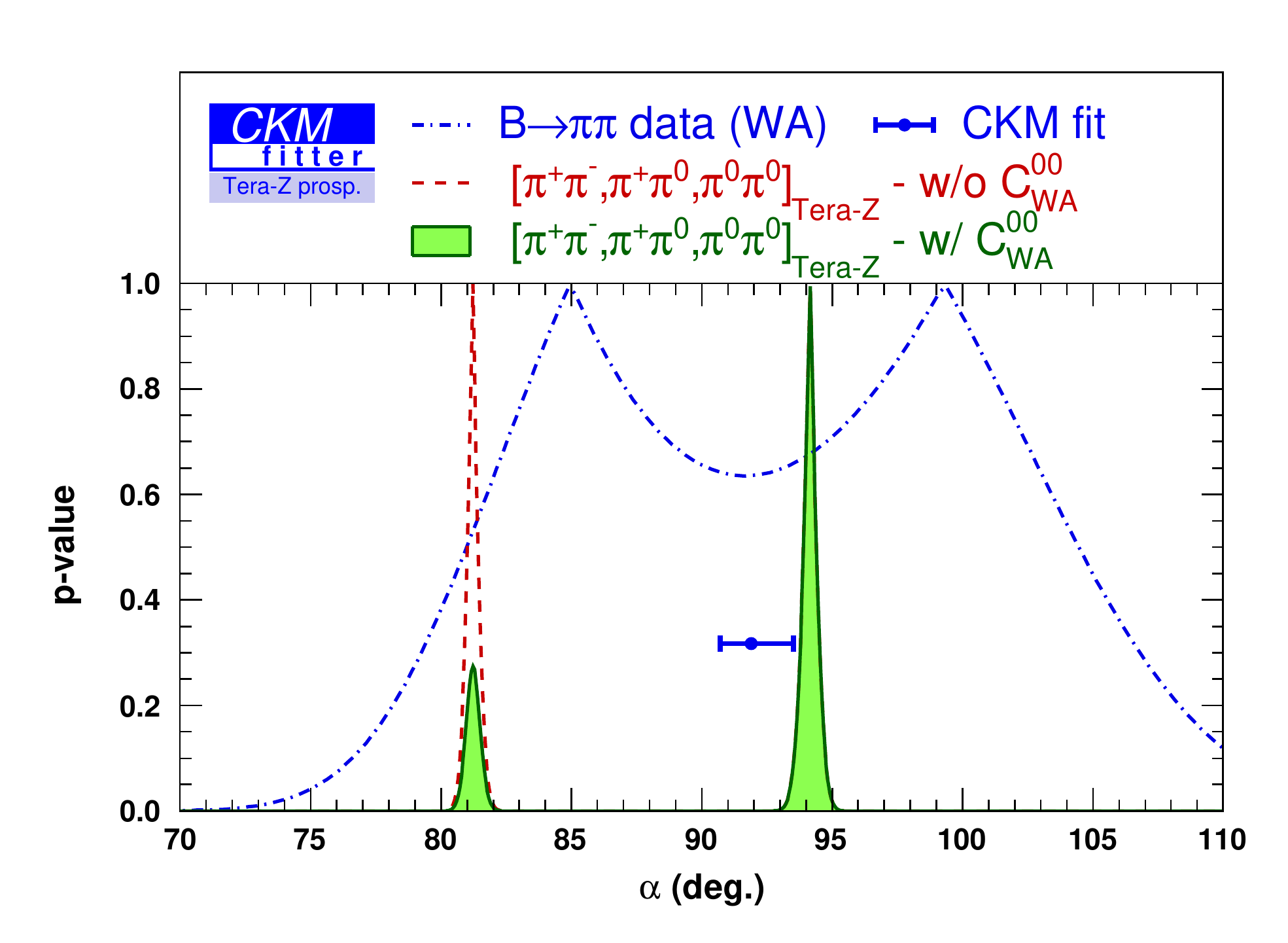}
\caption{Same plots as figure~\ref{fig:alphaTeraFull} for the alternative value $a_{CP}^{00}=0.840\pm 0.018$
}
\label{fig:alphaTeraFull-alt2}
\end{figure}

Let us now consider scenario 2 where charged modes $B^0\to \pi^+\pi^-$ and $B^+\to\pi^+\pi^0$ are also improved according to table~\ref{tab:SixParameterTZ} (using again the upper value for the statistical uncertainties). With our baseline value of $a_{CP}^{00}=0.950\pm 0.018$, we obtain the isospin triangles shown in figure~\ref{fig:IsospinTrianglesTeraFull}, leading to an improved resolution of the possible solutions for the $B$-meson isospin triangle, translating into a determination of $\alpha$ shown in figure~\ref{fig:alphaTeraFull}:
\begin{equation*}
 \text{Tera-$Z$ scenario 2} :  \alpha(\pi\pi)=(91.8\pm 0.4)^\circ~.
\end{equation*}
As seen in figure~\ref{fig:alphaTeraFull-noCP}, the $CP$ asymmetries $a_{CP}^{00}$ and $C_{CP}^{00}$ are central in this improvement. We can illustrate the dependence on the central values of the inputs with our two other choices for $a_{CP}^{00}$, shown in figures~\ref{fig:alphaTeraFull-alt1} and \ref{fig:alphaTeraFull-alt2}, leading to $\alpha(\pi\pi)=(99.2\pm 0.3)^\circ$ and $\alpha(\pi\pi)=(94.2\pm 0.4)^\circ$, for the central values $a_{CP}^{00}=0.53$ and 0.84 respectively. Depending on the value chosen, some of the mirror solutions disappear. All in all, we see that a very high precision can be reached through Tera-$Z$ $B\to\pi\pi$ measurements.
This precision of $0.4^\circ$ in $\alpha$ extracted from $\pi\pi$ mode at Tera-$Z$ is not only significantly higher than the current world average (13.6$^\circ$) including $\pi\pi$ data only, but also 5 times as high as that ($\sim$2$^\circ$) extrapolated from $\pi\pi$ data only at Belle II (50\,ab$^{-1}$)~\cite{BelleII2019}.
Even compared to the current precision of $\sim$4.2$^\circ$ combining $\pi\pi$, $\rho\rho$, and $\rho\pi$ modes, it is almost 10 times higher.



Given the small uncertainty of 0.3--0.4$^\circ$ obtained in scenario 2, we must stress that these results come from an approach based on isospin decomposition. The measurement precisions considered here means that the extraction of $\alpha$ may get limited by theoretical systematic uncertainties attached to the isospin limit considered, at the level of 1--2$^\circ$. As discussed in ref.~\cite{IsospinAna}, several isospin-breaking effects should be taken into account at this level of precision. If we focus on $B\to\pi\pi$ measurements, the precision on $\alpha$ is affected by theoretical uncertainties introduced by the $\Delta I=1/2$ electroweak penguin contributions~\cite{Buras:1998rb,Neubert:1998jq,Neubert:1998pt} (related to the differences of $u$ and $d$ charges) and $\pi^0-\eta-\eta'$ mixing~\cite{Gardner:1998gz,Zupan:2004hv} (related to the quark mass differences)~\cite{IsospinAna}. If the former can be estimated theoretically with a high precision, the latter may be improved by a better knowledge of $B^{+,0}\to \pi^{0,+}\eta(')$ modes. We leave the potential impact of Tera-$Z$ on this issue for future work.

From a broader perspective, the $\alpha$ sensitivity at Tera-$Z$ should be led by the three isospin-related $B\to \rho\rho$ decays. They have larger decay rates than $B\to \pi\pi$ in general. Moreover, at least two charged tracks will be produced in their four-pion final states, providing vertex information naturally. It is noteworthy that both $B^\pm \to \rho^\pm \rho^0$ and $B^0 \to \rho^+\rho^-$ give rise to final states containing both charged tracks and photons, making them ideal targets at Tera-$Z$. Currently, as can be seen from table~12 of ref.~\cite{IsospinAna}, the determination of $\alpha$ would be improved if more accurate $CP$ asymmetry measurements of $B\to\rho^0\rho^0$ are accessible. Progress in $B\to \rho\pi$ measurements is also possible, although the theoretical and experimental frameworks are more complicated. These results might help to solve some inconsistencies in the current data, in particular by reexamining $B^0\to \pi^+\pi^-\pi^0$~\cite{IsospinAna}. Considering all these modes is obviously a task beyond the scope of the present article, and we leave a complete evaluation of the Tera-$Z$ potential for the determination of the CKM angle $\alpha$ for future work.

\section{Dependence of the measurement precision on the detector performance}
\label{sec:Dependence}

\subsection{$b$-tagging}
\label{sec:btag}

Three different $b$-tagging conditions in table~\ref{tab:btagEff} are compared in this section.
The corresponding precisions of \Bospio\ and \Boseta\ are summarized in table~\ref{tab:btagAccuracy} when the ECAL energy resolution is fixed as $\frac{3\%}{\sqrt{E}} \oplus 0.3\%$.

\begin{table}[thp]
	\centering
	\small
	\begin{tabular}[t]{|c|ccc|}
		\hline
		$b$-tagging & $\epsilon_{b \to b}$ & $\epsilon_{c \to b}$ & $\epsilon_{uds \to b}$ \\
		\hline
		None 		& 100\%  & 100\%  & 100\% \\
		Baseline 	&  80\%  & 8.26\% & 0.85\% \\
		Ideal 		& 100\%  & 0\%    & 0\% \\
		\hline
	\end{tabular}
	\caption{Parameters of three $b$-tagging conditions. Symbols $\epsilon_{b \to b}$ and $\epsilon_{udsc \to b}$ represent the $b$-jet tagging efficiency and the background-jets mistagging rates, respectively.}
	\label{tab:btagEff}
\end{table}	
\begin{table}[thp]
	\centering
	\resizebox{1.\columnwidth}{!}{
		\begin{tabular}[t]{|c|ccccccccc|}
			\hline
			Signal & $\sigma_{m_B}$ (MeV) & $b$-tagging & Mass window (GeV) & \Bo & \Bs & $u\bar{u}$+$d\bar{d}$+$s\bar{s}$ & $c\bar{c}$ & $b\bar{b}$ & $\sqrt{S+B}/S$ (\%) \\
			\hline
			\multirow{3}*{\Bopio} & \multirow{3}*{30.25}
			& None & (5.222, 5.337) & 9.18$\times 10^{4}$ & 7$\times 10^{2}$ & 5.31$\times 10^{5}$ & 1.80$\times 10^{4}$ & 1.00$\times 10^{4}$ & 0.88 $\pm$ 0.03 \\
			~ & ~ & Baseline & (5.212, 5.347) & 7.59$\times 10^{4}$ & 9$\times 10^{2}$ & 5.5$\times 10^{3}$ & 1.6$\times 10^{3}$ & 8.7$\times 10^{3}$ & 0.40 $\pm$ 0.01 \\ 
			~ & ~ & Ideal & (5.205, 5.354) & 9.59$\times 10^{4}$ & 1.5$\times 10^{3}$ & 0 & 0 & 1.16$\times 10^{4}$ & 0.344 $\pm$ 0.004 \\
	 		\hline
	 		\multirow{3}*{\Bspio} & \multirow{3}*{30.21}
	 		& None & (5.322, 5.411) & 8.3$\times 10^{3}$ & 3.9$\times 10^{3}$ & 3.67$\times 10^{5}$ & 8.6$\times 10^{3}$ & 5.2$\times 10^{3}$ & 16 $\pm$ 1 \\
	 		~ & ~ & Baseline & (5.336, 5.397) & 2.8$\times 10^{3}$ & 2.5$\times 10^{3}$ & 2.4$\times 10^{3}$ & 5$\times 10^{2}$ & 2.2$\times 10^{3}$ & 4.0 $\pm$ 0.6 \\ 
	 		~ & ~ & Ideal & (5.336, 5.397) & 3.5$\times 10^{3}$ & 3.2$\times 10^{3}$ & 0 & 0 & 2.8$\times 10^{3}$ & 3.1 $\pm$ 0.2 \\ 
	 		\hline
	 		\multirow{3}*{\Boeta} & \multirow{3}*{33.30}
	 		& None & (5.228, 5.331) & 9$\times 10^{2}$ & 3.3$\times 10^{3}$ & 2.91$\times 10^{5}$ & 9.4$\times 10^{3}$ & 1.08$\times 10^{4}$ & 63 $\pm$ 3 \\
	 		~ & ~ & Baseline & (5.233, 5.326) & 7$\times 10^{2}$ & 2.1$\times 10^{3}$ & 2.3$\times 10^{3}$ & 7$\times 10^{2}$ & 8.0$\times 10^{3}$ & 17 $\pm$ 2 \\ 
	 		~ & ~ & Ideal & (5.232, 5.327) & 9$\times 10^{2}$ & 2.8$\times 10^{3}$ & 0 & 0 & 1.00$\times 10^{4}$ & 13 $\pm$ 1 \\
	 		\hline
	 		\multirow{3}*{\Bseta} & \multirow{3}*{33.26}
	 		& None & (5.315, 5.418) & 2$\times 10^{2}$ & 2.32$\times 10^{4}$ & 2.32$\times 10^{5}$ & 1.15$\times 10^{4}$ & 8.8$\times 10^{3}$ & 2.3 $\pm$ 0.1 \\
	 		~ & ~ & Baseline & (5.310, 5.423) & 2$\times 10^{2}$ & 1.92$\times 10^{4}$ & 2.2$\times 10^{3}$ & 1.0$\times 10^{3}$ & 7.4$\times 10^{3}$ & 0.90 $\pm$ 0.05 \\ 
	 		~ & ~ & Ideal & (5.310, 5.423) & 2$\times 10^{2}$ & 2.40$\times 10^{4}$ & 0 & 0 & 9.2$\times 10^{3}$ & 0.76 $\pm$ 0.03 \\ 
			\hline
		\end{tabular}
	}
	\caption{Measurement precisions of \Bopio, \Bspio, \Boeta, and \Bseta\ at different $b$-tagging conditions when ECAL energy resolution is $\frac{3\%}{\sqrt{E}} \oplus 0.3\%$.}
	\label{tab:btagAccuracy}
\end{table}

When no $b$-tagging is applied, the light-flavor \qq\ events overwhelm the other backgrounds and the signal. The baseline $b$-tagging can massively reduce the non-\bb\ background and improve the precision by a factor of 2--4. It also enhances the signal-to-background (S/B) ratios of \Bopio\ and \Bseta\ from $\mathcal{O}(10^{-1})$ to $\mathcal{O}(1)$. Compared to an ideal $b$-tagging performance, which can further improve the four measurement precisions by a factor of about 1.2, our baseline $b$-tagging is sufficient to suppress background candidates.

\subsection{ECAL energy resolution}
\label{sec:BMassReso}

In this case, we fix the $b$-tagging performance to the baseline one and compare the measurement precisions of \Bospio\ and \Boseta\ with different ECAL energy resolutions as shown in table~\ref{tab:Accuracy_ECALReso}. The scenario of $\frac{17\%}{\sqrt{E}} \oplus 1\%$ corresponds to the typical energy resolution of a regular ECAL at Tera-$Z$~\cite{CEPC_CDR_Phy,Aleksa:2021ztd}. In this case, the $B$-meson mass resolution worsens from $\sim$30\,MeV to $\sim$170\,MeV. The reconstruction efficiency and purity of \pio\ and $\eta$\ also drop when the ECAL energy resolution gets worse. Compared to the currently regular ECAL energy resolution, the proposed one can improve the measurement precisions of \Bospio\ and \Boseta\ by a factor of 3--5. Compared to \Boseta, \Bospio\ suffers less from the fake \pio\ background when the ECAL energy resolution gets worse.

\begin{table}[thp]
	\centering
	\resizebox{1.\columnwidth}{!}{
		\begin{tabular}[t]{|c|lcrrcc|}
			\hline
			ECAL energy resolution
			&  Channel  &  $\sigma_{m_{B}}$ (MeV) &  Signal  &  $q\bar{q}$ background  &  Background with false \pio($\eta$) & $\sqrt{S+B}/S$ (\%)   \\
			\hline
			\multirow{4}*{$\frac{3\%}{\sqrt{E}} \oplus 0.3\%$}
			&  \Bopio  &  30.25  &  7.59$\times 10^{4}$  &  1.58$\times 10^{4}$  &  7.52\% &  0.40 $\pm$ 0.01 \\
			~
			&  \Bspio  &  30.21  &  2.5$\times 10^{3}$  &  5.1$\times 10^{3}$  &  14.73\%  &  4.0 $\pm$ 0.6 \\
			~
			&  \Boeta  &  33.30  &  7$\times 10^{2}$  &  1.10$\times 10^{4}$  &  52.86\% &  17 $\pm$ 2 \\
			~
			&  \Bseta  &  33.26  &  1.92$\times 10^{4}$  &  1.06$\times 10^{4}$  &  65.25\%  &  0.90 $\pm$ 0.05 \\
			\hline
			\multirow{4}*{$\frac{17\%}{\sqrt{E}} \oplus 1\%$}
			&  \Bopio  &  166  &  5.77$\times 10^{4}$  &  3.81$\times 10^{5}$  &  4.04\%  &  1.15 $\pm$ 0.03 \\
			~
			&  \Bspio  &  165  &  2.2$\times 10^{3}$  &  1.43$\times 10^{5}$  &  5.74\%  &  20 $\pm$ 1 \\
			~
			&  \Boeta  &  170  &  3$\times 10^{2}$  &  6.82$\times 10^{4}$ &  88.27\%  &  85 $\pm$ 6 \\
			~
			&  \Bseta  &  174  &  8.3$\times 10^{3}$  &  4.92$\times 10^{4}$  &  86.30\%  &  2.9 $\pm$ 0.2 \\
			\hline
		\end{tabular}
	}
	\caption{Measurement precisions of \Bospio\ and \Boseta\ at different ECAL energy resolutions when using the baseline $b$-tagging.}
	\label{tab:Accuracy_ECALReso}
\end{table}

\begin{figure}[htbp]
	\centering
	\includegraphics[width=0.3\textwidth]{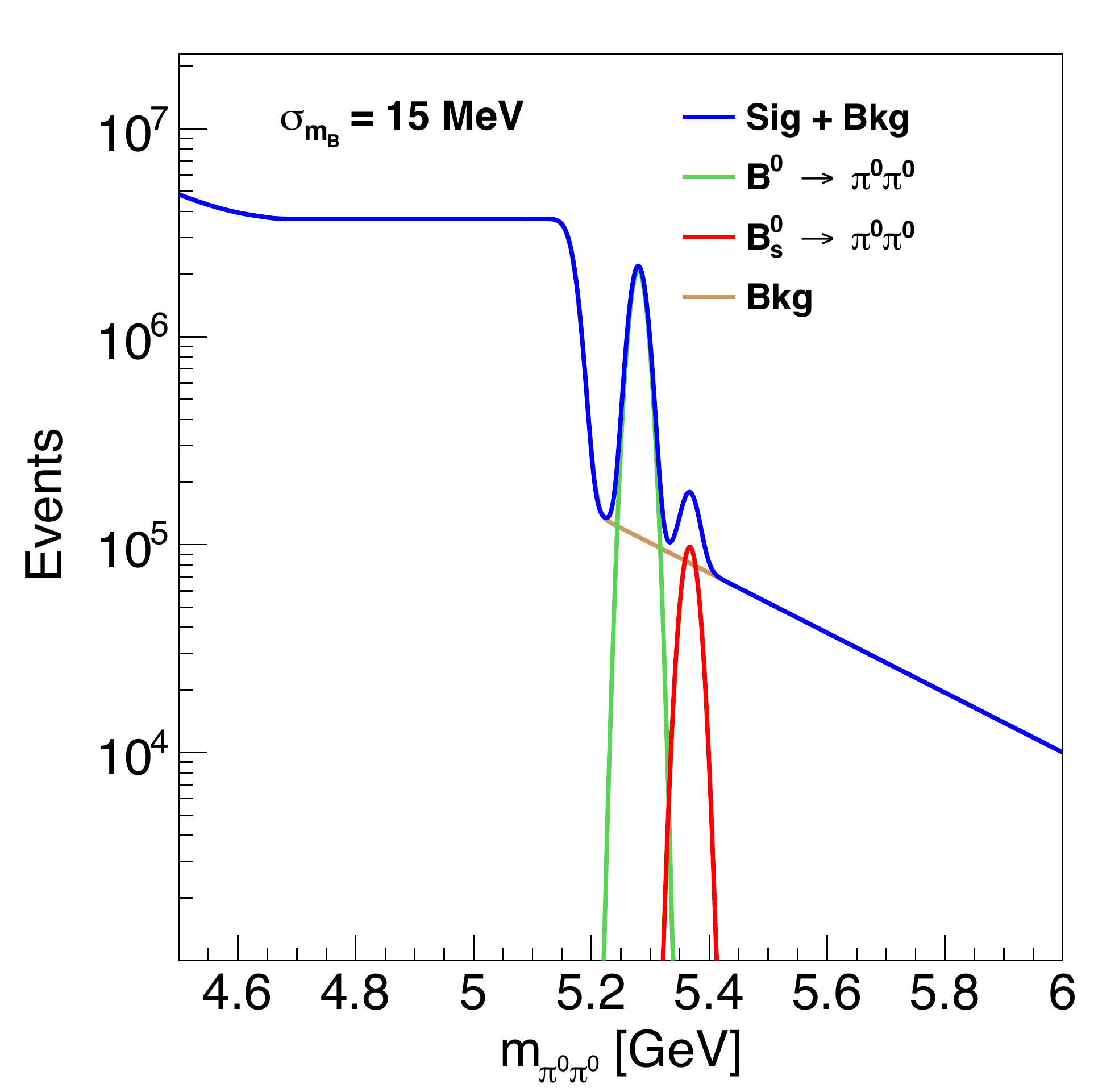}
	\includegraphics[width=0.3\textwidth]{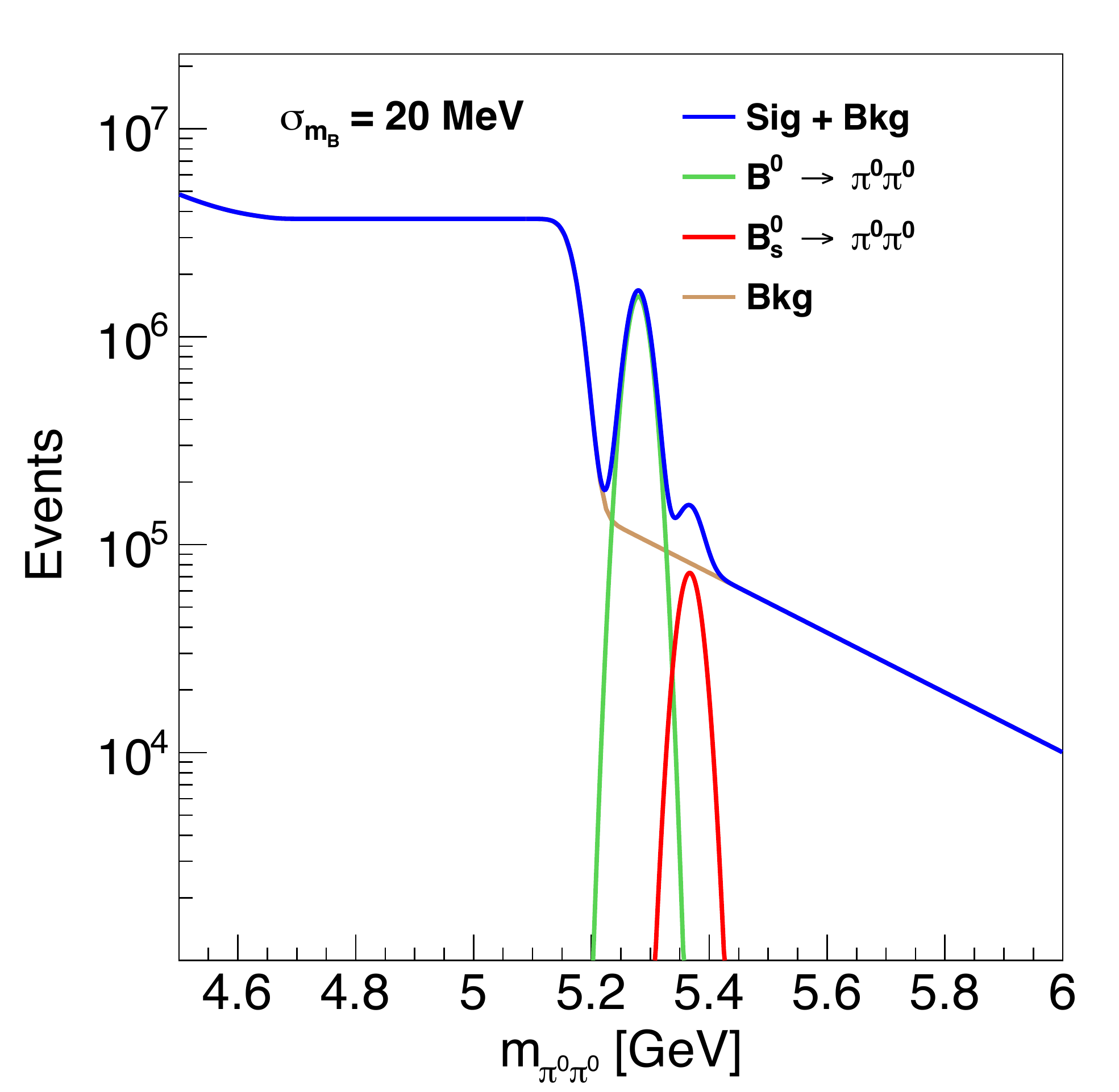}
	\includegraphics[width=0.3\textwidth]{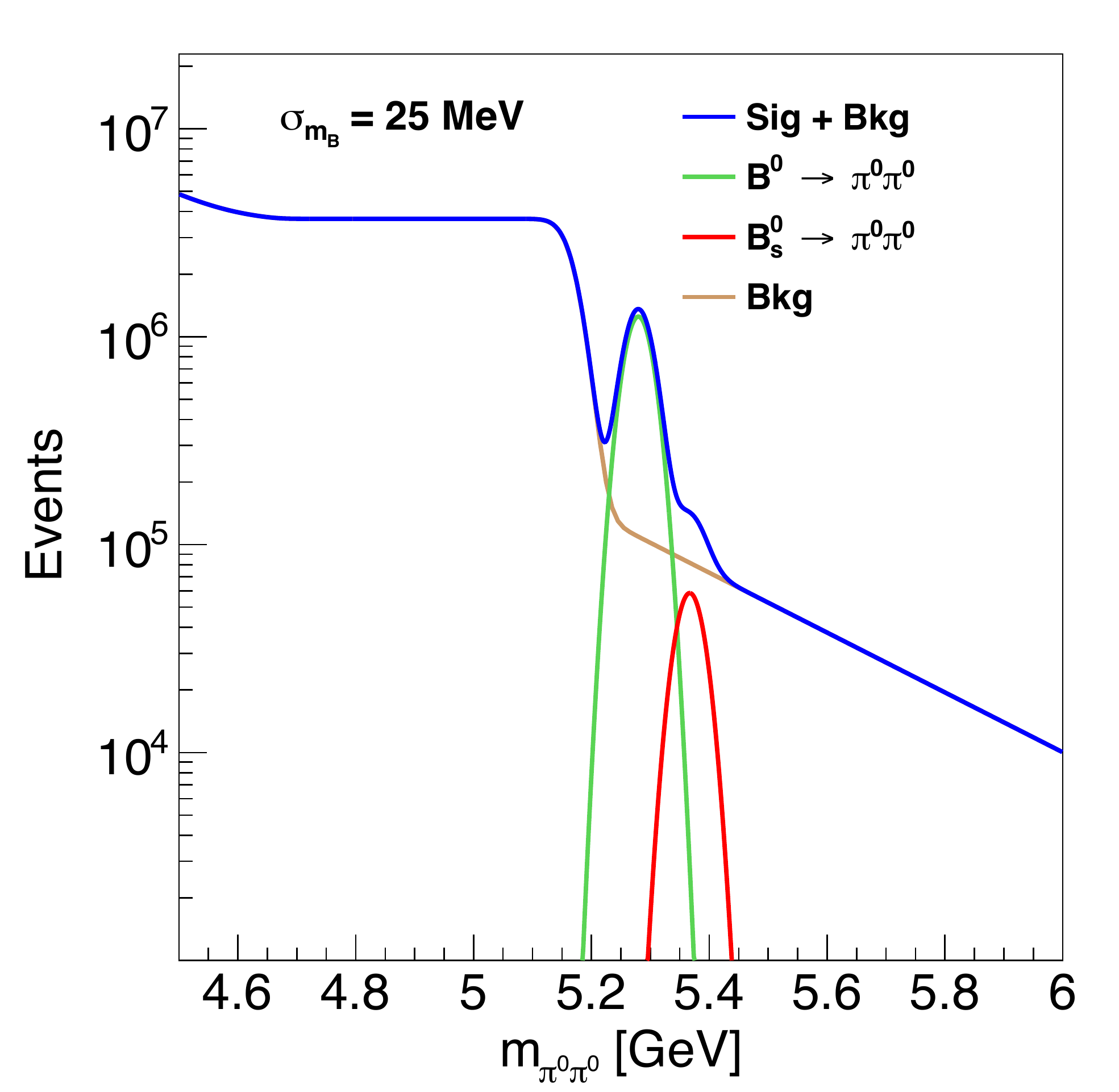}
	\includegraphics[width=0.3\textwidth]{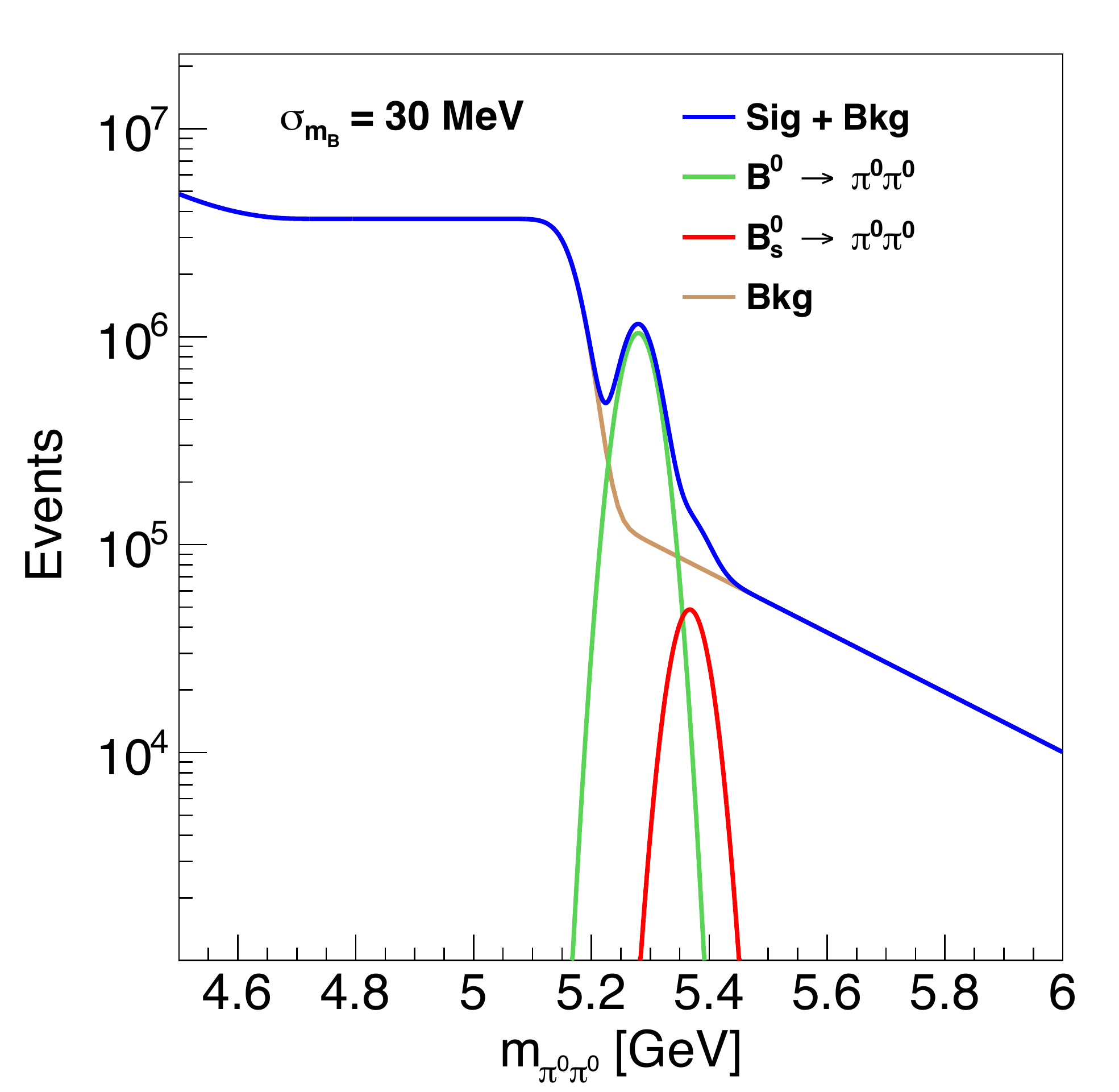}
	\includegraphics[width=0.3\textwidth]{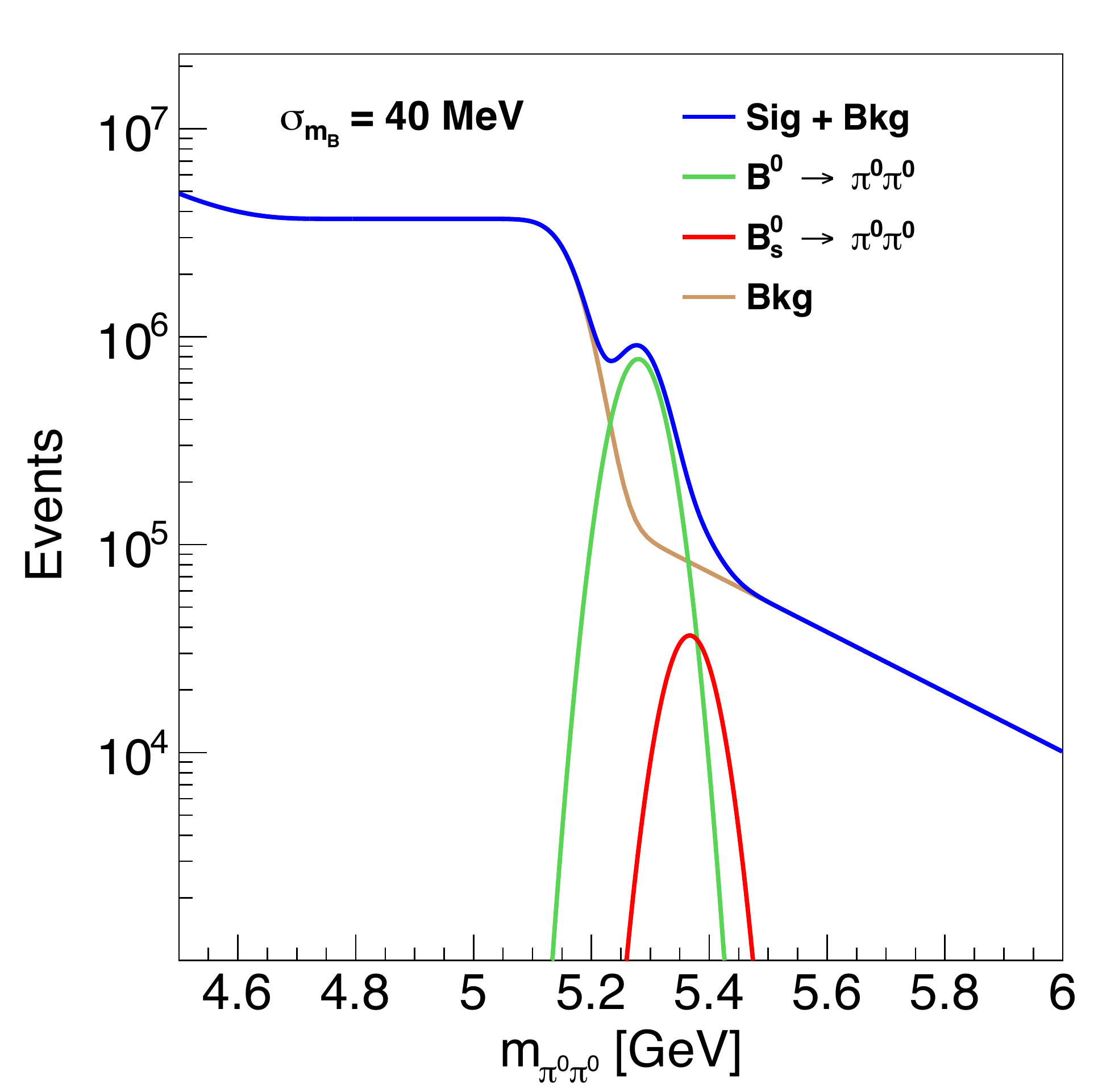}
	\includegraphics[width=0.3\textwidth]{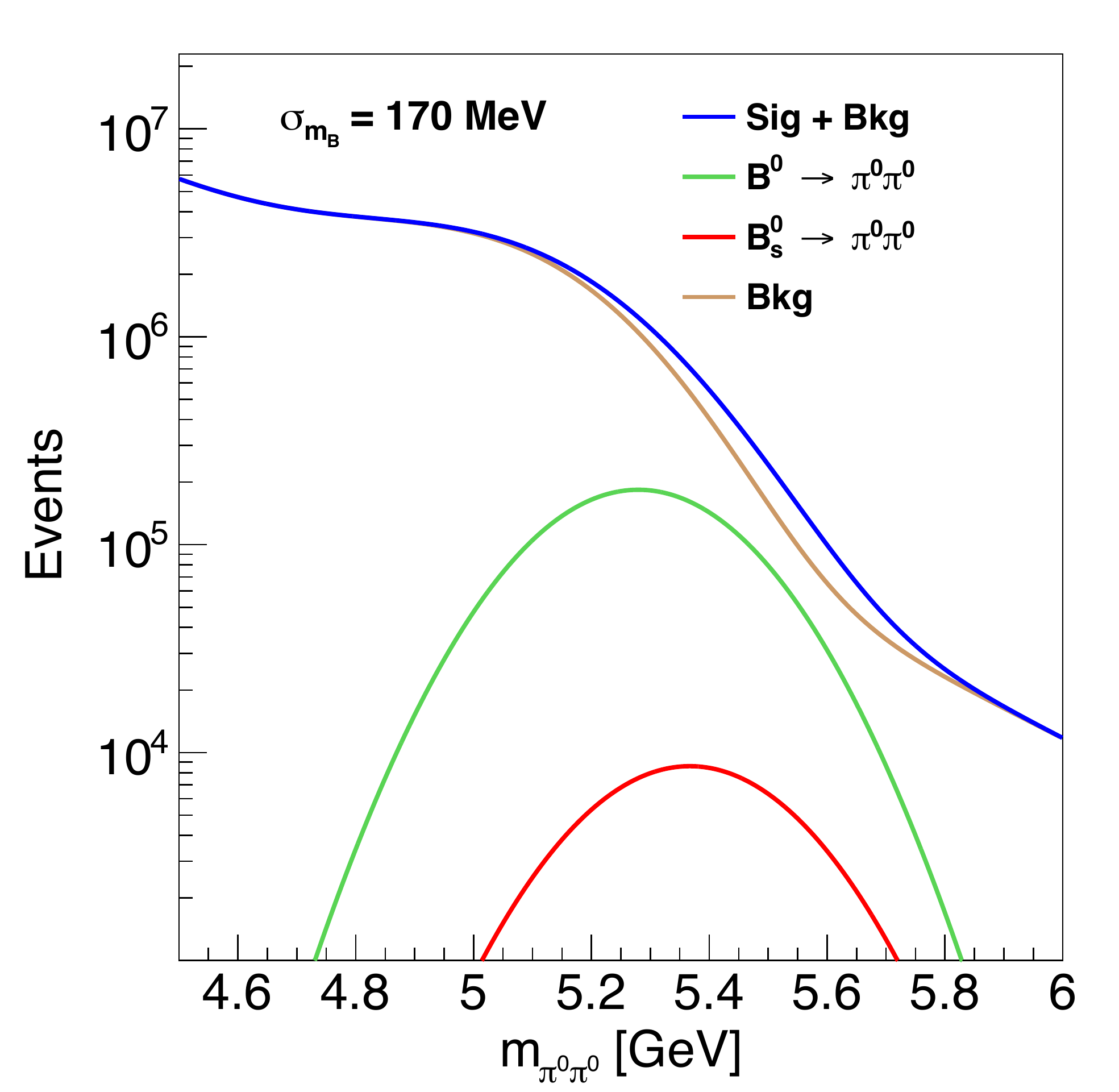}
	\caption{\MPioPio\ distributions of \Bopio, \Bspio, and \Zqq\ background at different $B$-meson mass resolutions when applying the baseline $b$-tagging.}
	\label{fig:SigBkgPDF}
\end{figure}

\begin{figure}[htbp]
	\centering
	\includegraphics[width=0.45\textwidth]{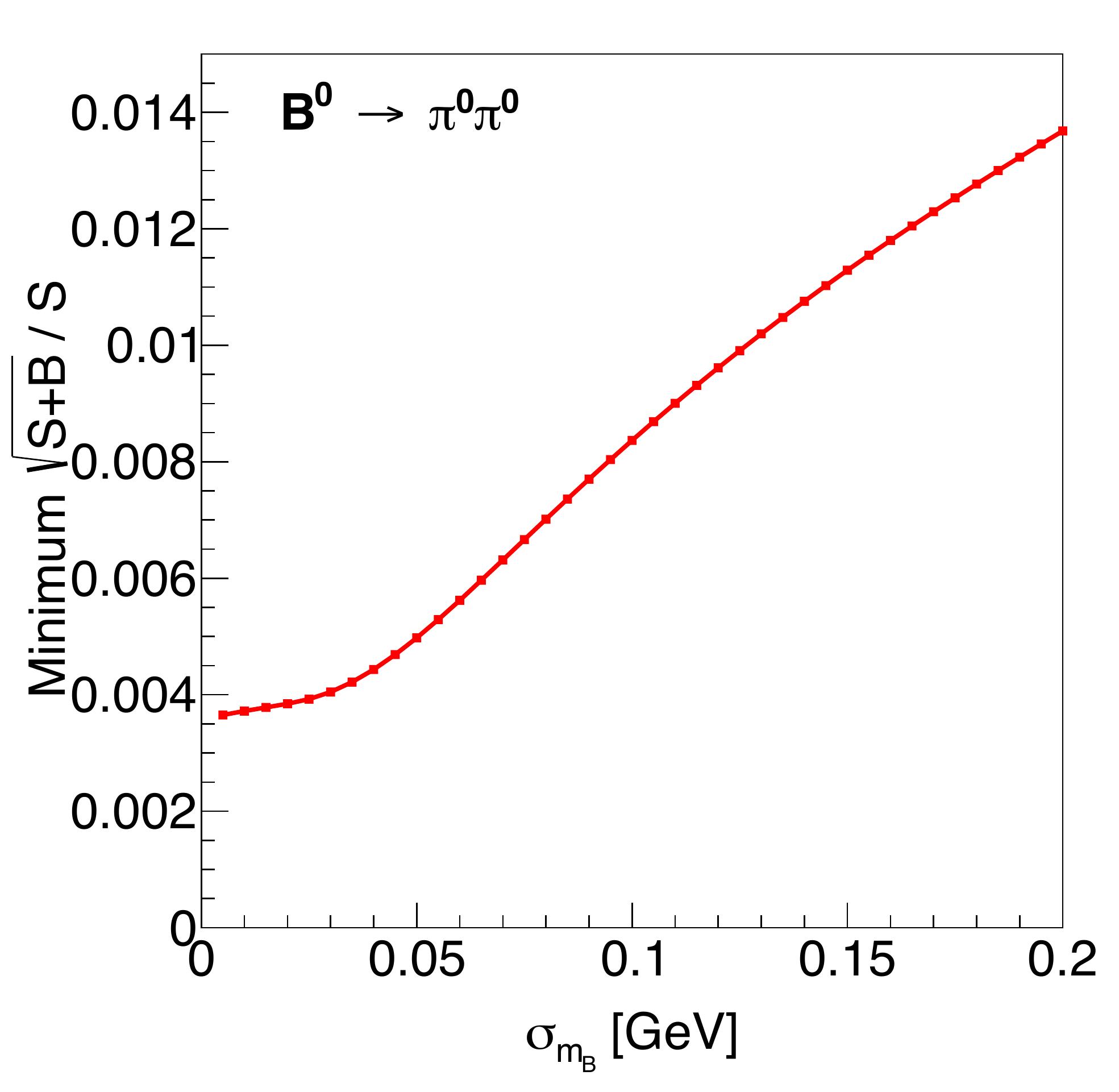}
	\includegraphics[width=0.45\textwidth]{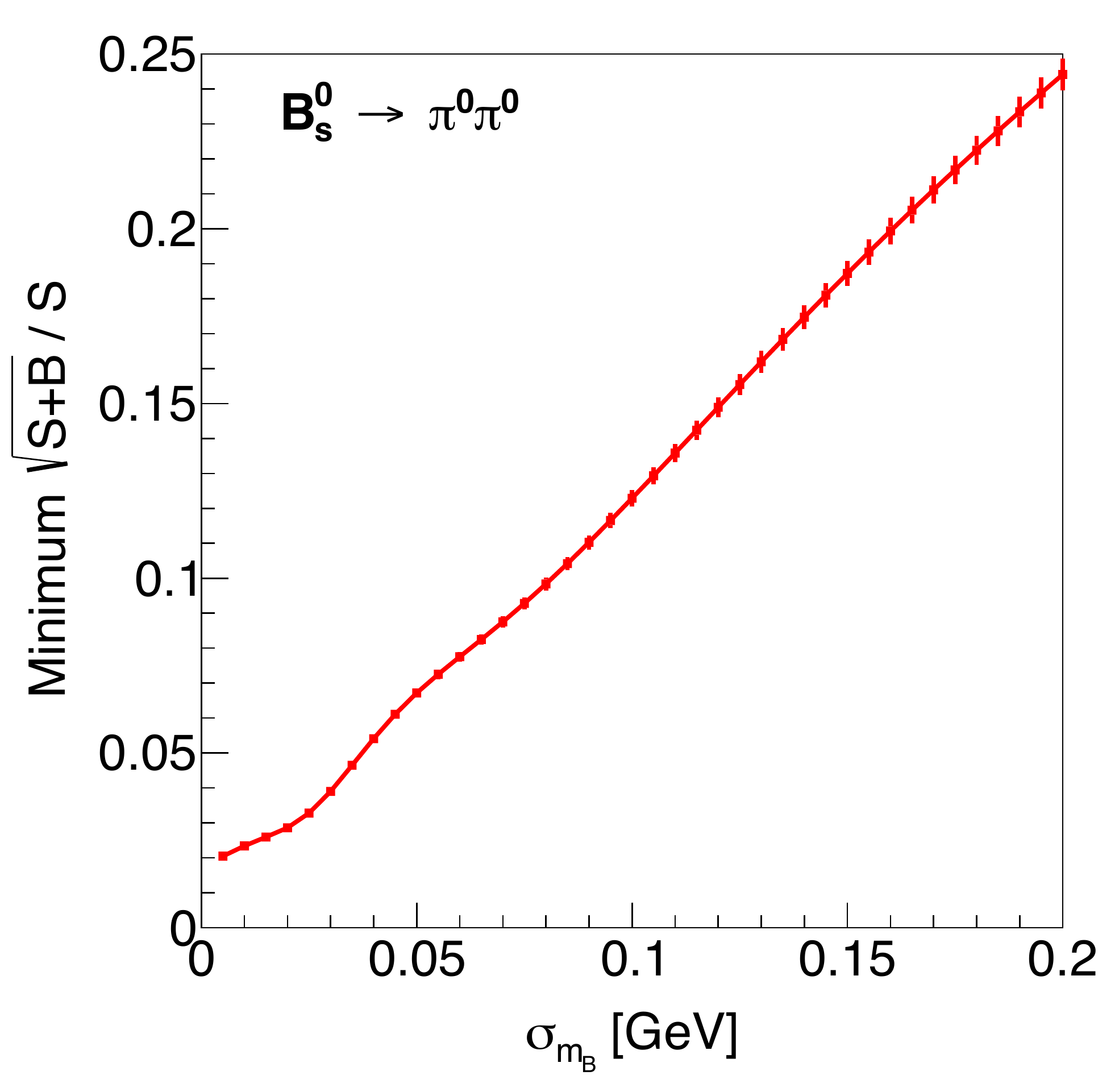}
	\caption{Measurement precision of $B^{0} \to \pi^{0}\pi^{0}$ (left) and $B^{0}_{s} \to \pi^{0}\pi^{0}$ (right) versus $B$-meson mass resolution \SigmaB.}
	\label{fig:AccuracyToyMC}
\end{figure}

Neglecting the effect from the fake \pio\ background, we develop a simplified model to evaluate the dependence of \Bospio\ measurement precision on the ECAL performance. We study the MCTruth \MPioPio\ distribution of the background after applying the selection criteria (except the invariant mass selection) used in table~\ref{tab:Pi0CutChain} and model it using the combination of an ARGUS function and an exponential function as described in section~\ref{sec:CutChain_Pi0}. The ECAL performance is parameterized as the Gaussian $B$-meson mass smearing \SigmaB, which is applied to the MCTruth background model. Two signal modes are described by two Gaussian distributions with a common standard deviation equals to \SigmaB\ and central values equal to $m_{B_{(s)}^0}$. The final \MPioPio\ distributions of \Bopio, \Bspio, and the background are derived by normalizing the above distributions to their respective yields at Tera-$Z$. In addition, the $b$-tagging efficiency of 80\% and the assumption of a constant signal selection efficiency of 50\% (including the \pio\ reconstruction efficiency of 98\% and the energy and angular selection efficiency of 52\%) are applied. Consequent distributions at different \SigmaB\ are shown in figure \ref{fig:SigBkgPDF}. Corresponding relative precisions are derived via the optimal mass window method mentioned in section~\ref{sec:CutChain_Pi0} and are plotted in figure \ref{fig:AccuracyToyMC}. For the reference detector setup with corresponding \SigmaB\ $\approx$ 30\,MeV, relative precisions of \Bopio\ and \Bspio\ are about 0.4\% and 4\%, respectively. For the typical regular ECAL energy resolution of $\frac{17\%}{\sqrt{E}} \oplus 1\%$ with corresponding \SigmaB $\approx$ 170\,MeV, the relative precisions of \Bopio\ and \Bspio\ are about 1.2\% and 21\%, respectively. These results are all consistent with the values given in table~\ref{tab:Accuracy_ECALReso}.

As the $B$-meson mass resolution further deteriorates, the sensitivities get even worse for both \Bopio\ and \Bspio, since the backgrounds are larger. The signal separation between two modes also weakens, making them pollute each other and further reducing their signal purities. For the precision of \Bopio, the influence from \Bspio\ is minor. The main contribution is from the \bb\ background, especially the three-body background. When \SigmaB\ increases to 20\,MeV, the three-body background becomes significant in the signal region of \Bopio\ and degrades the measurement sensitivity. The measurement of \Bspio\ suffers from both \Bopio\ and \qq\ background. When \SigmaB\ $<$ 15\,MeV, \Bo\ and \Bs\ are fully distinguishable. The main background is the combinatorial background from \qq\ events. As \SigmaB\ goes beyond 15\,MeV, \Bopio\ events start to pollute the \Bspio\ signal region. Finally, when \SigmaB\ > 40\,MeV, the three-body background is also involved. The different starting points of the contributions from \Bopio\ and the three-body background result in two inflection points on the \SigmaB-precision plot of \Bspio\ in figure~\ref{fig:AccuracyToyMC}.

\section{Summary and outlook}
\label{sec:Conclusion}

The Tera-$Z$ phase of future circular $e^+e^-$ colliders such as CEPC and FCC-ee, shows enormous potential in testing the SM and searching for new physics. In this paper, we focus on four neutral charmless $B$ decay channels, i.e. \Bopio, \Bspio, \Boeta, and \Bseta, to explore the potential of Tera-$Z$ in flavor physics. As a lepton collider, the clean environment at Tera-$Z$ allows precise measurements on these decay modes via their four-photon final states. During the process, the ECAL plays a crucial role in reconstructing the neutral final state $\pi^0$ and $\eta$ ($\to\gamma\gamma$). The separation between $B^0$ and $B^0_s$ in these cases also calls for an ECAL with high resolution.
We thus propose a reference ECAL energy resolution to achieve 30\,MeV $B$-meson mass resolution which can guarantee 3\,$\sigma$ separation between $B^0$ and $B^0_s$.
Furthermore, considering the $b$-jet tagging performance with an efficiency of 80\% and a purity of 90\%, we anticipate the relative precisions of signal strength measurements of \Bopio, \Bspio, \Boeta, and \Bseta\ at Tera-$Z$ to be 0.45\%, 4.5\%, 18\%, and 0.95\%, respectively.

We then discuss the measurements of $B^0\to\pi^0\pi^0$ and its two isospin-related modes, namely $B^0\to\pi^+\pi^-$ and $B^+\to\pi^+\pi^0$, in more detail. Branching ratios and $CP$ asymmetries of these three $B\to\pi\pi$ channels are necessary inputs to determine the CKM angle $\alpha$ ($\phi_2$). The current extraction of $\alpha$ from $B\to\pi\pi$ suffers from the relatively large uncertainty on $\pi^0\pi^0$ mode and the absence of measurement of the corresponding mixing-induced $CP$ asymmetry ($S_{CP}^{00}$), leading to multiple mirror solutions. Aside from the high \Bopio\ precision at Tera-$Z$, the time-integrated $CP$ asymmetry at the $Z$-pole uniquely includes the information of $S_{CP}^{00}$ due to the incoherent $b\bar{b}$ production from $Z$ decays. It thus provides an alternative and more precise way to extract $S_{CP}^{00}$ when combined with $B$-factory results, aside from the time-dependent measurement of \Bopio\ via low-statistics $\pi^0$ Dalitz decay or photon conversion.
More generally, the improved precision on $B\to\pi\pi$ measurements at Tera-$Z$ can be expected to help achieve unprecedented precision on $\alpha$.

With an effective tagging efficiency (power) of 15--25\% on the $b$-flavor charge tagging (jet charge measurement), the remarkable precision of 0.45\% on the $B^0\to\pi^0\pi^0$ branching ratio leads to statistical uncertainties of 0.014--0.018 on the time-integrated $CP$ asymmetry $a_{CP}^{00}$ (amounting to a linear combination of the direct $CP$ asymmetry $C_{CP}^{00}$ and the mixing-induced $CP$ asymmetry $S_{CP}^{00}$). Similar analyses on the other two $\pi\pi$ channels derive relative uncertainties of 0.18\% and 0.19\% on $\mathcal{B}(B^0\to\pi^+\pi^-)$ and $\mathcal{B}(B^+\to\pi^+\pi^0)$, as well as the statistical uncertainty of 0.004--0.005 on both $CP$ asymmetries $C_{CP}^{+-}$ and $S_{CP}^{+-}$. All these measurements at Tera-$Z$ are expected to achieve unprecedented improvements over the current world average. Compared to Belle II, Tera-$Z$ still improves the precision by a factor of 2--3. The anticipated precision on $C_{CP}^{+-}$ and $S_{CP}^{+-}$ at Tera-$Z$ is comparable to that at the upgraded LHCb.
 
We then consider these improvements in the framework of the CKM global fit and discuss the extraction of the CKM angle $\alpha$ from $\pi\pi$ modes under two scenarios. 
The results show that the sole improvement in $B^0\to\pi^0\pi^0$ observables (both $\mathcal{B}$ and $a_{CP}^{00}$) leads to a sharpening of the $\alpha$ determination from $\pi\pi$ modes alone, with a significant improvement in its precision from 13.6$^\circ$ (world average, $\pi\pi$ only) to 2--3$^\circ$.
The improvement in $a_{CP}^{00}$ proves essential for this reduction of the uncertainty on $\alpha$, while the combination of $a_{CP}^{00}$ from Tera-$Z$ and $C_{CP}^{00}$ from Belle (II) can already remove some of the ambiguities (mirror solutions) in $\alpha$. If these measurements are combined with inputs from charged-mode observables with improved precisions at Tera-$Z$, the landscape changes even more dramatically: the degeneracy among mirror solutions is further lifted, leading to an extraction of $\alpha$ from $\pi\pi$ mode at the level of $0.4^\circ$. 
This precision is then 5 times as high as the anticipated precision of $\sim$2$^\circ$ extrapolated from $\pi\pi$ data only at Belle II (50\,ab$^{-1}$).
Even compared to the current world average precision of $\sim$4.2$^\circ$ extracted from the combination of $\pi\pi$, $\rho\rho$, and $\rho\pi$ modes, the precision of 0.4$^\circ$ is almost 10 times higher.
Considering that the current direct determination of $\alpha$ is dominated by the $\rho\rho$ data, Tera-$Z$ has clearly a considerable potential in the determination of the $\alpha$ angle, which remains to be investigated beyond the scope of this paper.
In addition, the result in this paper is only the extrapolation based on the statistic of one Tera-$Z$.
According to the ongoing updates to the higher-luminosity accelerator design, which will result in more than three to five Tera-$Z$, we expect the future precision can be further improved.
For $\alpha$ extracted with a precision of a few tenths of a degree, it also proves necessary to reassess the theoretical systematic uncertainties associated with the isospin approach used to extract $\alpha$. Some of these systematic uncertainties can be evaluated using the isospin-breaking related modes $B^{+,0}\to \pi^{+,0}\eta(')$. 
In all these discussions, the input of $B\to \pi^0\pi^0$ $CP$ asymmetries, i.e. $a^{00}_{CP}$, $C^{00}_{CP}$, and $S^{00}_{CP}$, plays a central role and their precise measurements will impact significantly the determination of $\alpha$ from $\pi\pi$ data.


From the point of view of detector requirements, we further analyze the dependence of the signal strength precision of \Bospio\ and \Boseta\ on various detector performances. For these four channels produced in \bb\ events, the $b$-tagging is essential to reduce the non-\bb\ background. Compared to the case without $b$-tagging, the baseline $b$-tagging performance can improve the measurement precision of \Bospio\ and \Boseta\ by 2--4 times. The ECAL energy resolution is also a key detector parameter to efficiently reconstruct the neutral final state $\pi^0$ and $\eta$. It is crucial to the separation between $B^0$ and $B^0_s$, as well as the separation of \Bospio\ from the background of multi-body $b$-hadron decays. Compared to the currently typical regular ECAL energy resolution of $\frac{17\%}{\sqrt{E}} \oplus 1\%$ at Tera-$Z$, the proposed one can improve the measurement precision of \Bospio\ and \Boseta\ by a factor of 3--5.

In principle, the good EM resolution will have positive impact on almost all physics measurements with photon or $\pi^0$ in the final state, for example, the precision measurement of $H\to \gamma\gamma$ or searches for massive resonances via multiphoton final state. Since the photon is a critical ingredient of jets, improving the EM resolution also helps to improve the measurement of hadronic final states.
It must also be emphasized that our proposed reference calorimeter performance is indeed challenging for our current detector technology. Despite an excellent intrinsic resolution (almost 1.5--2 times better than any ECAL we have ever constructed), it also aims at a separation power comparable to that has been achieved through high granularity detector (like CALICE~\cite{CALICE_ECAL_2022} or CMS HGCAL~\cite{CMS_HGCAL_2017,CMS_HGCAL_TDR}). Of course, it is very challenging to design and commission such a detector. To achieve this, we need to rely on innovative detector and advanced reconstruction algorithm designs, future technology R\&D, and the continuous endeavour of our community.

In fact, the future electron-positron Higgs factory (such as CEPC, FCC, ILC, CLIC, etc.), has multiple scientific objectives and extremely demanding requirement on its detector system. A suited calorimeter shall provide excellent separation power, good intrinsic resolution, and low energy threshold. Each requirement is driven by a series of physics measurements (e.g. excellent separation power for measurements with jets and flavor physics with critical objects inside jets, good intrinsic resolution for relevant flavor physics, low energy threshold for QCD and flavor studies, etc.), while the final design of the whole detector shall optimize and balance all these requirements within the boundary condition, i.e., compatibility with the collision environment, construction cost, and technology robustness.

Our work thus illustrates the potential impact of the Tera-$Z$ on the precise measurements of \Bospio\ and \Boseta\ modes and their $CP$ asymmetries. It opens the possibility of testing the consistency of the SM with unprecedented precision and in a unique way. We believe that our analysis supports a strong physics case at Tera-$Z$ and shows its synergy with other experiments in the global picture of flavor physics.
As one case that represents the requirement for high intrinsic resolution of EM showers, our study also provides a clear physics case for the impact of good EM resolution.
We hope it can provide sufficient reference, together with benchmark studies regarding other requirements, to guide future detector design.

\appendix
\acknowledgments

We would like to thank Gang Li for the generator samples of the SM background and the full simulation study of the CEPC baseline $b$-tagging. 
We also thank Yudong Wang for the signal sample generation, and Xu-Chang Zheng for discussions on $B$ meson productions.
We are grateful to Shan Cheng and Chunhui Chen for their beneficial discussions on the status of $B\to\pi\pi$ study. 
This project is supported by the Institute of High Energy Physics, Chinese Academy of Sciences (E2545AU210, E15152U110). It has also received the support from the European Union’s Horizon 2020 research and innovation programme under the Marie Skodowska-Curie grant agreement No 860881-HIDDeN. L.~Li is supported by the DOE grant DE-SC-0010010. 

\bibliographystyle{JHEP}
\bibliography{reference}
\end{document}